\documentclass{article} % For LaTeX2e
\usepackage{iclr2023_conference,times}
\usepackage{float}
\usepackage{epsfig}
\usepackage{graphicx}
\usepackage{enumitem}

\usepackage{xspace}
\usepackage{graphics}
\usepackage{subfigure}
\usepackage{wrapfig}
\usepackage{multirow}
\usepackage{booktabs}
\usepackage{tablefootnote}
\newcommand{\tabincell}[2]{\begin{tabular}{@{}#1@{}}#2\end{tabular}} 
\usepackage{multirow,booktabs}
\usepackage{comment}
\def\eg{\textit{e.g.}}

\usepackage{amsmath,amsfonts,bm}

\def\eqref#1{equation~\ref{#1}}
\def\1{\bm{1}}

\DeclareMathAlphabet{\mathsfit}{\encodingdefault}{\sfdefault}{m}{sl}
\SetMathAlphabet{\mathsfit}{bold}{\encodingdefault}{\sfdefault}{bx}{n}

\newcommand{\name}{{SCALE-UP}}%h
\usepackage{hyperref}
\hypersetup{colorlinks,linkcolor={blue},citecolor={blue}} 
\usepackage{url}
\usepackage[capitalize]{cleveref}
\crefname{section}{Sec.}{Secs.}
\Crefname{section}{Section}{Sections}
\Crefname{table}{Table}{Tables}
\crefname{table}{Tab.}{Tabs.}

\usepackage{subfigure}
\usepackage{hyperref}
\usepackage{graphicx}
\usepackage{algorithm,algpseudocode,amssymb}

\usepackage[thmmarks,  thref, amsmath]{ntheorem}
\theoremseparator{.}

\newtheorem{theorem}{Theorem}

\usepackage[thmmarks,  thref, amsmath]{ntheorem}
\usepackage {xcolor}

\title{SCALE-UP: An Efficient Black-box Input-level Backdoor Detection via Analyzing Scaled Prediction Consistency}

\author{Junfeng Guo$^{1}$\footnotemark[1]~~\footnotemark[2]~, Yiming Li$^{2}$\footnotemark[1]~, Xun Chen$^{3}$\footnotemark[3]~, Hanqing Guo$^{4}$\footnotemark[2]~, Lichao Sun$^{5}$, Cong Liu$^{6}$  \\
$^{1}$Department of Computer Science, UT Dallas~\\$^2$Tsinghua Shenzhen International Graduate School, Tsinghua University \\$^{3}$Samsung Research America, Mountain View\\ $^{4}$Department of Computer Science, Michigan State University\\$^{5}$Department of Computer Science, Lehigh University \\$^{6}$Department of Electricity and Computer Engineering Department, UC Riverside\\
} 
\begin{document}

\maketitle

\renewcommand{\thefootnote}{\fnsymbol{footnote}}
\footnotetext[1]{The first two authors contributed equally to this paper.}
\footnotetext[2]{This work was done when Junfeng Guo and Hanqing Guo interned in Samsung Research America.}
\footnotetext[3]{
Corresponding Author: Xun Chen (e-mail: \href{mailto:xun.chen@samsung.com}{xun.chen@samsung.com}).
}
\begin{abstract}
Deep neural networks (DNNs) are vulnerable to backdoor attacks, where adversaries embed a hidden backdoor trigger during the training process for malicious prediction manipulation. These attacks pose great threats to the applications of DNNs under the real-world machine learning as a service (MLaaS) setting, where the deployed model is fully black-box while the users can only query and obtain its predictions. Currently, there are many existing defenses to reduce backdoor threats. However, almost all of them cannot be adopted in MLaaS scenarios since they require getting access to or even modifying the suspicious models. In this paper, we propose a simple yet effective black-box input-level backdoor detection, called SCALE-UP, which requires only the predicted labels to alleviate this problem. Specifically, we identify and filter malicious testing samples by analyzing their prediction consistency during the pixel-wise amplification process. Our defense is motivated by an intriguing observation (dubbed \emph{scaled prediction consistency}) that the predictions of poisoned samples are significantly more consistent compared to those of benign ones when amplifying all pixel values. Besides, we also provide theoretical foundations to explain this phenomenon. Extensive experiments are conducted on benchmark datasets, verifying the effectiveness and efficiency of our defense and its resistance to potential adaptive attacks. Our codes are available at \url{https://github.com/JunfengGo/SCALE-UP}.

\end{abstract}

\section{Introduction}
Deep neural networks (DNNs) have been deployed in a wide range of mission-critical applications, such as autonomous driving \citep{physGAN,grigorescusurvey,wendeep}, face recognition \citep{tangframe,lilearning,yanglarnet}, and object detection \citep{object_detection,zouobject,wangsalient}. In general, training state-of-the-art DNNs usually requires extensive computational resources and training samples. Accordingly, in real-world applications, developers and users may directly exploit third-party pre-trained DNNs instead of training their new models. This is what we called machine learning as a service (MLaaS).

%Deep neural networks (DNNs) have been deployed in a wide range of misson-critical applications, such as autonomous driving \citep{physGAN}, face recognition \citep{deepid} and object detection \citep{object_detection}. However, training a state-of-art DNN typically requires extensive computation resource and clean training data. Therefore, the real-world model user would like to use pre-trained DNN models from the third-party services ~\citep{AEVA,caffe_zoo,chen2019cloud,dong2021black} instead of training by themselves. 

However, recent studies \citep{DBLP:journals/corr/abs-1708-06733,goldblumdataset,survey} revealed that DNNs can be compromised by embedding adversary-specified hidden backdoors during the training process, posing threatening security risks to MLaaS. The adversaries can activate embedded backdoors in the attacked models to maliciously manipulate their predictions whenever the pre-defined trigger pattern appears. Users are hard to identify these attacks under the MLaaS setting since attacked DNNs behave normally on benign samples.

%Recent work~\cite{Trojannn,ISSBA,wanet,DBLP:journals/corr/abs-1708-06733} find that DNNs can be compromised by injecting specific backdoor triggers during the training phase. 
%As for the inference phase, inputs attached with the backdoor trigger would cause target misclassification to the infected DNNs. Meanwhile, the infected DNNs can achieve similar accuracy on benign inputs as the uninfected DNNs. Such backdoor attacks pose great threaten to the application of machine learning as a service~\citep{model_reuse,DBLP:journals/corr/abs-1708-06733,Trojannn}.

To reduce backdoor threats, there are many different types of defenses, such as model repairing \citep{lineural,wuadversarial,unlearn}, poison suppression \citep{durobust,anti,huangbackdoor}, and backdoor detection \citep{xiangpost,liucomplex,AEVA,rlbackdoor}. However, most of these defenses were designed under the white-box setting, requiring accessing or even modifying model weights. Accordingly, users cannot adopt them in MLaaS scenarios. Currently, there are also a few model-level \citep{huangone,dongblack,AEVA} and input-level \citep{li2021backdoor,deepweep,gaodesign} black-box backdoor defenses where users can only access final predictions. However, these defenses have some implicit assumptions of backdoor triggers ($e.g.$, a small static patch), leading to being easily bypassed by advanced backdoor attacks \citep{wanet,ISSBA}. %How to effectively defend against backdoor attacks under the black-box setting is still an important open question.
Their failures raise an intriguing question: \emph{what are the fundamental differences between poisoned and benign samples that can be exploited to design universal black-box backdoor detection?}

\begin{figure}[!t]
    \vspace{-3em}
    \centering
    \includegraphics[width=0.98\textwidth]{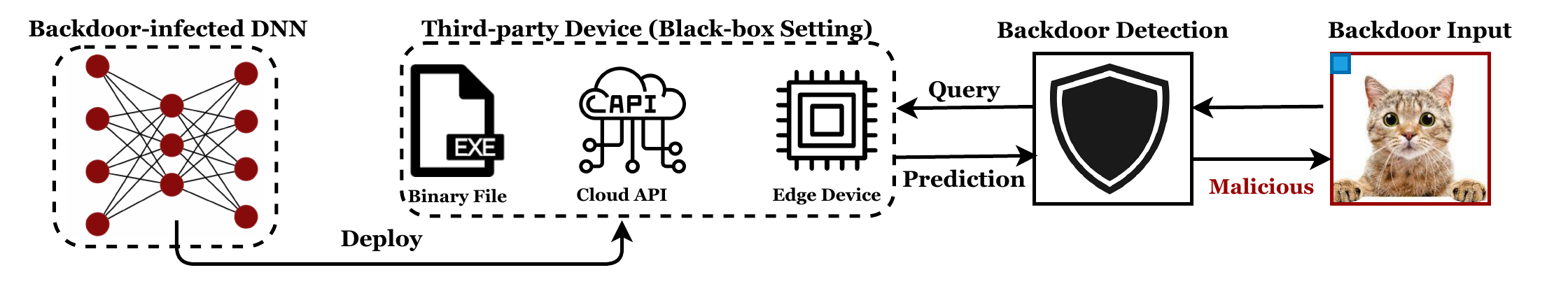}
    \vspace{-1em}
    \caption{An illustration of the black-box input-level backdoor detection.} 
    \label{fig:motivation}
    \vspace{-1em}
\end{figure}

In this paper, we focus on the \emph{black-box input-level backdoor detection}, where we intend to identify whether a given suspicious input is malicious based on predictions of the deployed model (as shown in \cref{fig:motivation}). This detection is practical in many real-world applications since it can serve as the `firewall' helping to block and trace back malicious samples in MLaaS scenarios. However, this problem is challenging since defenders have limited model information and no prior knowledge of the attack. Specifically, we first explore the pixel-wise amplification effects on benign and poisoned samples, motivated by the understanding that increasing trigger values does not hinder or even improve the attack success rate of attacked models (as preliminarily suggested in \citep{li2021backdoor}). We demonstrate that the predictions of attacked images generated by both classical and advanced attacks are significantly more consistent compared to those of benign ones when amplifying all pixel values. We refer to this intriguing phenomenon as \emph{scaled prediction consistency}. In particular, we also provide theoretical insights to explain this phenomenon. After that, based on these findings, we propose a simple yet effective method, dubbed scaled prediction consistency analysis (SCALE-UP), under both \emph{data-free} and \emph{data-limited} settings. Specifically, under the data-free setting, the SCALE-UP examines each suspicious sample by measuring its \textit{scaled prediction consistency} (SPC) value, which is the proportion of labels of scaled images that are consistent with that of the input image. The larger the SPC value, the more likely this input is malicious. Under the data-limited setting, we assume that defenders have a few benign samples from each class, based on which we can reduce the side effects of class differences to further improve our SCALE-UP.

In conclusion, our main contributions are four-fold. \textbf{1)} We reveal an intriguing phenomenon ($i.e.$, scaled prediction consistency) that the predictions of attacked images are significantly more consistent compared to those of benign ones when amplifying all pixel values. \textbf{2)} We provide theoretical insights trying to explain the phenomenon of scaled prediction consistency. \textbf{3)} Based on our findings, we propose a simple yet effective black-box input-level backdoor detection (dubbed `SCALE-UP') under both data-free and data-limited settings. \textbf{4)} We conduct extensive experiments on benchmark datasets, verifying the effectiveness of our method and its resistance to potential adaptive attacks.

\section{Related Work}
\label{sec:related_work}

\subsection{Backdoor Attack}
Backdoor attacks \citep{DBLP:journals/corr/abs-1708-06733,survey,hayasefew} compromise DNNs by contaminating the training process through injecting poisoned samples. These samples are crafted by adding adversary-specified trigger patterns into the selected benign samples. Backdoor attacks are stealthy since the attacked models behave normally on benign samples and the adversaries only need to craft a few poisoned samples. Accordingly, they introduce serious risks to DNN-based applications. In general, existing attacks can be roughly divided into two main categories based on the trigger property, including \textbf{1)} patch-based attacks and \textbf{2)} non-patch-based attacks, as follows:

\noindent \textbf{Patch-based Backdoor Attacks.} \citep{DBLP:journals/corr/abs-1708-06733} proposed the first backdoor attack, which was called BadNets. Specifically, BadNets randomly selected and modified a few benign training samples by stamping the trigger patch and changing their label with a pre-defined target label. The generated poisoned samples associated with the remaining benign samples will be released to users to train their models. After that, \citep{chen} first discussed the attack stealthiness and introduced trigger transparency, where they suggested that poisoned images should be indistinguishable compared with their benign version to evade human inspection. \citep{label_consis} argued that making trigger patches invisible is not stealthy enough since the ground-truth labels of poisoned samples are different from the target label. They designed the first clean-label backdoor attack where adversaries can only poison samples from the target class. Recently, \citep{li2021backdoor} proposed the first physical backdoor attack, where the location and appearance of the trigger contained in the digitized test samples may be different from that of the one used for training.

%Backdoor attacks\citep{Trojannn,DBLP:journals/corr/abs-1708-06733,label_consis,wanet,ISSBA,survey} compromise DNNs via contaminating the training dataset by injecting poisoned samples. These samples are crafted by adding adversary-specified trigger patterns into the selected benign training samples. Backdoor attacks are stealthy since the attacked models behave normally on benign samples while the adversaries only need to craft a few poisoned samples, posing serious risks to machine learning as a service (MLaaS).

%BadNets~\citep{DBLP:journals/corr/abs-1708-06733} first proposes to generate sparse patch-based trigger to cause misclassification on DNNs. \cite{chen} and \cite{tuap} further propose to perform backdoor attacks with a dense but invisible trigger. \cite{label_consis} proposes the label-consistent backdoor attack which contaminates the training data using correctly-labeled data. Most recently, a set of advanced backdoor attacks~\citep{wanet,ISSBA} are proposed, which can generate input-specific and invisible backdoor triggers to bypass detection~\citep{ISSBA,wanet}.

\noindent \textbf{Non-patch-based Backdoor Attacks.} Different from classical attacks whose trigger pattern is a small patch, recent advanced methods exploited non-patch-based triggers trying to bypass backdoor defenses. For example, \citep{zhaoclean} exploited full-image size targeted universal adversarial perturbation \citep{moosaviuniversal} as the trigger pattern to design a more effective clean-label backdoor attack. \citep{wanet} adopted image warping as the backdoor trigger, which deforms the whole image while preserving image content. Recently, \citep{ISSBA} proposed the first poison-only sample-specific trigger patterns, inspired by the DNN-based image steganography \citep{tancikstegastamp}. This attack broke the fundamental assumption ($i.e.$, the trigger is sample-agnostic) of most existing defenses and therefore could easily bypass them.

\subsection{Backdoor Defense}
Currently, there are many backdoor defenses to alleviate backdoor threats. In general, existing methods can be roughly divided into two main categories based on the defender's capacities, including \textbf{1)} white-box defenses and \textbf{2)} black-box defenses, as follows:

\noindent\textbf{White-box Backdoor Defenses.} In these approaches, defenders need to obtain the source files of suspicious models. The most typical white-box defenses are model repairing, aiming at removing hidden backdoors in the attacked DNNs. For example, \citep{liufine,wuadversarial} proposed to remove backdoors based on model pruning; \citep{lineural,xiaeliminating} adopted model distillation to eliminate hidden backdoors. There are also other types of white-box defenses, such as poison suppression \citep{durobust,anti,huangbackdoor} and trigger reversion \citep{nc,hutrigger,taobetter}. However, users cannot use them under the machine learning as a service (MLaaS) setting, where they can only obtain model predictions.

%\red{I will start from here later (Yiming, 0924)}

\noindent\textbf{Black-box Backdoor Defenses.} In these methods, defenders can only query the (deployed) model and obtain its predictions. Currently, there are two main types of black-box defenses, including \textbf{1)} model-level defenses \citep{huangone,dongblack,AEVA} and \textbf{2)} input-level defenses \citep{li2021backdoor,deepweep,gaodesign}. Specifically, the former ones intend to identify whether the (deployed) suspicious model is attacked while the latter ones detect whether a given suspicious input is malicious. In this paper, we focus on the input-level black-box defense since it can serve as the `firewall' helping to block and trace back malicious samples in MLaaS scenarios. However, as we will demonstrate in our experiments, existing input-level defenses can be easily bypassed by advanced backdoor attacks since they have some strong implicit assumptions of backdoor triggers. How to design effective black-box input-level backdoor detectors is still an important open question and worth further exploration.

%\noindent\textbf{Backdoor Defense.} Backdoor defense approaches can be categorized into \textit{model-level}, \textit{input-level} and \textit{training-set level}. Model-level defense approaches~\citep{nc,guo2019tabor,liu2019abs,wang2020practical,AEVA} aim to identify a given DNNs infected or not. \textit{training-set level} defenses~\citep{anti,dbd,unlearn,unlearn,du2020robust} aim to either sanitize the poisoned training dataset or eliminating the effect of backdoor samples. As for \textit{input-level} backdoor defenses~\citep{deepweep,rethinking,strip,frequency,ac} predict whether an input will trigger some Trojan behavior on an untrust DNN.  Neuron activation filtering~\citep{ac,nc} can identify the backdoor sample by comparing its corresponding neuron activation with benign samples. However, these approaches rely on the access to the parameters of the target DNN. STRIP~\citep{strip} assumes that the trigger is static across inputs thus identifies the backdoor sample via investigating the existence of input-agnostic features. ShrinkPad~\citep{rethinking} and DeepWeep~\citep{deepweep} propose to defend the backdoor samples via performing some relevant pre-processing techniques, such as flip, shrink and denoise. Frequency defense~\citep{frequency} finds that the backdoor trigger can become noticeable in the frequency domain. Therefore,\cite{frequency} proposes to distinguish the backdoor and benign samples in the frequency domain.  

\section{The Phenomenon of Scaled Prediction Consistency}
\label{sec:motication}
In this section, we explore the prediction behaviors of benign and poisoned samples generated by attacked DNNs since it is the cornerstone for designing black-box input-level backdoor defense. Before illustrating our key observations, we first review the general process of backdoor attacks.

\noindent\textbf{The Main Pipeline of Backdoor Attacks.} Let $\mathcal{D} = \{ (\bm{x}_i, y_i) \}_{i=1}^{N}$ represent a unmodified benign training set and $C: \mathcal{X} \rightarrow \mathcal{Y}$ is the deployed DNN, where $\bm{x}_i \in \mathcal{X}= [0,1]^{C\times W \times H}$ is the image, $y_i \in \mathcal{Y} = \{1,\ldots, K\}$ is its label, and $K$ is the number of different labels. The backdoor adversaries will select some benign samples ($i.e.$, $\mathcal{D}_s$) to generate their modified version by $\mathcal{D}_{m} = \left\{(\bm{x}', y_t)| \bm{x}' = \bm{x} + g(\bm{x}), (\bm{x},y) \in \mathcal{D}_s \right\}$, where $y_t$ is an adversary-specified target label and $g(\cdot)$ is a pre-defined poison generator.
For example, $g(\bm{x}) = \bm{m}\odot(\bm{t}-\bm{x})$ in BadNets \citep{DBLP:journals/corr/abs-1708-06733} and blended attack \citep{chen}, where $\odot$ represents the element-wise product, $m \in [0, 1]^{C\times W \times H}$ is a transparency mask, and $\bm{t} \in \mathcal{X}$ is the trigger pattern. Given $N_b$ benign samples and $N_p$ poisoned samples, the backdoor adversaries will train the attacked DNN $f(\cdot; \bm{\theta})$ based on the following optimization process (with loss $\mathcal{L}$):
\begin{equation}
    \label{eq:training}
    \min_{\bm{\theta}} \sum_{i=1}^{N_{b}} \mathcal{L}(f(\bm{x}_{i};\bm{\theta}),y_{i}) + \sum_{j=1}^{N_p} \mathcal{L}(f(\bm{x}'_{j};\bm{\theta}),y_t).
\end{equation}

As preliminarily demonstrated in \citep{li2021backdoor}, increasing the pixel value of backdoor triggers does not hinder or even improve the attack success rate. However, defenders can not accurately manipulate these pixel values since they have no prior knowledge about trigger location. Accordingly, we explore what will happen if we scale up all pixel values of benign and poisoned images.

\begin{figure}[!t]
\centering
\subfigure[Benign Model\label{fig:benign_emp}]{\includegraphics[width=0.3\textwidth]{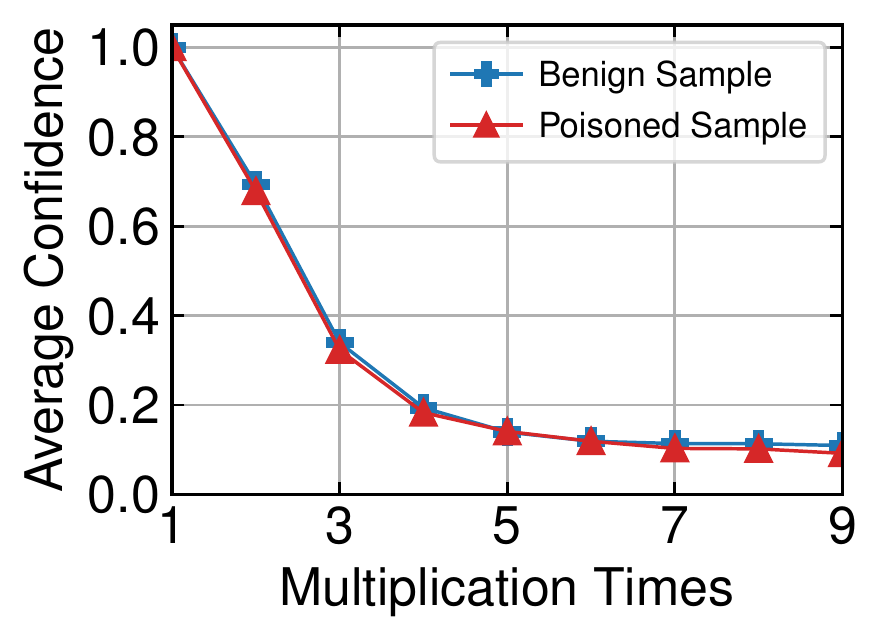}}\hspace{1em}
\subfigure[BadNets\label{fig:badnets_emp_emp}]{\includegraphics[width=0.3\textwidth]{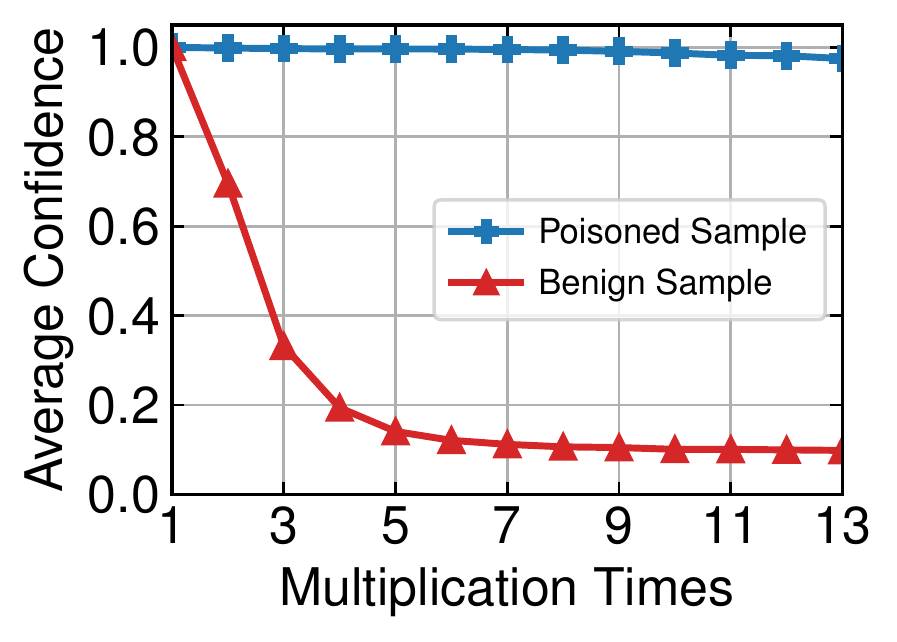}}\hspace{1em}
\subfigure[ISSBA\label{issba_emp}]{\includegraphics[width=0.3\textwidth]{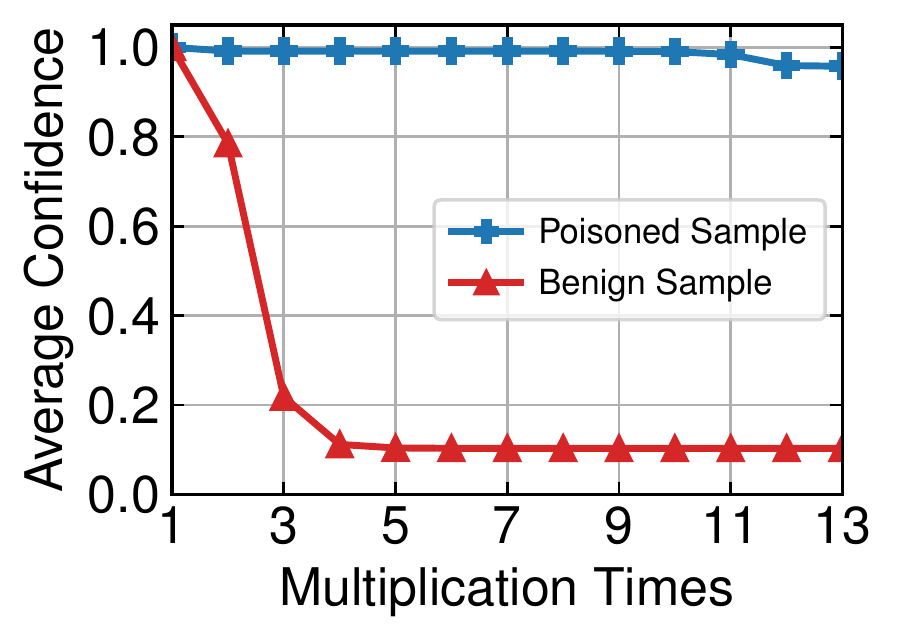}}
% \subfigure[Activation Similarity\label{infected_3d}]{\includegraphics[width=0.24\textwidth]{figure/empirical study/activation.pdf}}
\vspace{-0.5em}
\caption{The average confidence ($i.e.$, average probabilities on the originally predicted label) of benign and poisoned samples $w.r.t.$ pixel-wise multiplications under benign and attacked models.}
\vspace{-2mm}
\label{fig:intuition}
\end{figure}

\textbf{Settings.} In this section, we adopt BadNets \citep{DBLP:journals/corr/abs-1708-06733}) and ISSBA \citep{ISSBA} as the example for our discussion. They are representative of patch-based and non-patch-based backdoor attacks, respectively. We conduct experiments on the CIFAR-10 dataset \citep{CIFAR} with ResNet \citep{hedeep}. For both attacks, we inject a large number of poisoned samples to ensure a high attack success rate ($\geq 99\%$). For each benign and poisoned image, we gradually enlarge its pixel values with multiplication. We calculate the \emph{average confidence} defined as the average probabilities of samples on the originally predicted label. In particular, we select the label predicted upon the original sample as the originally predicted label for each varied sample and constrain all pixel values within $[0,1]$ during the 
multiplication process. More details are in our appendix.

%In the following experiments of this section, we implement a prevalent patch-based (\ie, BadNets~\citep{DBLP:journals/corr/abs-1708-06733}) and non-patch-based (\ie,ISSBA~\citep{ISSBA}) backdoor attacks to investigate our idea, respectively. We implement each attack on the ResNet-34~\citep{he2016deep} model and use CIFAR-10 as our experimental benchmark. For BadNets, we apply a 4x4 square consisting of random pixels as the trigger; As for ISSBA, we apply the default configurations and the invisible dynamic triggers as ~\cite{ISSBA}. For both attacks, we inject arbitrary amounts of poisoned samples to ensure the attack success rate ($ASR$) $\geq 99\%$.

\noindent \textbf{Results.} As shown in Figure \ref{fig:intuition}, the average confidence scores of both benign and poisoned samples decrease during the multiplication process under the benign model. In other words, the predictions of modified benign and poisoned samples are changed during this process. In contrast, poisoned and benign samples have significantly distinctive behaviors under attacked models. Specifically, the average confidence of benign samples decreases whereas that of poisoned samples is relatively stable with the increase of multiplication times. We call this phenomenon as \emph{scaled prediction consistency}.

%For each experiment, we first filter the samples classified incorrectly and then vary each backdoor/benign sample by gradually enlarging its pixel values by times. Notably, we select the label predicted upon the original sample as the target label for each varied sample and  constrain the pixel values within $[0,1]$ during the amplification process. The results are shown in \cref{fig:intuition}. From \cref{fig:intuition}, we observed that both the prediction confidence and the neuron activation similarity for backdoor inputs perform rather consistently compared with the benign samples against the amplification process under the infected model. As for the benign model, we find that the benign model performs similarly for the benign and backdoor samples, for which the prediction confidence drops drastically. We call it the \textit{Scaled Prediction Consistency Phenomenon}.        

% 
To further explain this intriguing phenomenon ($i.e.$, scaled prediction consistency), we exploit recent studies on neural tangent kernel (NTK) \citep{ntk} for analyzing the backdoor-infected models inspired by \citep{AEVA}, as follows: 

\begin{theorem}\label{lem1}
Suppose the poisoned training dataset consists of $N_b$ benign samples and $N_p$ poisoned samples, i.i.d. sampled from uniform distribution and belonging to $K$ classes. Assume that deep neural network $f(\cdot;\theta)$ be a multivariate kernel regression (RBF kernel) with the objective in \cref{eq:training}. For a given attacked sample $\bm{x}'= (\bm{1}-\bm{m})\odot \bm{x} + \bm{m}\odot \bm{t}$, we have: $\lim_{N_p\to N_b} C(n\cdot\bm{x}')=y_t, n
\geq1$.
\end{theorem}

\vspace{-0.5em}
In general, Theorem \ref{lem1} reveals that when the amount of poisoned samples closes to the benign samples or the attacked DNN over-fits the poisoned samples, it will still constantly predict the scaled attacked samples ($i.e.$, $n\cdot \bm{x}'$) as the target label $y_t$. Its proof can be found in \cref{sec:lemma1}. 

%Motivated by the \textit{Scaled Prediction Consistency Phenomenon}, we thus propose \textit{Scaled Prediction Consistency Analysis} approach.  

%%%%%%%%%%%%%分割线，下面是以前你写的

% We first investigate what happens to the prediction of the backdoor-infected model when the given inputs' pixel values (RGB format) increase by times.
% Since delivering a thoroughly theoretical analysis on DNNs is notoriously hard, we follow previous work~\citep{AEVA} to leverage recent studies on Neural Tangent Kernel (NTK) \citep{ntk} for analyzing the backdoor-infected models.

% The proof can be found in \cref{sec:lemma1}. Even though this Lemma can not directly explain the backdoor-infected model and leverage a basic backdoor attack~\citep{DBLP:journals/corr/abs-1708-06733} as previous work~\cite{AEVA}, but it can help us better understand such phenomenon.

% We also perform experiments to verify our intuition. We implement two popular backdoor attacks (\ie,BadNets~\citep{DBLP:journals/corr/abs-1708-06733} and ISSBA~\citep{ISSBA}), which represent a basic and advanced backdoor, respectively. We randomly select a label as the target label and use ResNet-34~\citep{he2016deep} to train the infected model on CIFAR-10 task, achieving attack success rate(ASR) $\geq 99\%$. The detailed configurations of this empirical study can be found in the \cref{sec:empirical_configuration}. 

%\section{Prediction Consistency Analysis for backdoor samples detection}

\section{Scaled Prediction Consistency Analysis (SCALE-UP)}
Motivated by the phenomenon of scaled prediction consistency demonstrated in the previous section, we propose a simple yet effective black-box input-level backdoor detection, dubbed scaled prediction consistency analysis (SCALE-UP), in this section.

%In this section, we introduce our proposed sample-wise backdoor detection framework, namely \name{}. We start by finding the connection between the trigger's pixel values and the prediction for a backdoor-infected DNN. The prediction consistency phenomenon is then revealed by both theoretical analysis and empirical studies. Inspired by empirical observations, the Scaled Prediction Consistency (SPC) measure is proposed, which is computed by the proportion of consistent prediction for inputs with  different increased magnitudes. The SPC score is then used to identify whether the given input is backdoored or not. We further improve \name{} via eliminating the variance of SPC scores across different classes under the scenario where limited data is accessible.

\subsection{Preliminaries}
\noindent \textbf{Defender's Goals.} In general, defenders have two main goals, including \emph{effectiveness} and \emph{efficiency}. Effectiveness requires that the detection method can accurately identify whether a given image is malicious or not. Efficiency ensures that detection time is limited and therefore the deployed model can provide final results on time after the detection and prediction process.

\noindent \textbf{Threat Model.} We consider backdoor detection under the black-box setting in machine learning as a service (MLaaS) applications. Specifically, defenders can only query the third-party deployed model and obtain its predictions. They do not have any prior information about the backdoor attack or the model. In particular, we consider two data settings, including \textbf{1)} \emph{data-free detection} and \textbf{2)} \emph{data-limited detection}. The former one assumes that defenders have no holding benign samples, while the latter one allows defenders to have a few benign samples from each class. Note that we only assume to have the predicted label instead of the predicted probability vector in our method.

%\noindent \textbf{Threat Model and Defender's Goal. }
%Our considered threat model contains two parts: the \textit{adversary} and the \textit{defender}. Consistent with previous work~\citep{DBLP:journals/corr/abs-1708-06733,Trojannn,chen,ISSBA}, the attacker can inject arbitrary amounts of poisoned samples in the training phase to ensure the attack efficacy. As for the defender, we consider a threat model with the weakest assumption, where both the poisoned training samples and target DNN's parameters are inaccessible. The defender can only obtain the final predictive label for each input. The defender treats the target black-box model as infected and aims to identify the backdoored sample at its inference phase. 

\subsection{Data-free Scaled Prediction Consistency Analysis}

As demonstrated in Section \ref{sec:motication}, we can use the average probability on the originally predicted label across its scaled images to determine whether a given suspicious image is malicious. In general, the larger the probability, the more likely the sample is poisoned. However, we can only obtain predicted labels while predicted probability vectors are inaccessible under our settings. Accordingly, we propose to examine whether the predictions of scaled samples are consistent, as follows:

Let $\mathcal{S}$ denotes a defender-specified scaling set ($e.g.$, $\mathcal{S}=\{3,5,7,9,11\}$). For a given input image $\bm{x}$ and the deployed classifier $C$, we define its \emph{scaled prediction consistency (SPC)} as the proportion of labels of scaled images that are consistent with that of the input image, $i.e.$,

\begin{equation}
    SPC(\bm{x}) = \frac{\sum_{n\in\mathcal{S}} \mathbb{I}\{C(n\cdot\bm{x})=C(\bm{x})\}}{|\mathcal{S}|},
\end{equation}
where $\mathbb{I}$ is the indicator function and $|\mathcal{S}|$ denotes the size of scaling set $\mathcal{S}$. In particular, we constrain $n\cdot\bm{x} \in [0,1]$ during the multiplication process.

Once we obtain the SPC value of suspicious input $\bm{x}$, we can determine it is malicious based on defender-specified threshold $T$. If $SPC(\bm{x})>T$, we deem it as a backdoor sample.

\begin{figure}[!t]
\centering
\vspace{-2em}
\subfigure[BadNets\label{fig:badnet_emp}]{\includegraphics[width=0.3\textwidth]{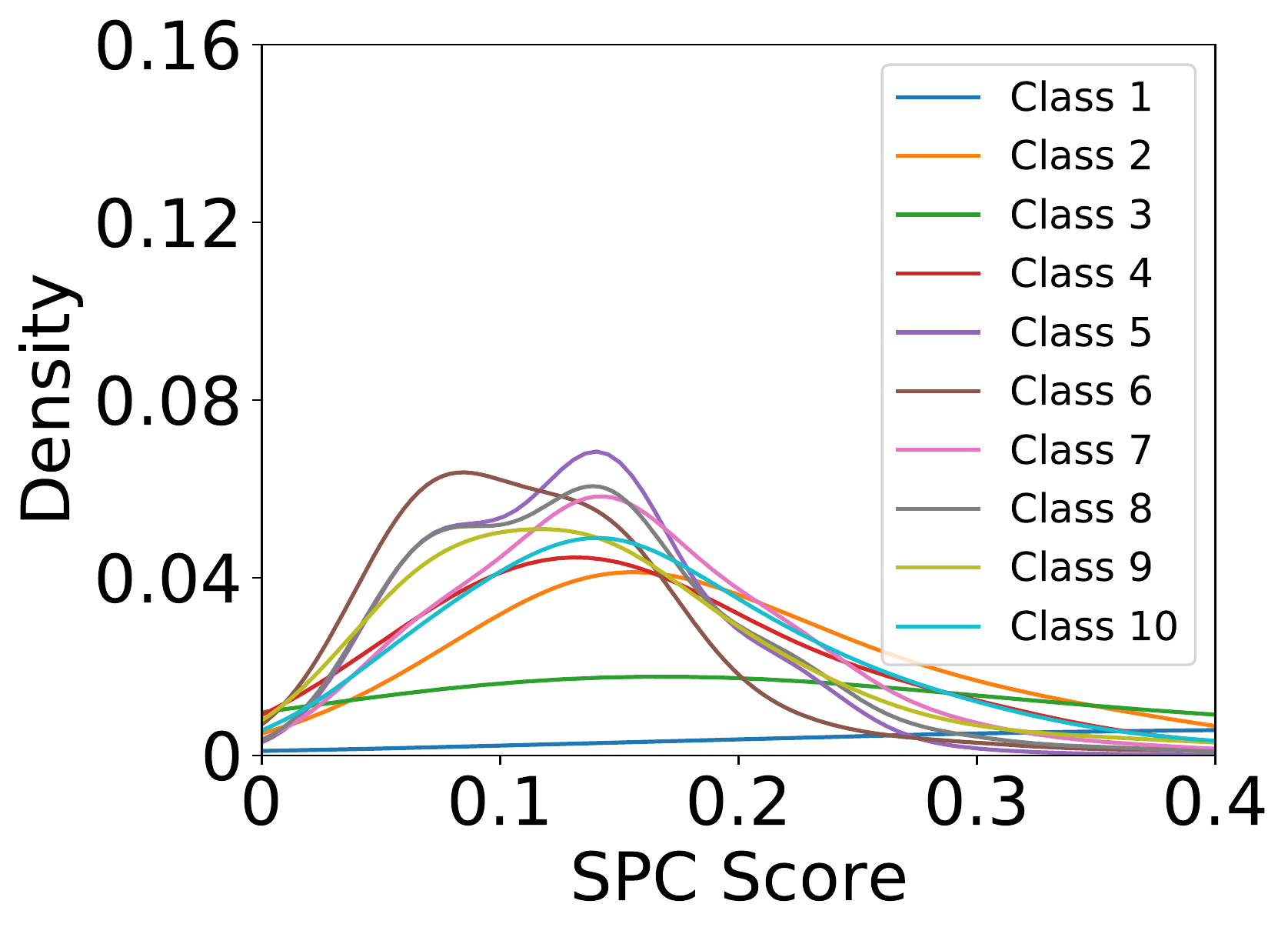}}\hspace{1em}
\subfigure[Label-Consistent\label{fig:lc_emp}]{\includegraphics[width=0.3\textwidth]{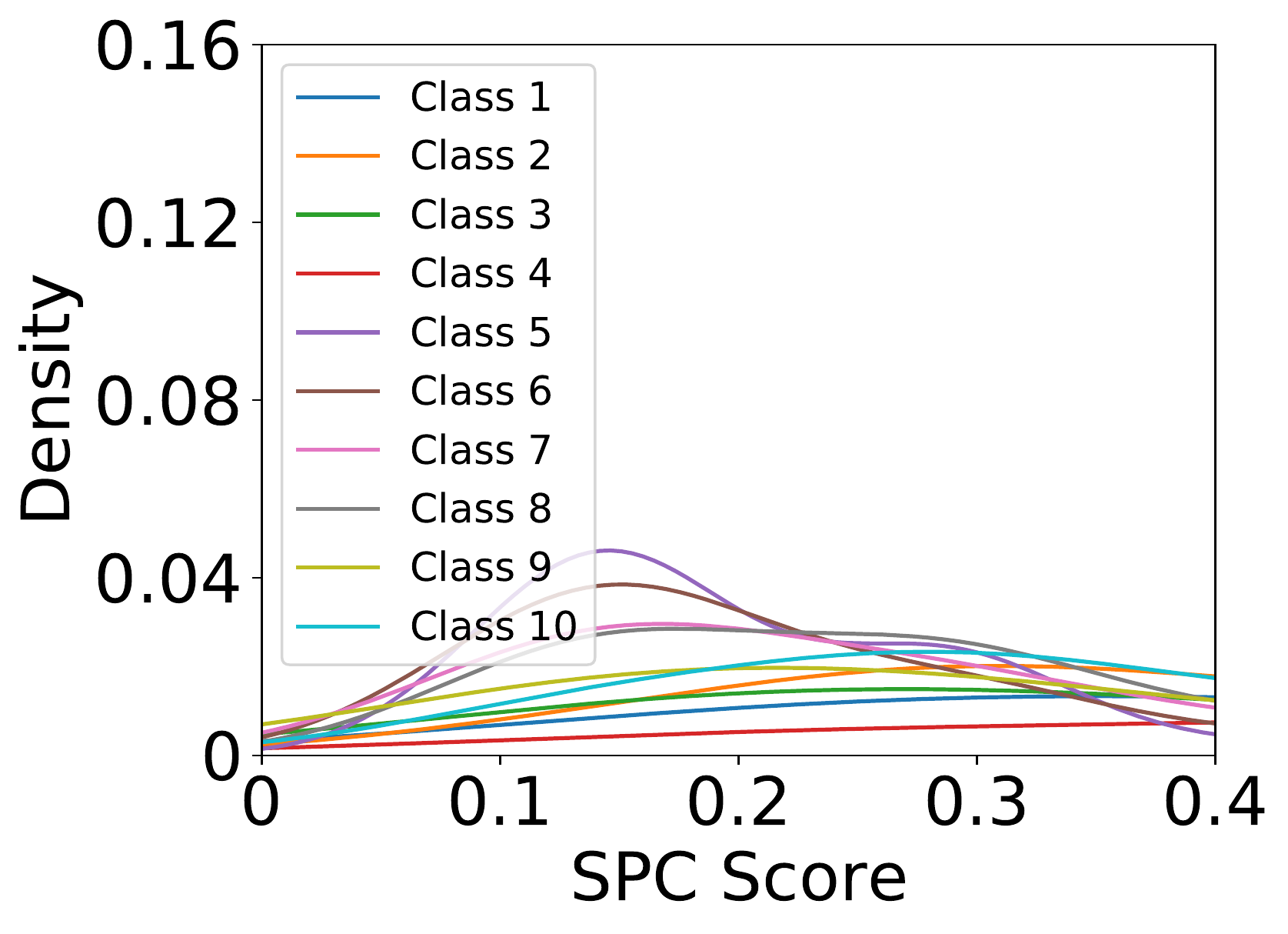}}\hspace{1em}
\subfigure[ISSBA\label{issba_density}]{\includegraphics[width=0.3\textwidth]{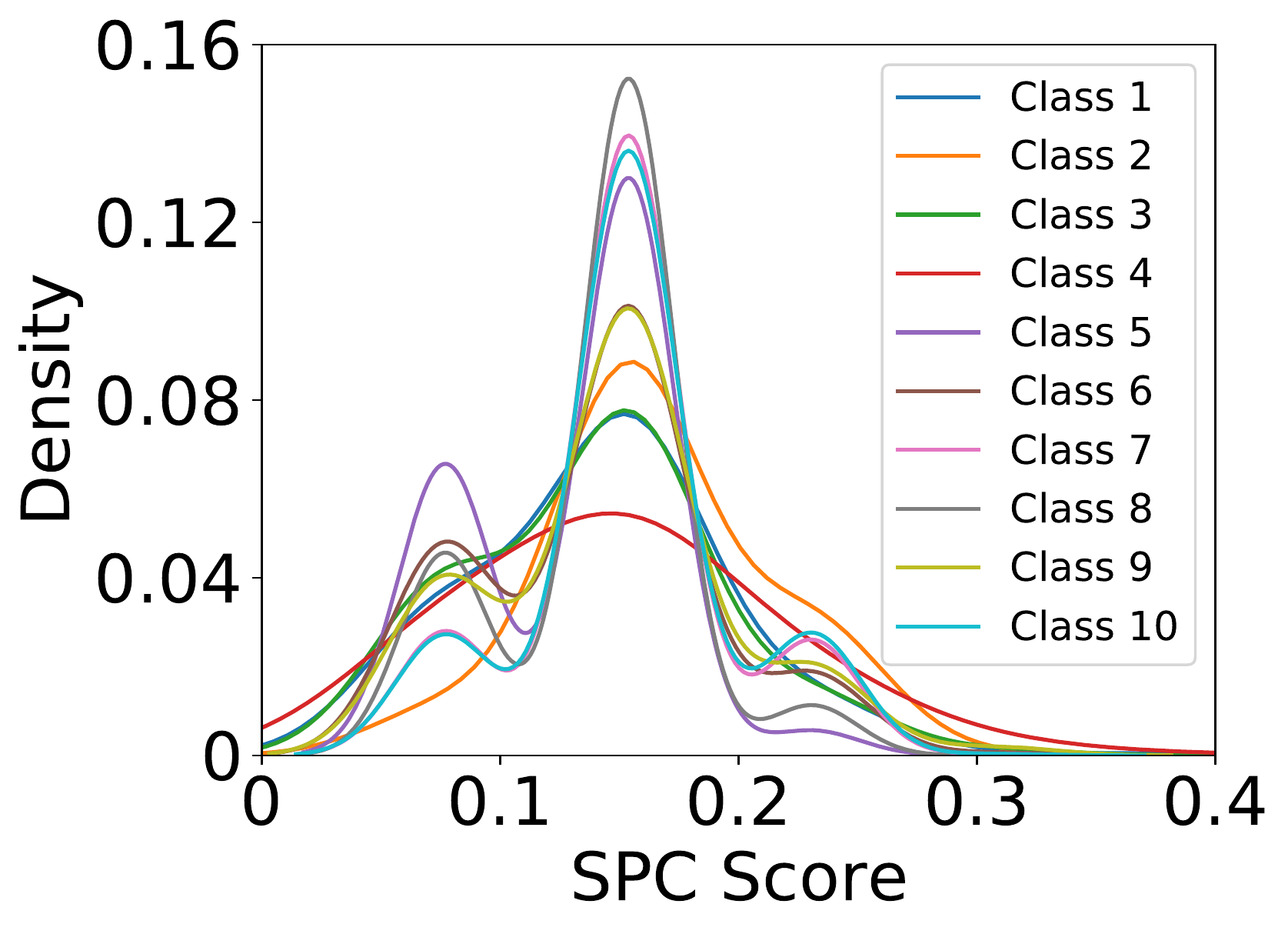}}
% \subfigure[Activation Similarity\label{infected_3d}]{\includegraphics[width=0.24\textwidth]{figure/empirical study/activation.pdf}}
\vspace{-0.5em}
\caption{The SPC scores of benign samples from different classes under attacked models.}
\vspace{-2mm}
\label{fig:statics}
\end{figure}

\subsection{Data-limited Scaled Prediction Consistency Analysis}

%\red{I will start from here later (Yiming0926)}

In our data-free scaled prediction consistency analysis, we treat all labels equally. However, we notice that the SPC values of benign samples under attacked models are different across classes (as shown in Figure \ref{fig:statics}). In other words, some classes are more consistent against image scaling compared to the remaining ones. These benign samples with have high SPC values may be mistakenly treated as malicious samples, leading to relatively low precision of our method.

In data-limited scaled prediction consistency analysis, we assume that defenders have a few benign samples from each class. This setting has been widely used in existing backdoor defenses \citep{lineural,AEVA,unlearn}. In this paper, we propose to exploit a set of statics summary ($i.e.$, mean $\mu$ and standard deviation $\sigma$) of these local benign samples to alleviate this problem, inspired by \citep{batch_normalization}. 
We first estimate the statics summary for SPC values on samples from different classes, based on which to normalize the SPC value of suspicious input images according to their predicted labels. Specifically, for each class $i$, we calculate its corresponding mean $\mu_{i}$ and standard deviation $\sigma_{i}$ based on samples $\bm{X}_{i}$ belonging to class $i$, as follows:
    \begin{equation}
        \mu_{i}  = \mathbb{E}_{\bm{x} \in \bm{X}_{i}} [SPC(\bm{x})], \quad \sigma_{i} = \sqrt{\mathbb{E}_{\bm{x} \in \bm{X}_{i}}[(\bm{x}-\mu_{i})^{2}]}.\vspace{-0.5em}
    \end{equation}

In the detection process, given a suspicious image $\bm{x}$ and the deployed model $C$, we normalize the SPC value generated by data-free SCALE-UP based on its predicted label $\hat{y} \triangleq C(\bm{x})$, as follows: 

\begin{equation}
        \label{eq:normalize}
       NSPC(\bm{x}) \triangleq SPC(\bm{x})-\frac{\mu_{\hat{y}}}{{\sigma_{\hat{y}}}}.\vspace{-0.5em}
\end{equation}

\begin{figure}[!t]
    \centering
    %\vspace{-2em}
    \includegraphics[width=0.98\textwidth]{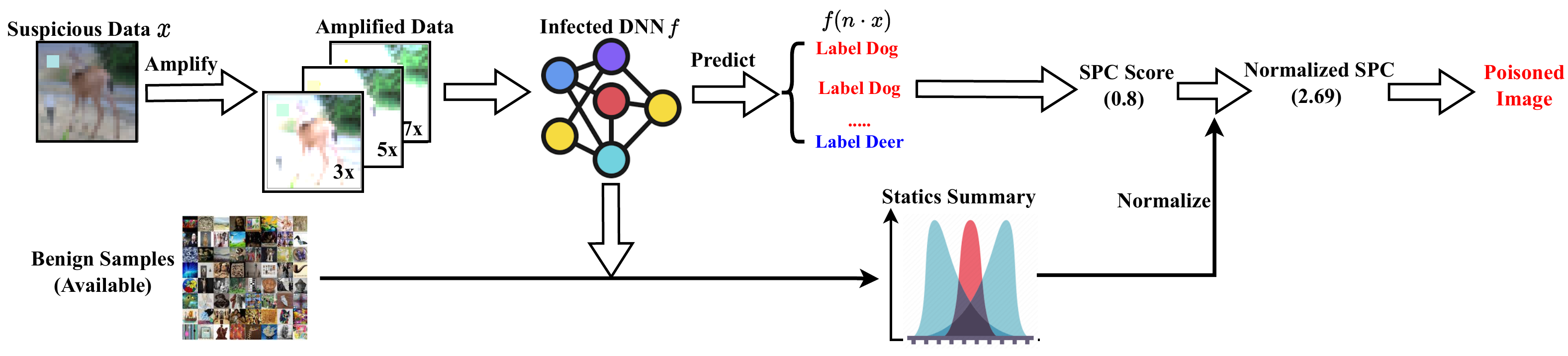} 
    \caption{The main pipeline of our (data-limited) SCALE-UP. For each suspicious input, it first generates a set of its amplified images. After that, it takes amplified images to query the deployed DNN and obtain their predicted labels. Thirdly, we compute and normalize the SPC value based on the results and that of some local benign samples. SCALE-UP treats the input as a malicious image if the normalized SPC value is greater than a defender-specified threshold $T$.}
    \label{fig:work_flow}
\end{figure}

Besides, for a more stable and effective performance, we automatically balance two terms in Eq.~(\ref{eq:normalize}) to make their values at the same level. The main pipeline of our method is summarized in Figure \ref{fig:work_flow}.

\begin{figure}[!t]
    \centering
    \includegraphics[width=0.98\textwidth]{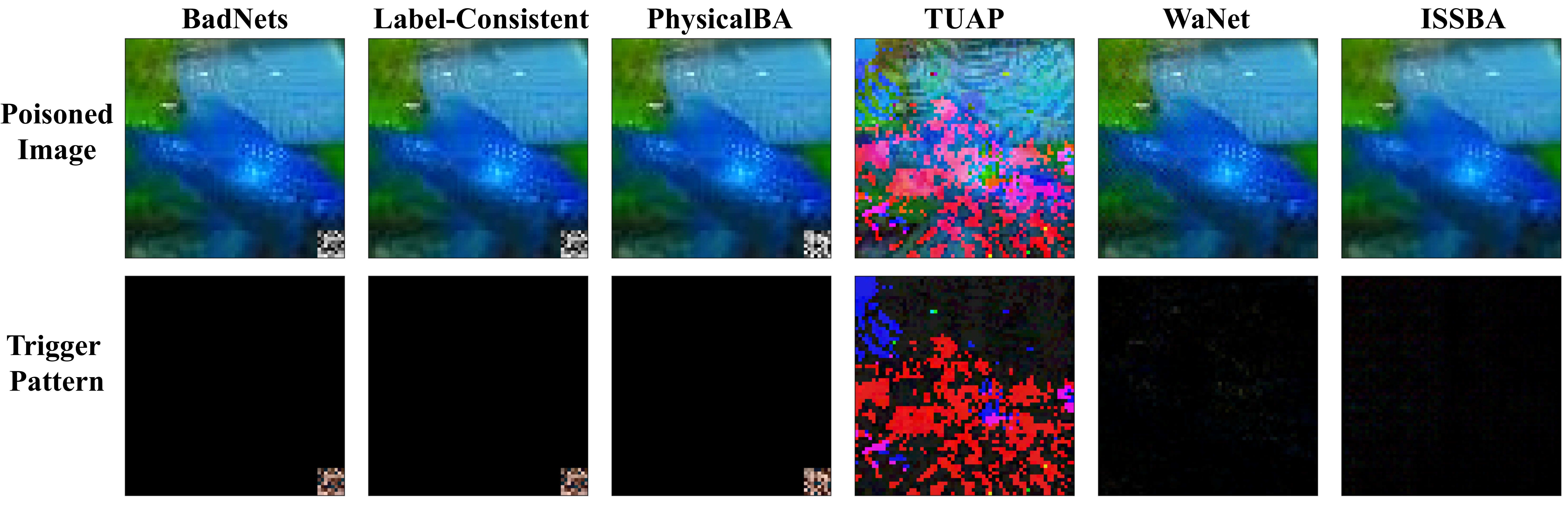}
    \vspace{-1mm}
    \caption{The demonstration of various trigger patterns of attacks used in our experiments.}
    \label{fig:trigger}
    \vspace{-2mm}
\end{figure}

\section{Experiments}
\label{sec:exp}

\subsection{Main Settings}
\noindent\textbf{Dataset and DNN Selection.} Following the settings in existing backdoor defenses, we conduct experiments on CIFAR-10 \citep{CIFAR} and (Tiny) ImageNet \citep{russakovskyimagenet} datasets with ResNet \citep{hedeep}. Please find more detailed information in our appendix.

\noindent \textbf{Attack Baselines.} In this paper, we evaluate our methods under six representative attacks, including \textbf{1)} BadNets \citep{DBLP:journals/corr/abs-1708-06733}, \textbf{2)} label consistent backdoor attack (dubbed `Label-Consistent') \citep{label_consis}, \textbf{3)} physical backdoor attack (dubbed `PhysicalBA') \citep{li2021backdoor}, \textbf{4)} 
clean-label backdoor attack with targeted universal adversarial perturbation (dubbed `TUAP') \citep{zhaoclean}, \textbf{5)} WaNet \citep{wanet}, and \textbf{6)} ISSBA \citep{ISSBA}. They are the representative of patch-based and non-patch-based backdoor attacks under different settings. For each attack, we randomly select the target label and inject sufficient poisoned samples to ensure the attack success rate $\geq 98\%$ while preserving the overall model performance. We implement these attacks based on the open-sourced backdoor toolbox \citep{libackdoorbox}. We demonstrate the trigger patterns of adopted attacks for Tiny ImageNet in Figure \ref{fig:trigger}. More detailed settings are in the appendix.

%\noindent\textbf{Attack methods.} We evaluate our approach under six attacks, which covers various types of backdoor attacks(\ie,static~\citep{DBLP:journals/corr/abs-1708-06733}, advanced input-specific~\citep{ISSBA,wanet}, label-consistent~\citep{label_consis}, smooth trigger~\citep{tuap} and physically robust attack~\citep{rethinking}). For BadNets~\citep{DBLP:journals/corr/abs-1708-06733} and PhysicalBA~\citep{rethinking} we use a 4x4 square as backdoor triggers, while for others~\citep{ISSBA,wanet,label_consis,tuap} we use their default less invisible triggers. The demonstration of various triggers is detailed in \cref{fig:trigger}.

%

\noindent \textbf{Defense Baselines.} In this paper, we focus on the backdoor detection under the black-box setting where defenders can only query the deployed model and obtain its predicted label. Accordingly, we compare our methods to ShrinkPad \citep{li2021backdoor}, DeepSweep \citep{deepweep}, and artifacts detection in the frequency domain (dubbed `Frequency') \citep{frequency}. We also compare our methods to STRIP \citep{gaodesign} that requires additional requirement ($i.e.$, obtaining predict probability vectors). We assume that defenders have 100 benign samples per class under the data-limited setting. Please find more defense details in our appendix.

\noindent \textbf{Settings for Evaluation Datasets.} Following the previous work \citep{leesimple}, we use a positive ($i.e.$, attacked) and a negative ($i.e.$, benign) dataset to evaluate each defense. Specifically, the positive dataset contains the attacked testset and its augmented version, while the negative dataset contains a benign testset and its augmented version. The augmented datasets are created by adding small random noise to their original version. The noise magnitude is set to 0.05. In particular, adding these random noises will not significantly affect the attack success rate and the benign accuracy of deployed models. The introduction of the augmented datasets is to prevent evaluated defenses from over-fitting the benign or the poisoned testsets.

%\noindent\textbf{Baseline methods.} 
%Since we focus on defending backdoor attacks in the black-box setting, therefore we only compare our approach with the black-box defense approach. We compare our approach with four different input-level backdoor defenses: STRIP~\citep{strip}, ShrinkPad~\citep{rethinking}, Frequency~\citep{frequency} and DeepSweep~\citep{deepweep}. To evaluate each approach, we use a positive(infected) and a negative(benign) dataset following previous work~\citep{lee2018simple}. Specifically, the positive dataset contains a backdoored testset and its corresponding augmented dataset. While, the negative dataset contains a benign testset and its corresponding augmented dataset. The augmented dataset is created by adding certain magnitude of random noise to the original dataset. Notably, such added random noise will not affect the attack efficacy or accuracy for backdoored or benign testset under target model. The magnitude of added random noise is set around 0.05. The introduction of the augmented datasets is used to prevent the evaluated approach over-fitting the benign and backdoored testsets.  

\noindent \textbf{Evaluation Metrics.} Following existing detection-based backdoor defenses \citep{gaodesign,AEVA}, we adopt the area under receiver operating curve (AUROC) \citep{fawcettintroduction} to evaluate defense effectiveness, while use the inference time for evaluating efficiency. In general, the higher the AUROC and the lower the inference time, the better the backdoor detection.

%\noindent\textbf{Evaluation metrics.} Following previous work~\citep{strip,AEVA}, we use \textit{The Area under Receiver Operating Curve (AUROC)} and \textit{Inference Time} to evaluate the performance and efficiency of our approach, respectively. The details for each metric are described in Appendix.

\begin{wrapfigure}{!t}{0.30\textwidth}\vspace{-4mm}
  \begin{center}
  \vspace{-0.5em}
    \includegraphics[width=0.30\textwidth]{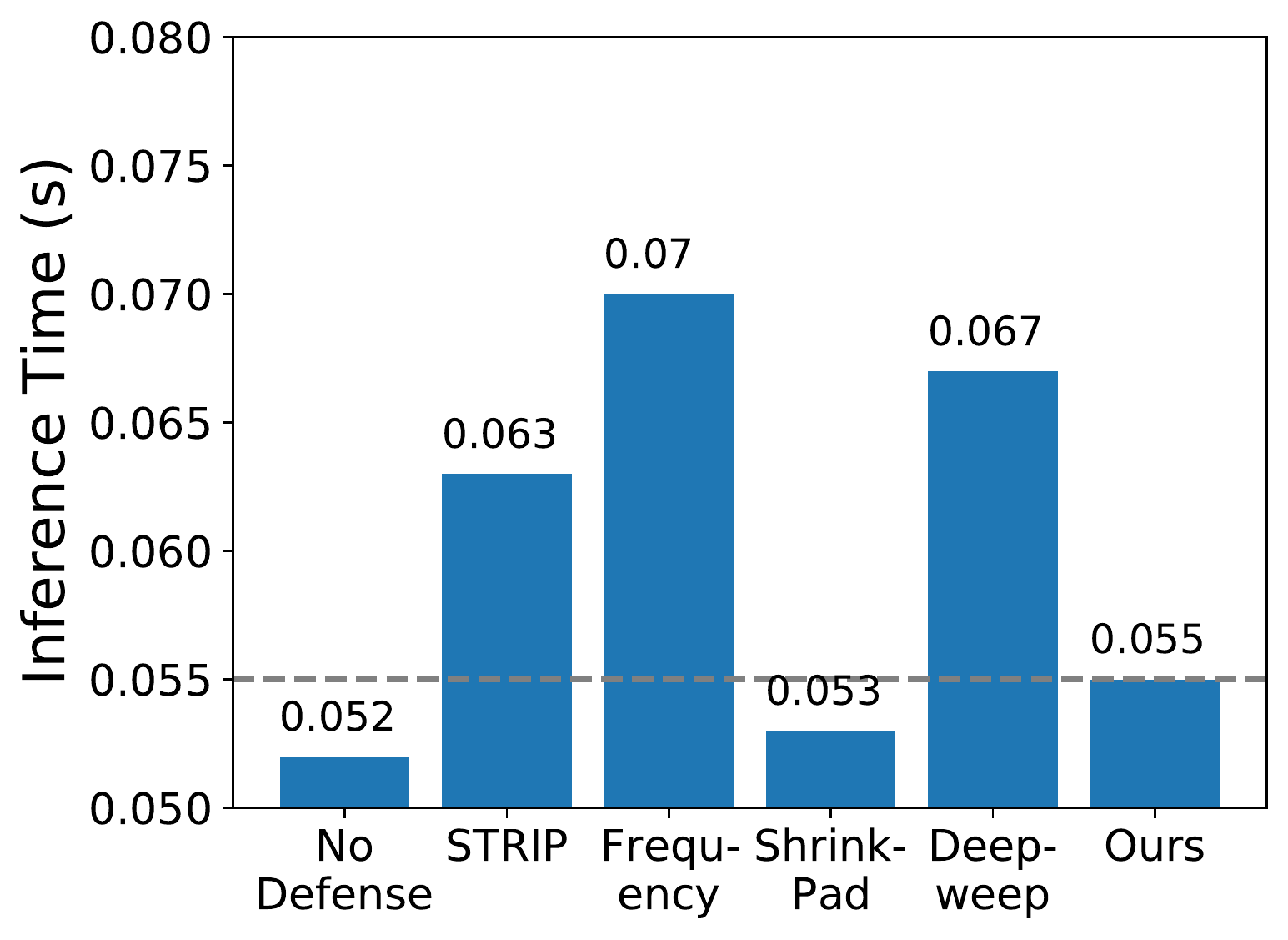}
  \end{center}
  \vspace{-1.3em}
  \caption{The inference time on the CIFAR-10 dataset.}
 \label{fig:inference}
 \vspace{-0.3em}
\end{wrapfigure}

\begin{table}[!t]
    \centering
    %\vspace{-2em}
    \caption{The performance (AUROC) on the CIFAR-10 dataset. Among all different methods, the best result is marked in boldface while the value with underline denotes the second-best result. The failed cases ($i.e.$, AUROC $<0.55$) are marked in red. Note that STRIP require obtaining predicted probability vectors while other methods only need the predicted labels.}
         \scalebox{0.89}{
        \begin{tabular}{c|cccccc|c}
    \toprule
                  \tabincell{c}{Attack$\rightarrow$\\ Defense$\downarrow$}      &                   BadNets& Label-Consistent  & PhysicalBA    & TUAP &   WaNet                      & ISSBA    & \textbf{Average}                       
                         \\ \hline
STRIP   &   \textbf{0.989}                            & 0.941                             &  \textbf{0.971}    & 0.671  & \color{red}0.475  & \color{red}0.498      & 0.758                      \\ \hline
ShrinkPad      &  0.951                           &  \textbf{0.957} & 0.631                          & \textbf{0.869} & \color{red}{0.531}  &        \color{red}0.513        & 0.742             \\ 
DeepSweep       & 0.967                           &  0.921                          & 0.946 & 0.743 &  \color{red}0.506  &        0.729         &       0.802 \\
 
Frequency   & 0.891                        &   0.889                            & 0.881 & \underline{0.851} &   \color{red}0.461  & \color{red}0.497        & 0.745    \\
 \hline
Ours (data-free) & \underline{0.971}                        &   0.947                      & 0.969  & 0.816& \underline{0.918}&\textbf{0.945}                &  \underline{0.928}   

\\
Ours (data-limited)                &  \underline{0.971}                      &   \underline{0.954}      & \underline{0.970}  &  0.830   &\textbf{0.925} &            \textbf{0.945}   & \textbf{0.933}      

\\ \bottomrule
    \end{tabular}}
    \centering
    % For each attack approach, we build 60 infected models for each task.}
    %\vspace{-4mm}
     \label{table:result_cifar10}
    \vspace{-2mm}
\end{table}

\begin{table}[!t]
    \centering
    \caption{The performance (AUROC) on the Tiny ImageNet dataset. Among all different methods, the best result is marked in boldface while the value with underline denotes the second-best result. The failed cases ($i.e.$, AUROC $<0.55$) are marked in red. Note that STRIP require obtaining predicted probability vectors while other methods only need the predicted labels.}
         \scalebox{0.89}{
        \begin{tabular}{c|cccccc|c}
    \toprule
                  \tabincell{c}{Attack$\rightarrow$\\ Defense$\downarrow$}      &                   BadNets& Label-Consistent  & PhysicalBA    & TUAP &   WaNet                      & ISSBA    & \textbf{Average}                       
                         \\ \hline
STRIP   &   \textbf{0.959}                            & \textbf{0.939}                             &  \textbf{0.959}    & 0.638  & \color{red}0.501  & \color{red}0.471      & 0.745                      \\ \hline
ShrinkPad      &  0.871                           &  \underline{0.938} & 0.672                          & \textbf{0.866} & \color{red}{0.498}  &        \color{red}0.492        & 0.737            \\ 
DeepSweep       & \underline{0.951}                          &  0.930                          & \underline{0.939} & 0.759 &  \color{red}0.503  &        0.714         &       0.799 \\
 
Frequency   & 0.864                       &   0.859                            & 0.864 & \underline{0.837} &   \color{red}0.428  & \color{red}0.540        & 0.732    \\
 \hline
Ours (data-free) & 0.936                       &   0.904                      & \underline{0.939}  & 0.763& \underline{0.943}& \underline{0.948}                &  \underline{0.905}   

\\
Ours (data-limited)                &  0.947                     &   0.911      & \underline{0.939}  &  0.763   &\textbf{0.946} &            \textbf{0.949}   & \textbf{0.909}      

\\ \bottomrule
    \end{tabular}}
    \centering
    %\vspace{-4mm}
     \label{table:result}

\end{table}

%\red{I will start from here later (Yiming0927)}

\newpage

\subsection{Main Results}
As shown in Table \ref{table:result_cifar10}-\ref{table:result}, \emph{all baseline detection methods fail in defending against some evaluated attacks}. Specifically, they have relatively low AUROC in detecting advanced non-patch-based attacks ($i.e.$, WaNet and ISSBA). This failure is mostly because these defenses have some implicit assumptions ($e.g.$, the trigger pattern is sample-agnostic or static) about the attack, which are not necessarily true in practice.  In contrast, \emph{our methods reach promising performance in all cases on both datasets}. For example, the AUROC of our data-limited SCALE-UP is 0.5 greater than all baseline defenses in detecting WaNet on the Tiny ImageNet dataset. Even under the classical patch-based attacks ($i.e.$, BadNets, Label-Consistent, and PhysicalBA), the effectiveness of our methods is on par with or better than all baseline defenses. Our methods are even better than STRIP, which requires obtaining predicted probability vectors instead of predicted labels. We also provide the ROC curves of defenses against all attacks in Appendix \ref{ap:ROC}. These results verify the effectiveness of our defenses.

Besides, we also calculate the inference time of all defenses under the same computational facilities. In particular, we calculate the inference time of methods requiring to obtain the predictions of multiple images by feeding them simultaneously (in a data batch) into the deployed model instead of predicting them one by one. Besides, we only report the inference time of our SCALE-UP under the data-limited setting, since both of them have very similar running times. As shown in Figure \ref{fig:inference}, our method requires fewer inference times compared to almost all baseline defenses. The only exception is ShrinkPad, whereas its effectiveness is significantly lower than that of our method. Our detection is approximately $5\%$ slower compared with the standard inference process without any defense. These results show the efficiency of our SCALE-UP detection.

%We first investigate the overall performance of \name{} against 6 different backdoor attacks under two popular benchmarks. The results are shown in \cref{table:result_cifar10} and \cref{table:result}. We find that \name{} performs effective against all evaluated attacks with $\geq 0.867$ on average AUROC. We also find that in most cases, \name{} performs better with limited labeled data. We also compare \name{} with four existing defenses. The results for CIFAR-10 and TinyImageNet are shown in \cref{table:result_cifar10} and \cref{table:result}. As shown in these tables, \name{} can achieve comparable performance against existing input-level defense approaches on several basic input-static backdoors (\eg, BadNets, TUAP, etc). However, \name{} is the only existing approach performs effectively against the advanced input-specific backdoor attacks (\ie,  WaNet, ISSBA). This is because, existing defense approaches (\eg,STRIP, ThinkPad, etc) only focus on the properties of basic backdoor attacks, thus fail against the recent advanced backdoor attacks. As an inference-phase defense approach, we also evaluate the efficiency of \name{} compared with existing defense approaches. The results are shown in \cref{fig:inference}. We implement each approach under the platform with one NVIDIA GPU 1080 Ti and an Intel(R) Xeon(R) CPU E5-2650 v4 @ 2.20GHz and enable parallel computation. The detailed settings and measurements are included in the Appendix. We find that our approach runs merely $5\%$ slower compared with the normal model (baseline).  

%\input{preliminary}

\subsection{Discussion}
In this section, we discuss whether our method is still effective under different (adversarial) settings.

%After investigating the overall performance of \name{}, we here conduct 10 additional ablation studies to further evaluate the robustness of \name{}.

% \begin{wrapfigure}{!t}{0.309\textwidth}
%     \vspace{-4mm}
%   \begin{center}
%     \includegraphics[width=0.309\textwidth]{figure/experiments/trigger_size_2.pdf}
%   \end{center}
%   \vspace{-4mm}
%   \caption{\small The impact of trigger
% size on detection accuracy.}

% \vspace{-3mm}
%  \label{fig:triggerasize}
% \end{wrapfigure}

\subsubsection{Defending against Attacks with Larger Trigger Sizes}
Recent studies \citep{deepweep} revealed that some defenses ($e.g.$, ShrinkPad) may fail in detecting samples with a relatively large trigger size. In this part, we use two patch-based attacks ($e.g.$, BadNets and PhysicalBA) on the Tiny ImageNet dataset for discussion. As shown in Figure \ref{fig:triggerasize}, our methods have high AUROC values ($>0.93$) across different trigger sizes under both data-free and data-limited settings, although there are some mild fluctuations. These results verify the resistance of our SCALE-UP detection to adaptive attacks with large trigger patterns.

%\noindent\textbf{The impact for the Trigger size.}
%We test \name{} on Tiny-ImageNet with varied trigger sizes. Since most backdoor attacks are either constrained to some specific triggers~\citep{tuap,ISSBA,wanet} or sensitive to the trigger size~\citep{label_consis}, we choose BadNets~\citep{DBLP:journals/corr/abs-1708-06733} and PhysicalBA~\citep{rethinking} to evaluate the sensitivity of \name{} on the trigger sizes. We test \name{} using ResNet-34 and the implementation of each attack as well as \name{} is consistent with \cref{sec:exp}. The results are shown in \cref{fig:triggerasize}. As seen in \cref{fig:triggerasize}, we find that \name{} performs resilient to different trigger sizes (ranges from $8\text{x}8$ to $40\text{x}40$) with AUROC $\geq 93\%$ in both data-free and-limited settings. Such results are also consistent with our theoretical analysis.

% \begin{wrapfigure}{!t}{0.309\textwidth}\vspace{-6mm}
%   \begin{center}
%     \includegraphics[width=0.309\textwidth]{figure/experiments/adap2.pdf}
%   \end{center}
%   \vspace{-5mm}
%   \caption{\small The performance of backdoor attacks and SCALE-UP with varying poison rates.}
% \label{fig:adapt_1}
%  \vspace{-2mm}
% \end{wrapfigure}

\subsubsection{The Resistance to Potential Adaptive Attacks}
Most recently, \citep{qirevisiting} demonstrated that reducing the poisoning rate is a simple yet effective method to design adaptive attacks for detection-based defenses, since it can reduce the differences between benign and poisoned samples. Motivated by this finding, we first explore whether our SCALE-UP methods are still effective in defending against attacks with low poisoning rates. We adopt BadNets on the CIFAR-10 dataset as an example for our discussions. In particular, we report the results of all poisoned testing samples and those that can be predicted as the target label, respectively. We design this setting since attacked models may still correctly predict many poisoned samples even if they contain trigger patterns when the poisoning rate is relatively low.  

\begin{figure}[!t]
\centering
\subfigure[\label{fig:triggerasize}]{\includegraphics[width=0.31\textwidth]{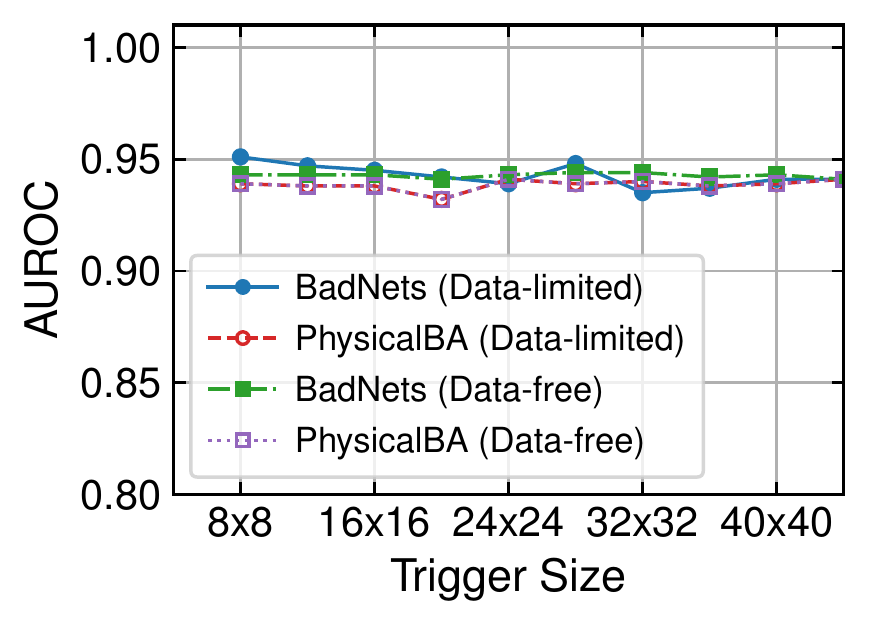}}\hspace{1em}
\subfigure[\label{fig:adapt_1}]{\includegraphics[width=0.31\textwidth]{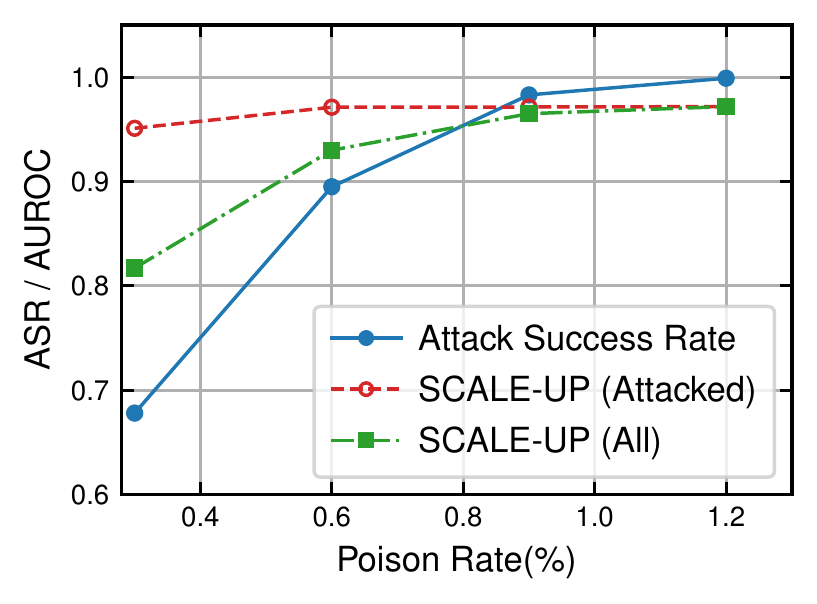}}\hspace{1em}
\subfigure[\label{fig:adapt_robust}]{\includegraphics[width=0.31\textwidth]{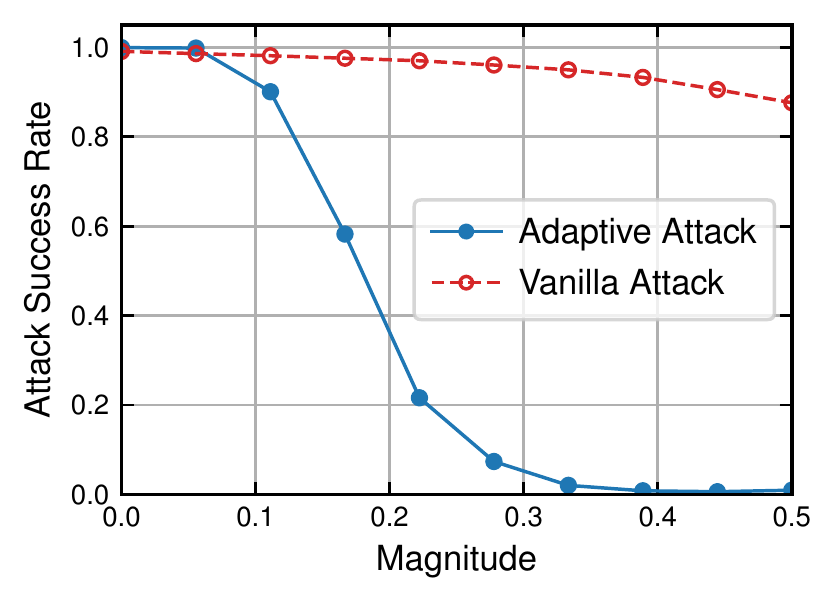}}
% \subfigure[Activation Similarity\label{infected_3d}]{\includegraphics[width=0.24\textwidth]{figure/empirical study/activation.pdf}}
\vspace{-3mm}
\caption{The results of additional experiments in our discussion. \textbf{(a)} The performance of our methods under attacks with different trigger sizes. \textbf{(b)} The attack performance and the defense effectiveness on all poisoned testing samples and those that can successfully attack the deployed model. \textbf{(c)} The effectiveness of adaptive and vanilla backdoor attacks on poisoned samples with random noise under different magnitudes.}
% \red{Figure.(a): The impact of trigger
% size on detection accuracy. Figure (b): The performance of backdoor attacks and \name{} with varying poison rates. Figure (c): The robustness of the adaptive and basic backdoor attacks.}}
%\vspace{-2em}
\vspace{-3mm}
\label{fig:discussion}
\end{figure}

As shown in Figure \ref{fig:adapt_1}, the attack success rate (ASR) increases with the increase of the poisoning rate. Our method can still correctly detect poisoned samples that can successfully attack the deployed model even when the poisoning rate is set to $0.4\%$ where the ASR is lower than 70\%. In these cases, the AUROC $>0.95$. Besides, our method can still reach promising performance (AUROC $>0.8$) in detecting all poisoned samples. These results verify the resistance of our defense to adaptive attacks with low poisoning rates, where attacked models don't over-fit backdoor triggers.

%\noindent\textbf{The resistance to potential adaptive attacks.}We first evaluate the efficacy of \name{} on models infected with varying poison rates, which is a basic approach for evaluating the robustness of backdoor defense. The results are shown in \cref{fig:adapt_1}. Since a low poison rate ($\leq 0.6\%$) would affect the efficacy of backdoor samples with a low attack success rate, we test \name{} under both effective backdoor samples and all backdoor samples. The results show that \name{} can still  be effective in detecting effective backdoor samples with AUROC $\geq 95\%$. Such results also imply that \name{} can perform effectively against backdoor attacks even if the infected model doesn't over-fit the backdoor triggers.
% \begin{wrapfigure}{!t}{0.309\textwidth}
% \vspace{-3mm}
%   \begin{center}
%   \includegraphics[width=0.309\textwidth]{figure/experiments/adap.pdf}
%   \end{center}
%   \vspace{-5mm}
%   \caption{\small The robustness of the adaptive and basic backdoor attacks.}
%  \label{fig:adapt_robust}
%  \vspace{-3mm}
% \end{wrapfigure}
To further evaluate the resistance of our SCALE-UP to potential adaptive methods, we evaluate it under the worst scenario, where the backdoor adversaries are fully aware of our mechanism. Specifically, we design a strong adaptive attack by introducing an additional defense-resistant regularization term to the vanilla attack illustrated in \cref{eq:training}. This regularization term is used to prevent scaled poisoned samples $n\cdot \bm{x}'_{j}$ being predicted as the target label $y_t$, as follows:
\vspace{-2mm} 
\begin{equation}
\label{eq:adapt}
       \min_{\bm{\theta}} \sum_{i=1}^{N_{b}} \mathcal{L}(f(\bm{x}_{i};\bm{\theta}),y_{i}) + \sum_{j=1}^{N_p} \mathcal{L}(f(\bm{x}'_{j};\bm{\theta}),y_t)+
        \sum_{j=1}^{N_p} \mathcal{L}(f(n \cdot \bm{x}'_{j};\bm{\theta}),y_j).
\end{equation}

%Since attackers typically inject backdoor attacks in the training phase, we thus consider a strong adaptive attack from the perspective of optimization. Specifically, during the training phase, we manipulate the typical training process for performing backdoor attacks (\cref{eq:training}) by adding an additional regularization term as:
%\vspace{-2mm}
%\begin{equation}
%\label{eq:adapt}
%        \min_{\bm{\theta}} \sum_{i=1}^{N_{b}} \mathcal{L}(f(\bm{x}_{i};\bm{\theta}),y_{i}) + \sum_{j=1}^{N_p} \mathcal{L}(f(\bm{x}'_{j};\bm{\theta}),y_t)+
%        \sum_{j=1}^{N_p} \mathcal{L}(f(\bm{n\cdot}\bm{x}'_{j};\bm{\theta}),y_{i})
%    \vspace{-4mm}
%\end{equation}

Similar to previous experiments, we adopt BadNets to design the adaptive attack on the CIFAR-10 dataset. As we expected, this method can bypass our detection resulting in a low AUROC ($i.e.$, 0.467). However, \emph{the adaptive attack would make the poisoned samples significantly more vulnerable to small random Gaussian noises}. As shown in Figure \ref{fig:adapt_robust}, random noises with a small magnitude ($<0.3$) will significantly reduce the attack success rate of the adaptive attack, while having minor adverse effects on the vanilla attack. In other words, defenders can easily adopt random noises to defend against this adaptive attack. We speculate that its vulnerability is mostly because the regularization term significantly constrains the generalization of attacked DNNs on the poisoned samples. We will further explore its intrinsic mechanism in our future work.

\subsubsection{The Effectiveness of Scaling Process}\label{sec:noise}
Technically, the scaling process in our SCALE-UP detection can be regarded as a data augmentation method generating different modified versions of the suspicious input image. It naturally raises an intriguing question: \emph{If other augmentation methods are adopted, is our SCALE-UP detection still effective?} Since flip operations and frequency domain analysis have been adopted in  \citep{li2021backdoor,frequency} for defense and proved to have minor benefits to detecting advanced backdoor attacks \citep{ISSBA,wanet}, we investigate the effectiveness of adding increasing magnitudes of random noise. Due to the limitations of space, we include the detailed experimental design and evaluation in Appendix \ref{ap:scale_eff}.

\section{Conclusion}
In this paper, we proposed a simple yet effective black-box input-level backdoor detection (dubbed SCALE-UP) that can be used in real-world applications under the machine learning as a service (MLaaS) setting. Our method was motivated by an intriguing new phenomenon (dubbed \emph{scaled prediction consistency}) that the predictions of poisoned samples are significantly more consistent compared to those of benign ones when amplifying all pixel values. We also provided theoretical foundations to explain this phenomenon. In particular, we designed our SCALE-UP detection method under both data-free and data-limited settings. Extensive experiments on benchmark datasets verified the effectiveness and efficiency of our method and its resistance to potential adaptive attacks. 

%In this paper, we propose a simple yet effective black-box input-level backdoor detection requiring only the predicted labels to alleviate this problem. Specifically, we identify and filter malicious testing samples by analyzing their prediction consistency during the pixel-wise amplification process. Our defense is motivated by an intriguing phenomenon (dubbed \emph{scaled prediction consistency}) that the predictions of attacked samples are significantly more consistent compared to those of benign ones when amplifying all pixel values. Besides, we also provide theoretical insight to explain this phenomenon. Extensive experiments are conducted on benchmark datasets, verifying the effectiveness and efficiency of our defense and its resistance to potential adaptive attacks.

%This paper addresses the black-box input-level backdoor detection for machine learning as a service setting. We propose the \name{} algorithm, which is based on analyzing the prediction consistency for inputs during the pixel-wise amplification process. Extensive experiments demonstrate the efficacy of \name{} across popular tasks and state-of-the-art backdoor attacks.

\clearpage
\section*{Ethics Statement}
DNNs have been widely and successfully adopted in many mission-critical applications. Accordingly, their security is of great significance. The existence of backdoor threats raises serious concerns about using third-party models under the machine learning as a service (MLaaS) setting. In this paper, we propose a simple yet effective black-box input-level backdoor detection. Accordingly, this work has no ethical issues since it does not reveal any new security risks and is purely defensive. However, we need to notice that our methods can only be used to filter poisoned testing samples whereas they do not reduce the intrinsic backdoor vulnerability of deployed models. Our defense also couldn't recover trigger patterns. People should not be too optimistic about eliminating backdoor threats. We will further improve our method by exploring how to recover triggers.  
\section*{Acknowledgment}
This work is mainly supported by Samsung Research America, Mountain View, CA and partially supported by NSF CNS 2135625, CPS 2038727, CNS Career 1750263, and a Darpa Shell grant.

\section*{Reproducibility Statement}
We have provided detailed information on datasets, models, training settings, and computational facilities in our main manuscript and appendix. The codes for reproducing our main experiments are also open-sourced at \url{https://github.com/JunfengGo/SCALE-UP}.

%We detail the implementation, configuration and descriptions for our approach, evaluated datasets, comparison baselines, and evaluation metrics. Codes for \name{} is also open-sourced as a supplementary file.

\bibliography{iclr2022_conference}
\bibliographystyle{iclr2023_conference}
\clearpage
\appendix
\section*{Appendix}
\setcounter{theorem}{0}

\section{The Omitted Proof of Theorem 1}\label{sec:lemma1}

\begin{theorem}
Suppose the poisoned training dataset consists of $N_b$ benign samples and $N_p$ poisoned samples, i.i.d. sampled from uniform distribution and belonging to $K$ classes. Assume that deep neural network $f$ adopt RBF kernel and cross-entropy loss with the optimization objective in Eq.\ref{eq:training}. For a given attacked sample $\bm{x}'= (\bm{1}-\bm{m})\odot \bm{x} + \bm{m}\odot \bm{t}$, we have: $\lim_{N_p\to N_b} C(n\cdot\bm{x}')=y_t, n
\geq1$.
\end{theorem}

\paragraph{Proof of Theorem 1:}

Following~\citep{AEVA}, we have the regression solution for NTK is:
\begin{equation}
    \phi_t(\cdot)=\frac{\sum_{i=1}^{N_b}K(\cdot, \bm{x_i})\cdot y_{i}+\sum_{i=1}^{N_p}K(\cdot, \bm{x'_{i}})\cdot y_{t}}{\sum_{i=1}^{N_b}K(\cdot, \bm{x_i})+\sum_{i=1}^{N_p}K(\cdot, \bm{x'_{i}})},
\end{equation}
where $\phi_{t}(\cdot) \in \mathbb R$ is the predictive probability output of $f(\cdot;\theta)$ for the target class $t$ and $y_{i}$ is the corresponding one-hot label. $K(\bm{x},\bm{x_{i}})=e^{-2\gamma||\bm{x}-\bm{x_{i}}||^{2}}$ ($\gamma >0$). Since the training samples are evenly distributed, there are $\frac{N_b}{k}$ benign samples belonging to $y_t$. Without loss of generality,  we assume the target label $y_{t}=1$ while others are 0. Then the regression solution can be converted to:

\begin{equation}
    \phi_t(\cdot)=\frac{\sum_{i=1}^{N_b/k}K(\cdot, \bm{x_i})+\sum_{i=1}^{N_p}K(\cdot, \bm{x'_{i}})}{\sum_{i=1}^{N_b}K(\cdot, \bm{x_i})+\sum_{i=1}^{N_p}K(\cdot, \bm{x'_{i}})},\label{eq:kernel2}
\end{equation}
For a given backdoored sample $\bm{x'} = (1-m)\odot \bm{x}+m\odot\bm{t}$, we can simplify \cref{eq:kernel2} as:
\begin{equation}
    \phi_t(\bm{x'})\geq \frac{\sum_{i=1}^{N_p}K(\bm{x'}, \bm{x'_{i}})}{\sum_{i=1}^{N_b}K(\bm{x'}, \bm{x_i})+\sum_{i=1}^{N_p}K(\bm{x'},\bm{x'_{i}})},
\end{equation}
we here remove the term $\sum_{i=1}^{N_b/k}K(\bm{x'}, \bm{x_i})$. This is because $\bm{x'}$ typically don't belong to the target $y_{t}$ and $\sum_{i=1}^{N_b/k}K(\bm{x'}, \bm{x_i}) << \sum_{i=1}^{N_p}K(\bm{x'}, \bm{x'_{i}})$, otherwise the attacker has no incentive to craft poisoned sample. 

When $N_p$ close to $N_b$, which implies that the poisoning rate close to $50\%$, the attacker can achieve the optimal attack efficacy~\citep{Trojannn,DBLP:journals/corr/abs-1708-06733,ISSBA}. Given $K(\bm{x},\bm{x_{i}})=e^{-2\gamma||\bm{x}-\bm{x_{i}}||^{2}}$ ($\gamma >0$), if $N_p = N_b$, we have:

\begin{equation}\label{eq:le}
        \phi_t(n\cdot \bm{x'})=\frac{\sum_{i=1}^{N_p}K(n\cdot \bm{x'}, \bm{x'_{i}})}{\sum_{i=1}^{N_b}K(n\cdot \bm{x'}, \bm{x_i})+\sum_{i=1}^{N_p}K(n\cdot \bm{x'}, \bm{x'_{i}})}.
\end{equation}

If $n=1$, we can easily obtain that:

\begin{flalign}
        &\sum^{N_p}_{i=1} e^{-2\gamma||(1-m)\odot(\bm{x}-\bm{x_{i}})||^{2}} - e^{-2\gamma||(1-m)\odot(\bm{x}-\bm{x_{i}})+m\odot(\bm{t}-\bm{x_{i}})||^{2}}\\
         &=\sum^{N_p}_{i=1} e^{-2\gamma||(1-m)\odot(\bm{x}-\bm{x_{i}})||^{2}}(1-e^{-2\gamma||m\odot(\bm{t}-\bm{x_{i}})||^{2}})>0.
\end{flalign}

Since the internal term $(1-e^{-2\gamma||m\odot(\bm{t}-\bm{x_{i}})||^{2}})$ can be always larger than 0, thus it is clear that $f(\bm{x'})=y_t$, which is also consistent with the practice.

However, when $n>1$, to compare  $\sum_{i=1}^{N_p}K(n\cdot \bm{x'}, \bm{x'_{i}})$ and $\sum_{i=1}^{N_b}K(n\cdot \bm{x'}, \bm{x_i})$, we have:
\begin{flalign}
\label{eq:fix}
    &\sum_{i=1}^{N_o}K(n\cdot \bm{x'}, \bm{x'_{i}}) - \sum_{i=1}^{N_b}K(n\cdot \bm{x'},\bm{x_i}) \\
        &=\sum^{N_b}_{i=1} e^{-2\gamma||(1-m)\odot(n\cdot \bm{x}-\bm{x_{i}})+m\odot(n-1)\bm{t}||^{2}} - e^{-2\gamma||(1-m)\odot(n\cdot \bm{x}-\bm{x_{i}})+m\odot(n\cdot\bm{t}-\bm{x_{i}})||^{2}}\\
        &=\sum^{N_b}_{i=1} e^{-2\gamma||(1-m)\odot(n\cdot \bm{x}-\bm{x_{i}})+m\odot(n-1)\bm{t}||^{2}} (1-e^{-2\gamma(||m\odot(n\cdot\bm{t}-\bm{x_{i}})||^{2}-||m\odot(n-1)\bm{t})||^{2})}).\\
\end{flalign}

Regarding the internal term $||m\odot(n\cdot\bm{t}-\bm{x_{i}})||^{2}-||m\odot(n-1)\bm{t})||^{2}$, we have:

\begin{flalign}
    &||m\odot(n\cdot\bm{t}-\bm{x_{i}})||^{2}-||m\odot(n-1)\bm{t})||^{2}\\
    &=\sum_{j,k \in~trigger} ||(n-1)\cdot\bm{t}_{j,k}+(\bm{t}_{j,k}-\bm{x_{i,j,k}})||^{2}-||(n-1)\cdot\bm{t}_{j,k}||^{2}\\
    &=\sum_{j,k \in ~trigger} \delta_{i,j,k}^{2}+2(n-1)\cdot\delta_{i,j,k}\bm{t}_{j,k} , 
\end{flalign}
where $\delta_{i,j,k}$ is the pixel-level residue between the trigger and benign samples. We assume that the $\delta_{i,j,k}\bm{t}_{j,k}$ close to a zero mean for inputs, thus we can rewrite~Eq.(\ref{eq:fix}) as follows:

\begin{flalign}\label{eq:n}
    &\sum_{i=1}^{N_p}K(n\cdot \bm{x'}, \bm{x'_{i}}) - \sum_{i=1}^{N_b}K(n\cdot \bm{x'},\bm{x_i}) \\
    &\approx \sum^{N_p}_{i=1} e^{-2\gamma||(1-m)\odot(n\cdot \bm{x}-\bm{x_{i}})+m\odot(n-1)\bm{t}||^{2}} (1-e^{-2\gamma\sum_{j,k \in trigger}{\delta_{i,j,k}^{2}}})\\
    &>0.
\end{flalign}
Put \cref{eq:n} and \cref{eq:le} together, we know that $\phi_{t}(n\cdot \bm{x'}) \geq 0.5$, as $N_{p} \to N_{b}$, we have:
    $$\lim_{N_p\to N_b} C(n\cdot \bm{x'_{t}})=y_t, n\geq1.$$

\hfill$\square$

\section{The Detailed Configurations of the Empirical Study}
\label{sec:empirical_configuration}
We adopt BadNets \citep{DBLP:journals/corr/abs-1708-06733}) and ISSBA \citep{ISSBA} as the example for our discussion. They are the representative of patch-based and non-patch-based backdoor attacks, respectively. We conduct experiments on the CIFAR-10 dataset \citep{CIFAR} with ResNet-34 \citep{hedeep}. For both attacks, we inject a large number of poisoned samples to ensure a high attack success rate ($\geq 99\%$). For each benign and poisoned image, we gradually enlarge its pixel values with multiplication. We calculate the \emph{averaged confidence} defined as the averaged probabilities of samples on the originally predicted label. In particular, we select the label predicted upon the original sample as the originally predicted label for each varied sample and constrain all pixel values within $[0,1]$ during the multiplication process. In particular, we follow previous works~\citep{DBLP:journals/corr/abs-1708-06733,ISSBA} to implement the backdoor attacks. Specifically, the trigger for BadNets is a $4\times4$ square consisting of random pixel values; the trigger of ISSBA is generated via DNN-based image steganography \citep{tancikstegastamp}. Both attacks are implemented via \texttt{BackdoorBox} \citep{libackdoorbox}.

\section{The Details for Training Attacked Models}
We train backdoor-infected models using BackdoorBox~\citep{libackdoorbox}. We set the training epoch as 200 and the poisoning rate as $5-10\%$ for each attack to ensure a high attack success rate. In particular, except for PhysicalBA, we don't involve additional data augmentation in training infected models as we want to better reveal the properties of various backdoor approaches. For each infected model, we randomly select infected labels to ensure their predictions on benign inputs are similar to the benign models, which ensures the stealthiness of backdoor attacks. Regarding the data-limited scenario and ablation study, we intentionally affect multiple labels to ensure the infected models similar to benign models except for the Trojan behaviors. Specifically, we inject less amount($\leq 5\%$) of poisoned samples to affect labels other than the target label. This is because previous work~\citep{AEVA} found that certain dense backdoor attacks (\eg, ISSBA, WaNet) would make the infected DNNs sensitive to noisy or out-of-distribution samples on CIFAR-10 dataset. Accordingly,they are less stealthy and are easy to be detected during the sampling process of the data-limited setting and settings of our ablation study. As such, in these settings, to evaluate \name{} in a rather practical scenario, we train infected DNNs to have similar behaviors on noisy samples as the benign DNNs. The details for each dataset are included in Table \ref{table:tab1}.    

\begin{table}[!t]
\centering
\caption{Detailed information about the adopted datasets. }
\begin{tabular}{cccc}
\toprule

Dataset         & \# Classes & Image Size & \# Training Images \\ \midrule
CIFAR-10     & 10           & $3 \times 32\times 32$      & 50,000                \\ 
Tiny ImageNet & 200          & $3 \times 64\times 64$      & 1,000,000              \\ \bottomrule
\end{tabular}
\label{table:tab1}
\end{table}

\subsection{The Accuracy and Attack Success Rate (ASR) for Evaluated Models}
 \label{sec:ASR_info}
 The accuracy and ASR for the evaluated models for each task in included in Table ~\ref{table:over}.
 
 \begin{table}[H]
\centering
\caption{The BA and ASR for the evaluated models on each dataset.}
\begin{tabular}{cccc}
\toprule
\multirow{2}{*}{Task$\downarrow$~Model$\rightarrow$} & \multicolumn{2}{c}{Infected Model} & \multirow{2}{*}{Normal Model Accuracy} \\ \cline{2-3}
              & BA & ASR &         \\ \hline
CIFAR-10         & $\geq 90.04\%$                 &$ \geq97.7\%$              & $\geq92.31\%$ \\ \hline
Tiny ImageNet        & $\geq36.98\%$                 & $\geq97.22\%$             & $\geq40.11\%$ \\ \bottomrule

\end{tabular}

\begin{table}[H]
\caption{\small The performance of six defense baselines against partial backdoor attacks}
\label{table:result_partial}
\centering
\scalebox{0.8}{
\begin{tabular}{c|c|c|c|c|c|c}
\toprule
Task$\downarrow$  ~ Attack $\rightarrow$      & STRIP & ShrinkPad & Frequency & DeepSweep & Ours (data-free) & Ours (data-limited) \\ \hline
CIFAR-10     & 0.617              & 0.949     & 0.891     & 0.967    & 0.971            & 0.971               \\ \hline
Tiny ImageNet & 0.601              & 0.868     & 0.861     & 0.951    & 0.936            & 0.971               \\ \bottomrule
\end{tabular}}
\end{table}
\label{table:over}
\end{table}

\section{The Detailed Configurations for Baseline Defenses}

\begin{itemize}
    \item \textbf{STRIP:} We implement STRIP following their official open-sourced codes\footnote{ \url{https://github.com/garrisongys/STRIP.git}}. 
    \item \textbf{ShrinkPad:} We implement ShrinkPad following their official open-sourced codes\footnote{\url{https://github.com/THUYimingLi/BackdoorBox.git}}.
    \item\textbf{Frequeny:} We implement Frequency approach following their official codes\footnote{\url{https://github.com/YiZeng623/frequency-backdoor.git}}.
    \item\textbf{DeepSweep:} We implement DeepSweep using \texttt{Scipy} package to remove the high-frequency noise and use \texttt{torchvision.transforms} and \texttt{keras.preprocess} packages to conduct transformation to inputs. Notably, we don't apply \texttt{finetune} process within DeepSweep since we only focus on the black-box detection scenarios. 
\end{itemize}

\section{The Descriptions for Main Evaluation Metric}

\begin{itemize}
    \item The receiver operating curve (ROC) shows the trade-off between detection
the success rate for poisoned samples and detection error rate for benign samples across different decision
thresholds $T$ under infected-DNNs.
    \item Inference Time: we implement each approach
under the platform with one NVIDIA GPU 1080 Ti and a Intel(R) Xeon(R) CPU E5-2650 v4 @
2.20GHz with batch size $=1$. We test the inference time of each approach with an average of 1,000 runs.
\end{itemize}

\section{Settings For Measuring the Inference Time}

Since we focus on defending against backdoor attacks in the inference phase, we here measure the inference time by:
\begin{itemize}
    \item Identifying whether the input sample is poisoned or not.
    \item If the input is a benign sample, we next should use the target model to predict it. 
\end{itemize}
For STRIP, ShrinkPad, DeepSweep, and \name{}, we leverage the target model's prediction on the (augmented) inputs for defense purpose, which means the input can be identified and predicted at the same time. As for Frequency, which leverages a secondary neural network to predict the frequency domain of each given input. However, if the input is identified as benign, the target DNNs should also deliver prediction on it. We here assume the benign and poisoned samples have equal possibilities. Therefore, we calculate the inference time for Frequency as follows:
\begin{equation}
    \text{time} = \text{TIME}(\text{Frequency(input)})+0.5\cdot \text{TIME}(\text{DNN(input)}). 
\end{equation}
While for other approaches, we measure their inference time via:
\begin{equation}
    \text{time} = \text{TIME(DNN((Augumented) INPUT)))}.
\end{equation}

In particular, we calculate the inference time of methods required to obtain the predictions of multiple images by feeding them simultaneously (in a batch) into the deployed model instead of predicting them one by one. This approach is feasible since defenders can easily and efficiently obtain all of them before feeding them into the deployed model.

\section{Performance Under Multiple-Backdoor Triggers Within a Single Infected Label}

\begin{figure}[!t]
\centering
\vspace{-2em}
\subfigure[Trigger I\label{fig:t1}]{\includegraphics[width=0.24\textwidth]{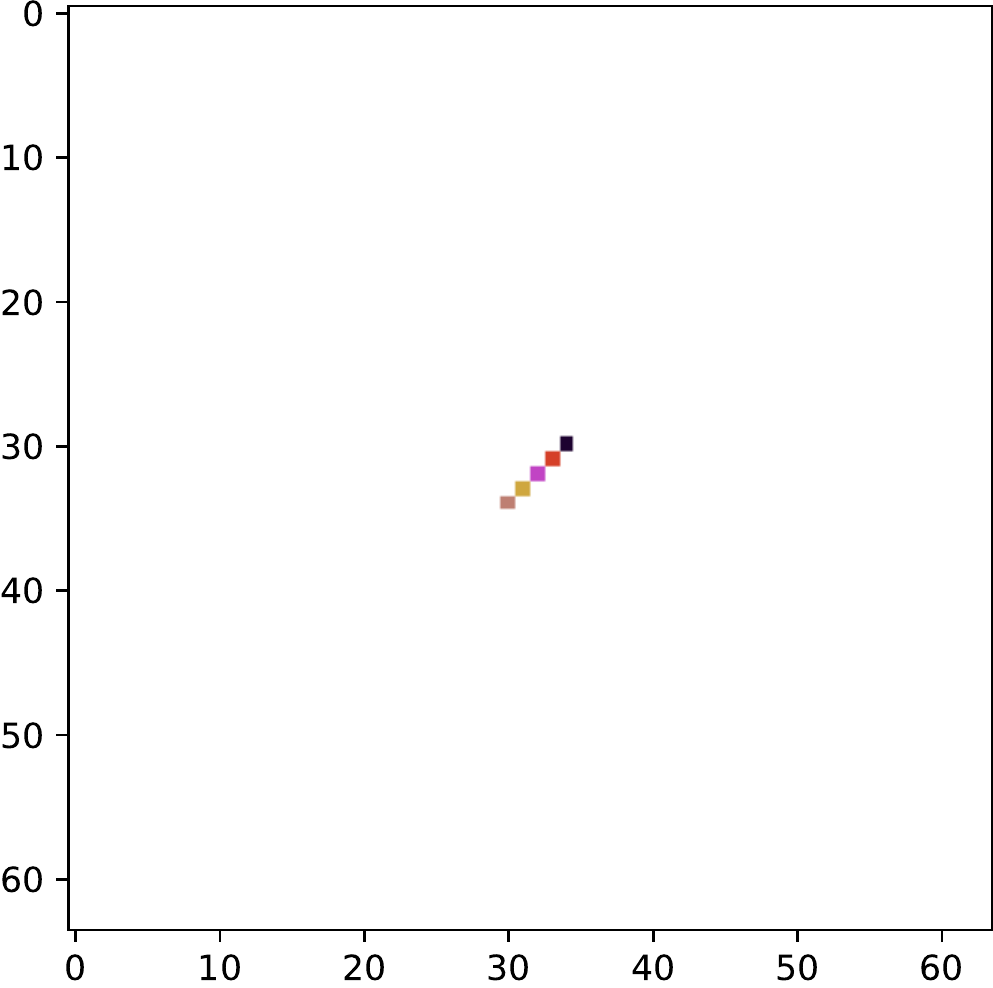}}
\hspace{0.5em}
\subfigure[Trigger II\label{tirggerII}]{\includegraphics[width=0.24\textwidth]{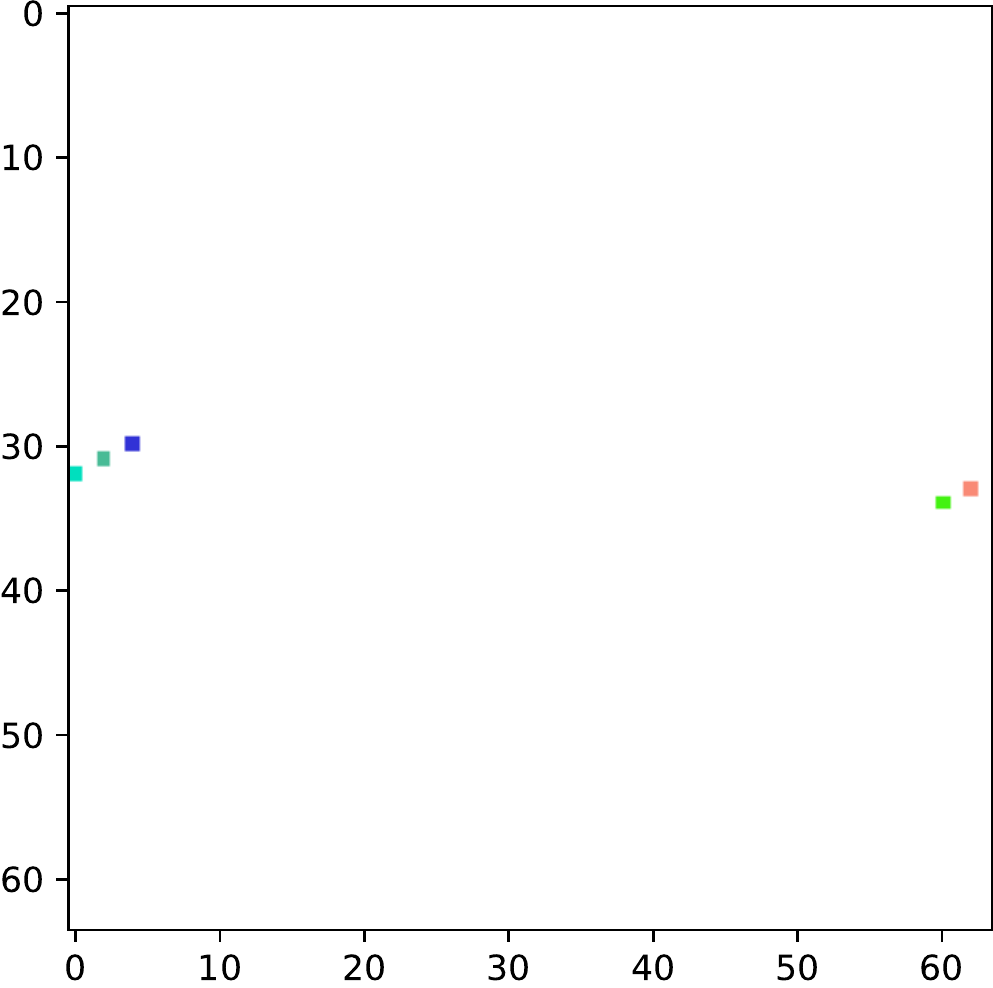}}
\hspace{0.5em}
\subfigure[Trigger III\label{fig:t2}]{\includegraphics[width=0.24\textwidth]{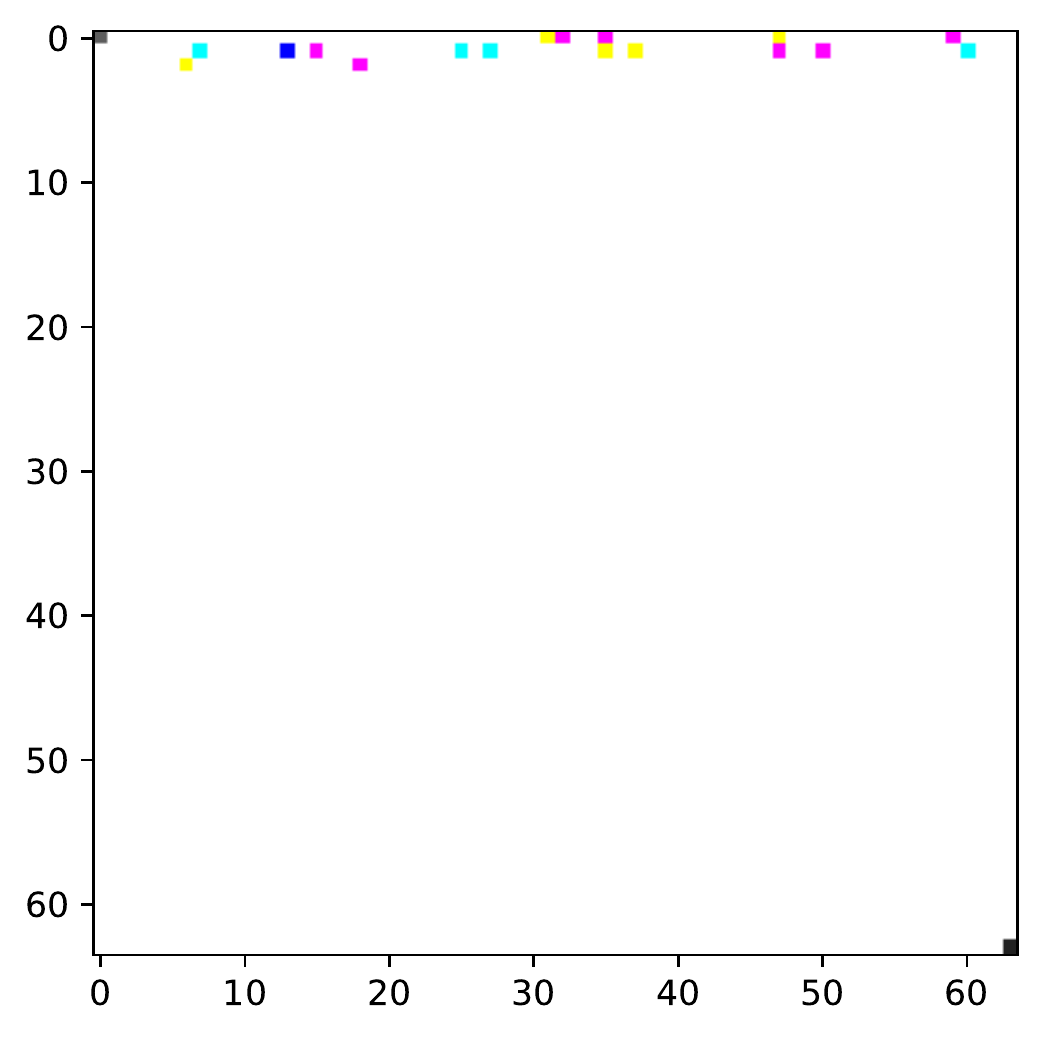}}
\vspace{-2mm}
\caption{The demonstration of dynamic triggers.}
\vspace{-3mm}
\label{fig:multi_label}
\end{figure}

\begin{figure}[!t]
\centering
\subfigure[CIFAR-10\label{fig:cifar10_mull}]{\includegraphics[width=0.48\textwidth]{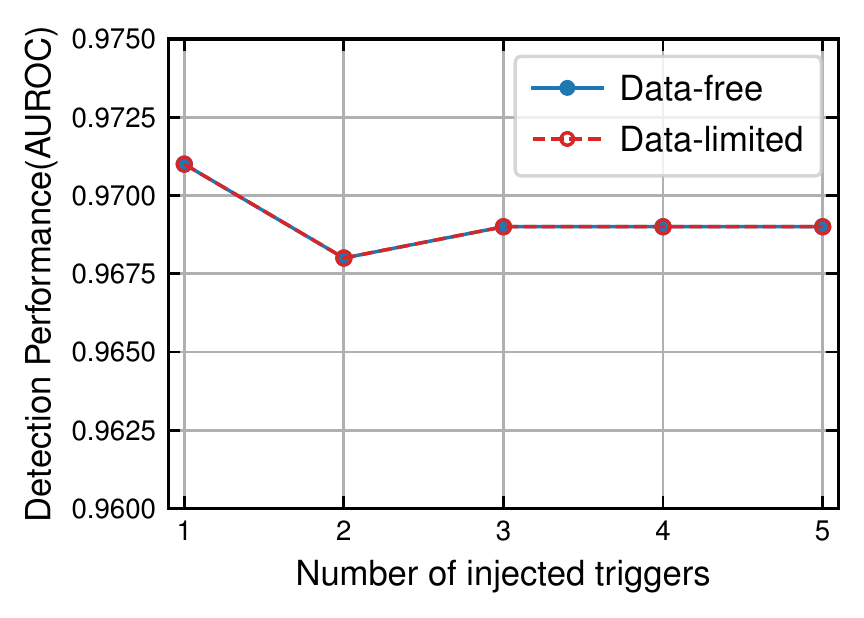}}
\subfigure[TinyImageNet\label{tinyimagenet_mul}]{\includegraphics[width=0.48\textwidth]{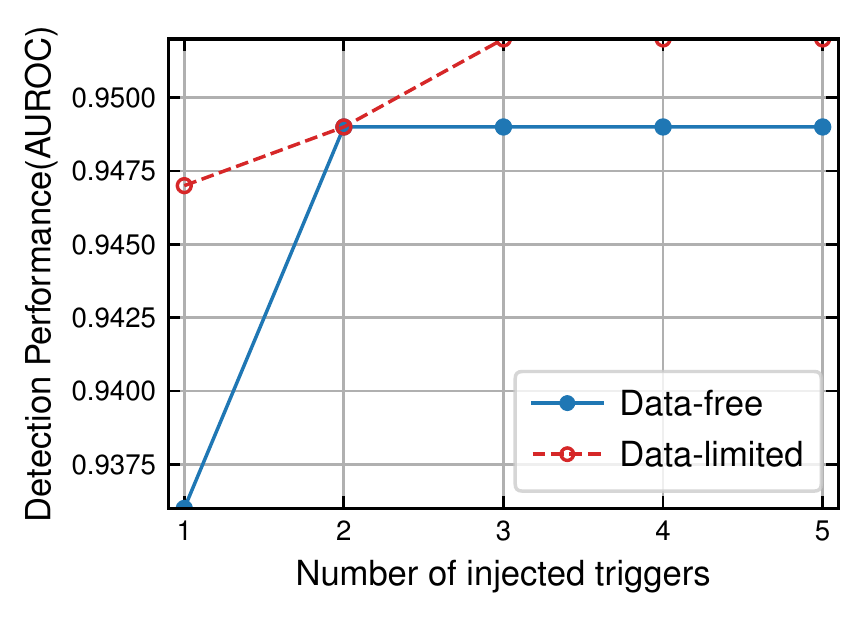}}
\vspace{-2mm}
\caption{The average results for multiple triggers within a single label.}
\vspace{-3mm}
\label{fig:multi_label_per}
\end{figure}

\begin{wrapfigure}{h}{0.30\textwidth}\vspace{-4mm}
  \begin{center}
  \vspace{-4mm}
    \includegraphics[width=0.30\textwidth]{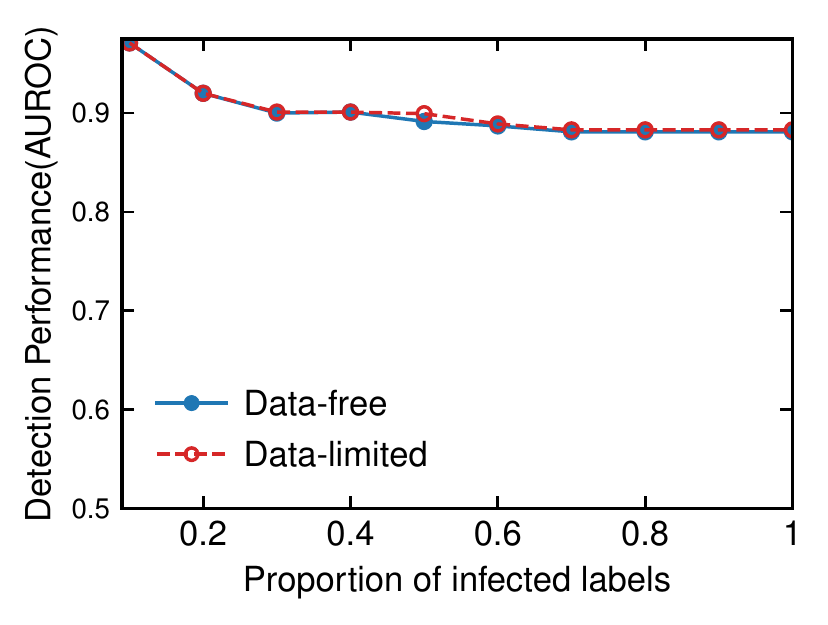}
  \end{center}
  \vspace{-4mm}
  \caption{The performance of \name{} for multiple infected labels. }%The results are averaged over the number of infected labels.}
 \label{fig:multi_labels}
 \vspace{-1em}
\end{wrapfigure}

Consistent with~\citep{AEVA,nc}, we also evaluate the efficacy of \name{} under a more challenging scenario where multiple backdoors are embedded within a single target label. We randomly select a label as the  infected label and inject various types of poisoned samples in the training phase. We inject arbitrary amounts of poison samples for each backdoor trigger to ensure the attack efficacy $ASR \geq 99\%.$ The demonstrations for used backdoor triggers are shown in Figure \ref{fig:multi_label}. Under such considered scenario, we evaluate our \name{} on CIFAR-10 and Tiny ImageNet datasets using ResNet-34.  

As shown in Figure \ref{fig:multi_label_per}, \name{} performs resilient to the increasing number of injected backdoor triggers. This may be caused by infected models already generalized for backdoor triggers.

\section{The Performance Against Multiple Infected Labels}

We also test \name{} under the scenario where the suspicious model has multiple infected labels. Under this scenario, we test \name{} on CIFAR-10. This is because models on Tiny ImageNet would be prone to multiple infected labels, as reported by~\citep{AEVA}, affecting more than $14\%$ labels can make the accuracy significantly drop$\geq 3\%$. We implement BadNets as backdoor attacks, the trigger size is $4\times 4$. The results are shown in Figure \ref{fig:multi_labels}.
These results show that affecting multiple infected labels could slightly reduce the performance of \name{}. Besides, the data-limited scenario performs better than the data-free scenario. However, even with $100\%$ labels are infected, \name{} can still perform effectively with AUROC $\geq 0.883$.

% \begin{figure}[H]
% \centering
% \subfigure[CIFAR-10\label{infected_3d}]{\includegraphics[width=0.4\textwidth]{figure/experiments/mul_infected.pdf}}
% \vspace{-2mm}
% \caption{The performance of \name{} for multiple infected labels. The results are averaged over the number of infected labels}
% \vspace{-3mm}
% \label{fig:multi_labels}
% \end{figure}  
\begin{figure}[!t]
\centering
\vspace{-1em}
\subfigure[BadNets\label{fig:badnet_impact_n}]{\includegraphics[width=0.47\textwidth]{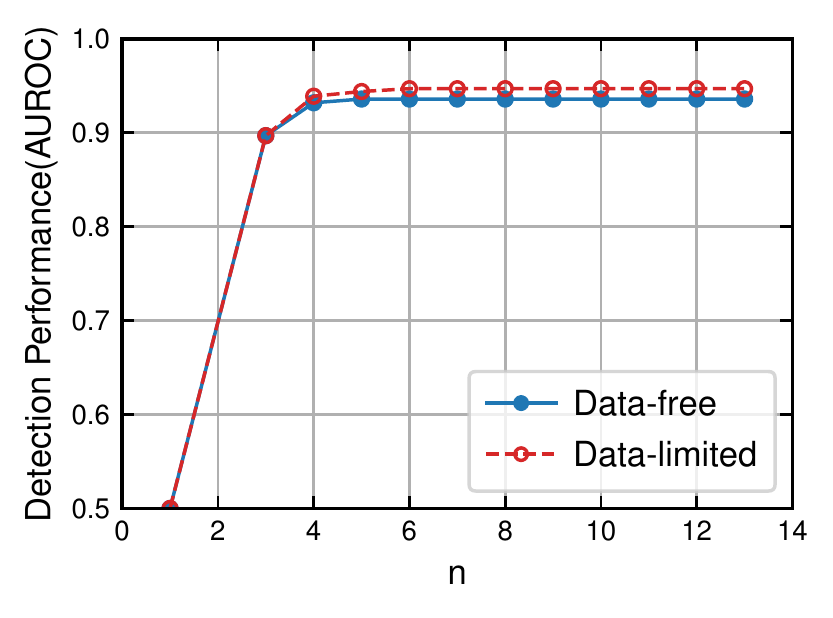}}\hspace{2em}
\subfigure[TUAP\label{tuap_impact_n}]{\includegraphics[width=0.47\textwidth]{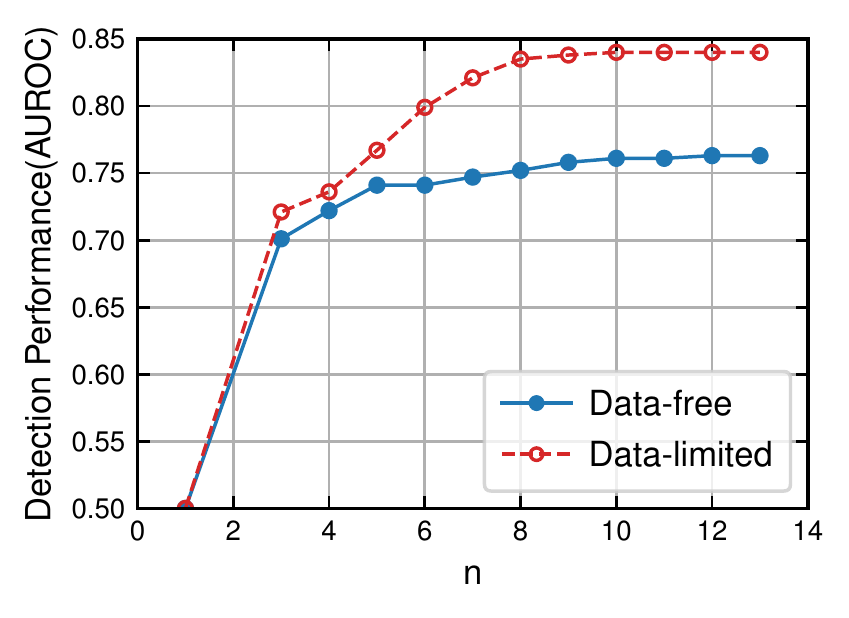}}
\subfigure[Label-Consistent\label{n_lc}]{\includegraphics[width=0.47\textwidth]{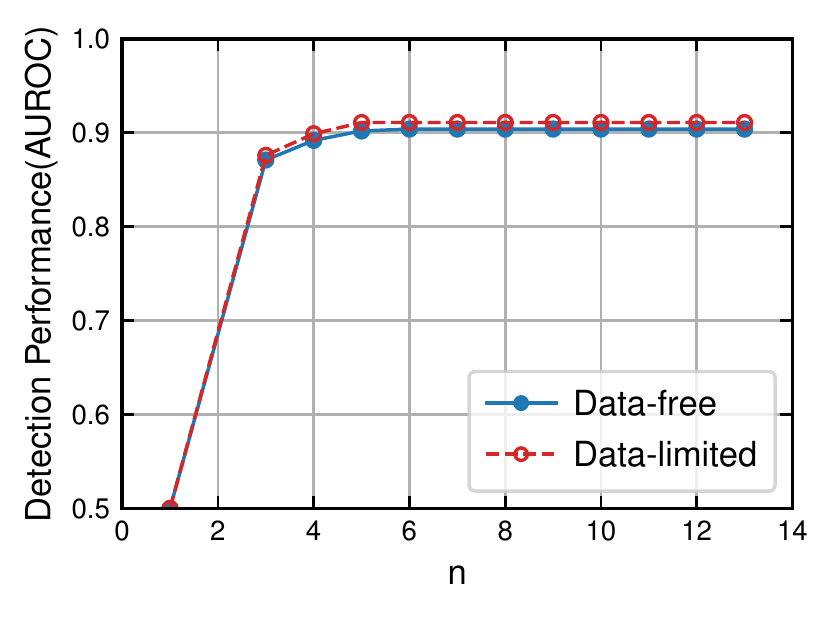}}\hspace{2em}
\subfigure[PhysicalBA\label{physicalBA}]{\includegraphics[width=0.47\textwidth]{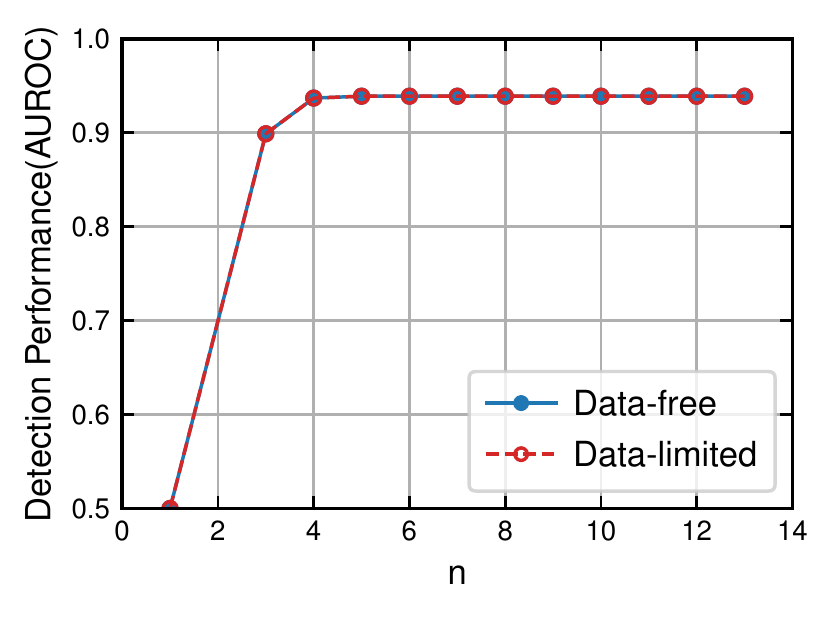}}
\subfigure[ISSBA\label{issba_n_impact}]{\includegraphics[width=0.47\textwidth]{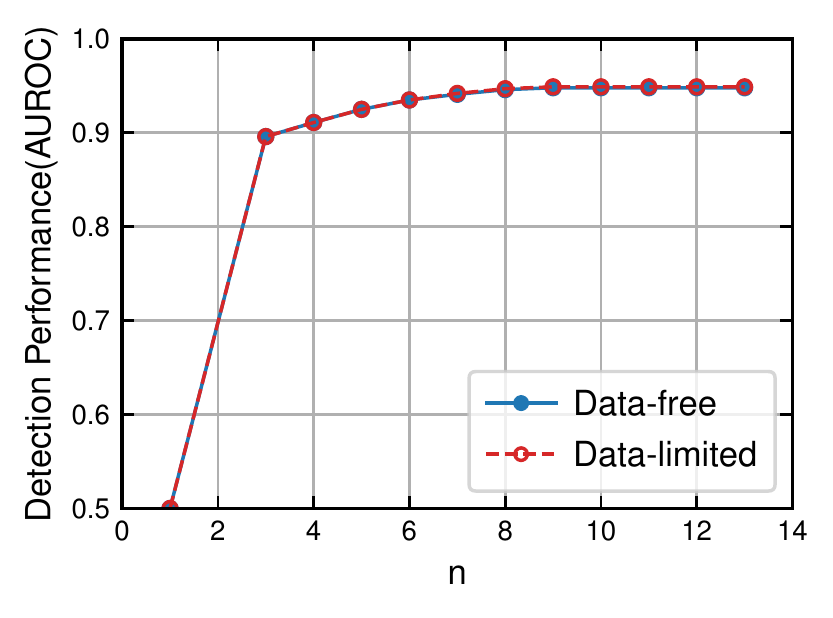}}\hspace{2em}
\subfigure[WaNet\label{wanet_n_impact}]{\includegraphics[width=0.47\textwidth]{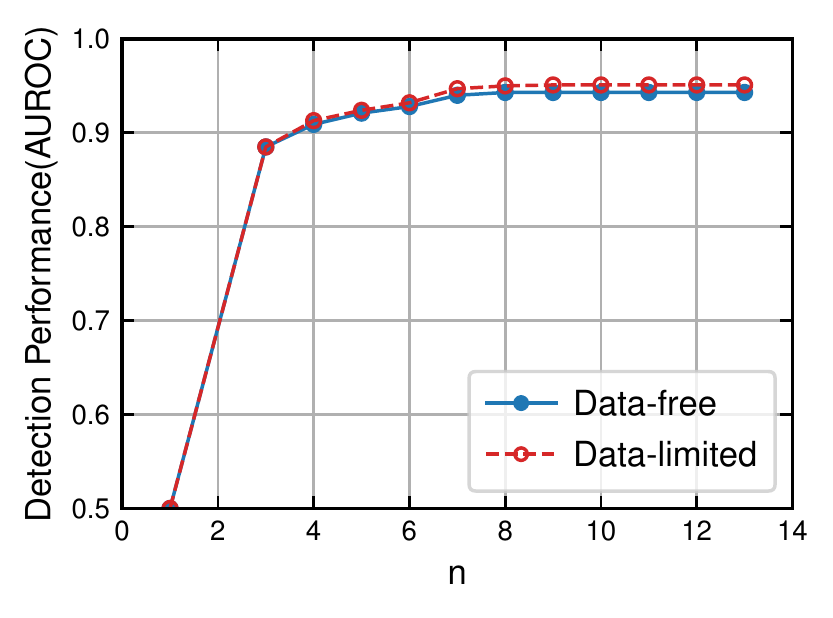}}
\vspace{-0.8em}
\caption{The impact for the coefficients $n$.}
\vspace{-1em}
\label{fig:impact_n}
\end{figure}  

\section{Impacts for the Number of Coefficients}
We test \name{} on six attacks with varying $n$. We here use ResNet-34 on TinyImageNet for evaluation. As shown in Figure \ref{fig:impact_n}, we find that the performance of \name{} increases along with $n$ increasing. Moreover, we find that \name{} performs more sensitive on $n$ for the TUAP attack compared with other attack techniques. Moreover, we find that \name{} performs similarly sensitive on $n$ in both data-limited and data-free settings. In most settings, with $n\geq 11$ \name{} can achieve optimal performance on six different attacks.

\section{Impact for the Size of Local Samples}

\begin{figure}[!t]
\centering
\subfigure[BadNets\label{fig:badnets_size}]{\includegraphics[width=0.42\textwidth]{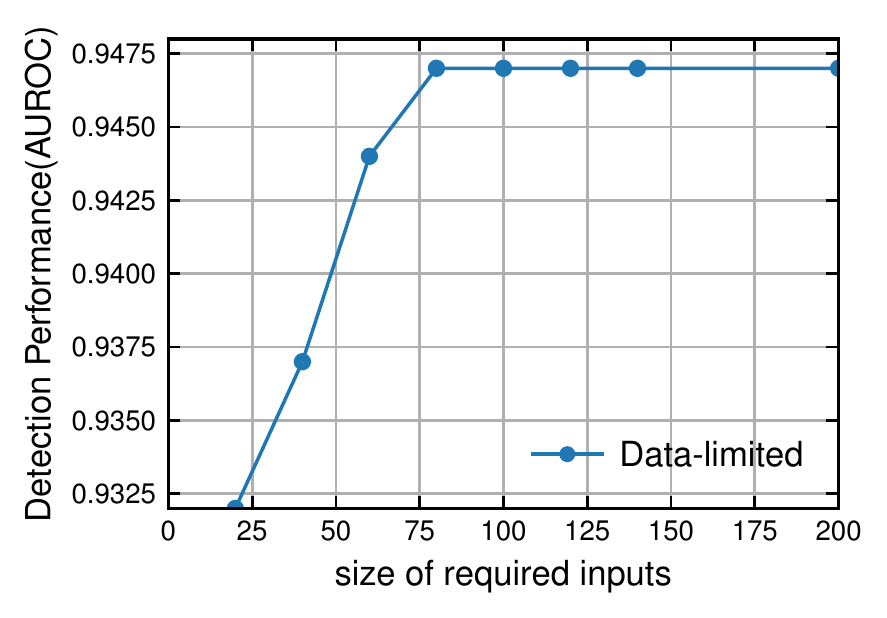}} \hspace{1.5em}
\subfigure[TUAP\label{tuap_size}]{\includegraphics[width=0.42\textwidth]{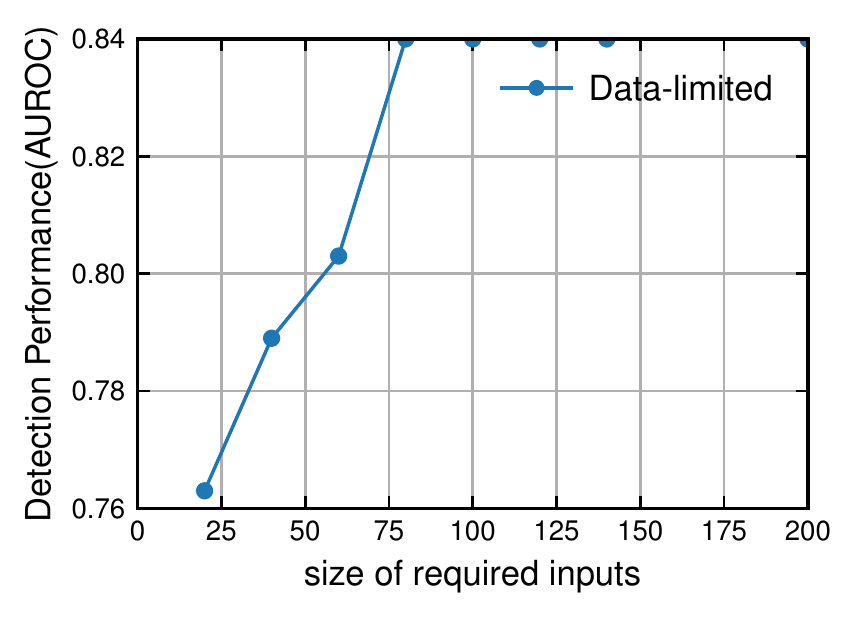}}
\subfigure[Label-Consistent\label{size_lc}]{\includegraphics[width=0.42\textwidth]{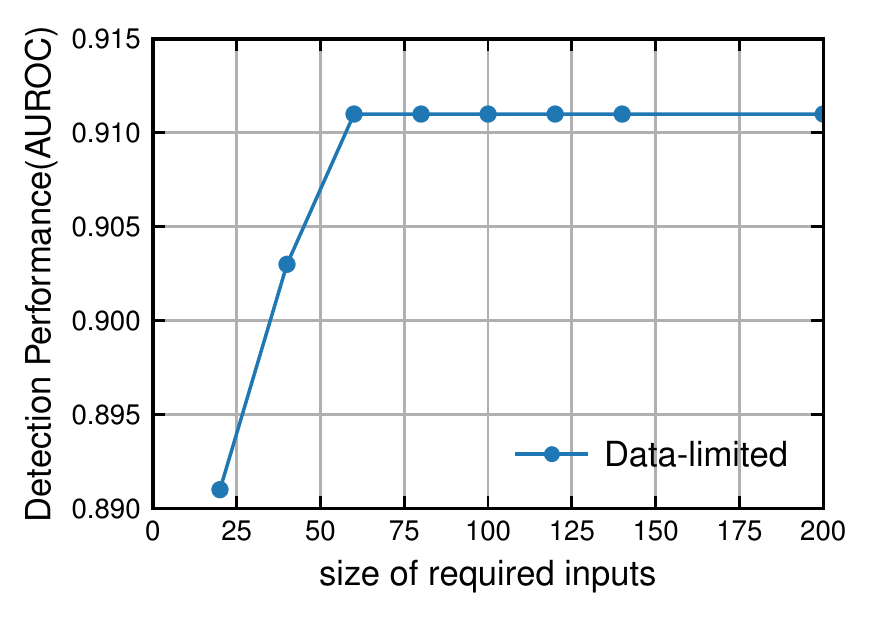}} \hspace{1.5em}
\subfigure[PhysicalBA\label{physicalBA_size}]{\includegraphics[width=0.42\textwidth]{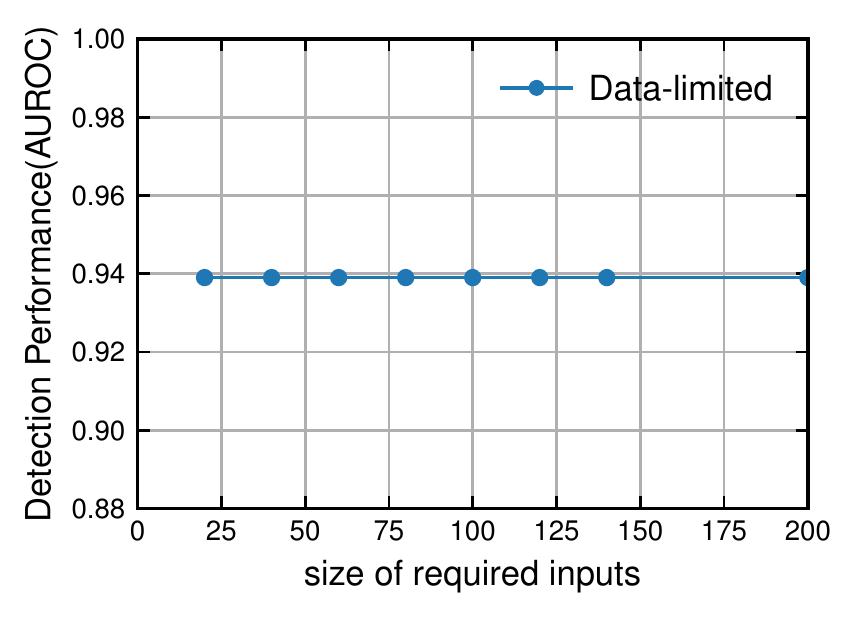}}
\subfigure[ISSBA\label{issba_size}]{\includegraphics[width=0.42\textwidth]{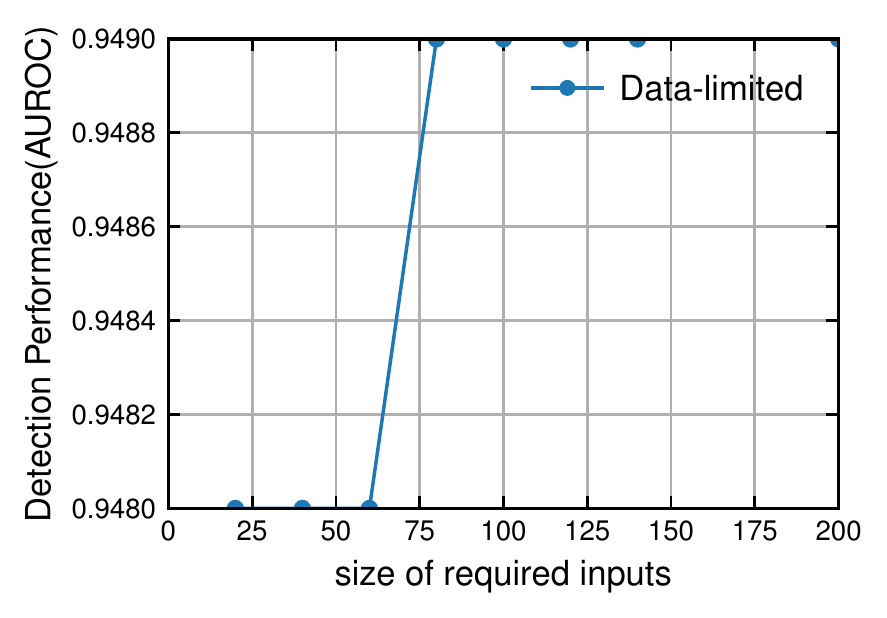}} \hspace{1.5em}
\subfigure[WaNet\label{wanet_size}]{\includegraphics[width=0.42\textwidth]{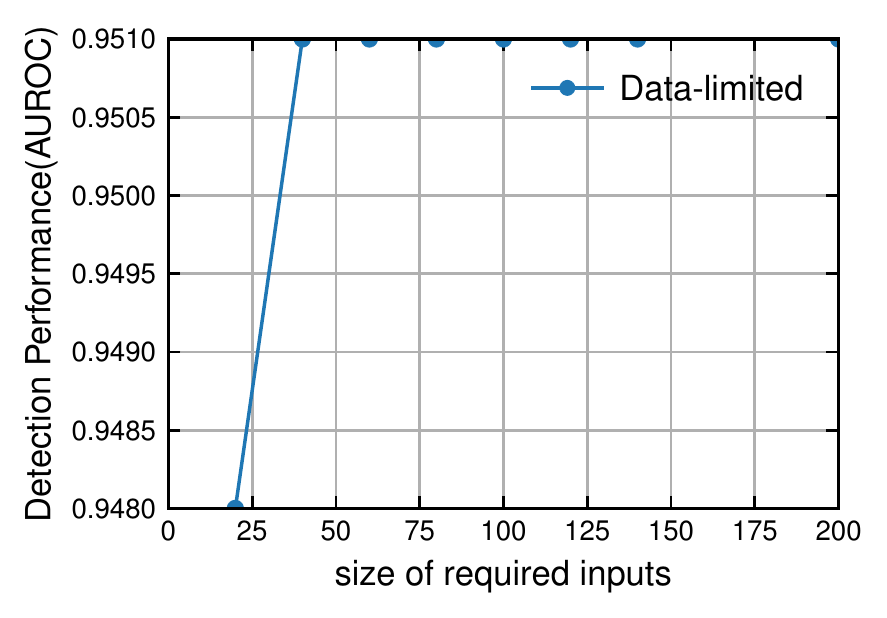}}
\vspace{-2mm}
\caption{The impact for the size of required inputs.}
\vspace{-3mm}
\label{fig:impact_size}
\end{figure}

We also test the sensitivity of \name{} on the size of local samples per label under the data-limited setting. We test \name{} on Tiny ImageNet using ResNet-34 against six attacks. The results are shown in Figure \ref{fig:impact_size}. We can see that with the size of local samples increases, the performance of \name{} improves and achieves optimal performance when the size $\geq 100.$

\section{Performance under Source-label-specific Backdoor Scenarios}

The Source-label-specific (Partial) backdoor scenarios is that the backdoor attacks can perform effectively when it applies to images of a certain specific class. Such a scenario makes backdoor attacks very hard to detect~\citep{nc,gaodesign}, thus the attacker may have a great incentive to implement such a backdoor attack in the real world. Therefore, we evaluate \name{} under such a practical scenario and compare \name{} with previous work. We test \name{} using ResNet-34 on CIFAR-10 and Tiny ImageNet.
As shown in Table \ref{table:result_partial}, we find that  most defense approaches perform resilient against the partial backdoor attack except STRIP~\citep{gaodesign}. This is because STRIP assumes the trigger can perform effectively across various images. Under this scenario, \name{} can outperform all baseline defenses.

\begin{figure}[!ht]
\centering
\subfigure[BadNets\label{fig:badnets_robust}]{\includegraphics[width=0.42\textwidth]{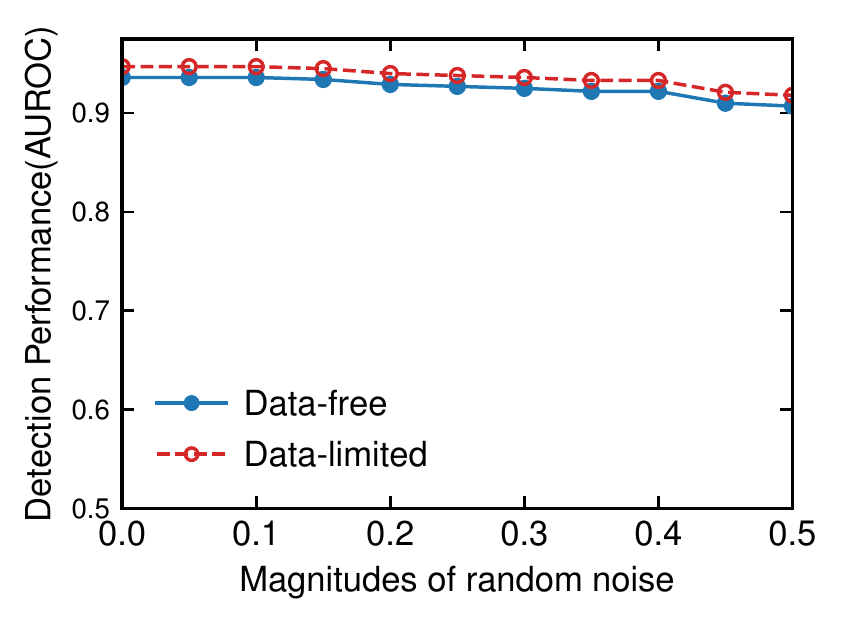}} \hspace{1.5em}
\subfigure[PhysicalBA\label{physical_robust}]{\includegraphics[width=0.42\textwidth]{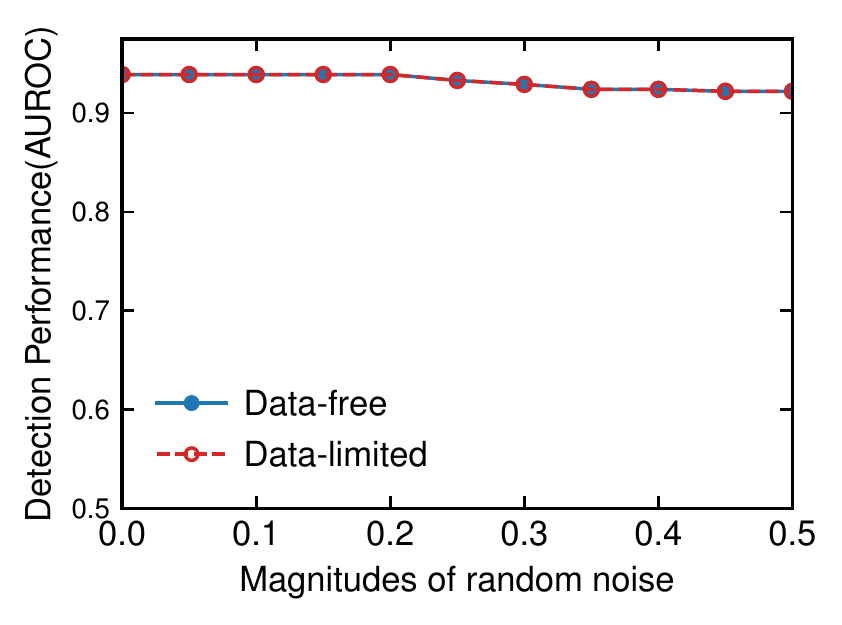}}
\vspace{-2mm}
\caption{The robustness of \name{}.}
\vspace{-3mm}
\label{fig:robustness}
\end{figure}

\section{The Robustness of \name{}}

Since \name{} is an inference-phase backdoor defense approach, it is necessary to investigate the robustness of \name{} on benign and poisoned samples. Following previous work~\citep{durobust}, we evaluate the robustness of our approach by testing different magnitudes of noisy inputs. Notably, we test \name{} on benign and poisoned samples, respectively, which is because they exhibit different robustness under random noise as we show in Section~\ref{sec:noise}. Moreover, the magnitudes of added random noise ensure the classification accuracy and attack success rate for benign and poisoned samples. The results are shown in Figure \ref{fig:robustness}. We test our approach using ResNet-34 on the TinyImageNet task. The noise is randomly sampled from Gaussian distribution and we intentionally filter the failed poisoned samples. We only test BadNets and PhysicalBA since only these two attacks perform robustness against random noise, as illustrated in Section \ref{sec:noise}. We find that our approach is robust against noisy poisoned samples.

% \begin{figure}[h]
% \centering
% \subfigure[BadNets\label{fig:badnets_robust}]{\includegraphics[width=0.3\textwidth]{figure/experiments/robust/robust_BadNets.pdf}}
% \subfigure[PhysicalBA\label{physical_robust}]{\includegraphics[width=0.3\textwidth]{figure/experiments/robust/robust_phy.pdf}}
% \vspace{-2mm}
% \caption{The robustness of \name{}.}
% \vspace{-3mm}
% \label{fig:robustness}
% \end{figure}

\begin{table}[!t]
    \centering
    \caption{The performance (AUROC) on the Tiny ImageNet dataset under VGG-19. Among all different methods, the best result is marked in boldface while the value with underline denotes the second-best result. The failed cases ($i.e.$, AUROC $<0.55$) are marked in red. Note that STRIP requires obtaining predicted probability vectors while other methods only need the predicted labels.}
         \scalebox{0.88}{
        \begin{tabular}{c|cccccc|c}
    \toprule
                  \tabincell{c}{Attack$\rightarrow$\\ Defense$\downarrow$}      &                   BadNets& Label-Consistent  & PhysicalBA    & TUAP &   WaNet                      & ISSBA    & \textbf{Average}                       
                         \\ \hline
STRIP   &   \textbf{0.941}                            & \underline{0.908}                             &  \textbf{0.941}    & 0.576  & \color{red}0.521  & \color{red}0.489      &   0.729                    \\ \hline
ShrinkPad      &  0.857                          &  \textbf{0.919} & 0.631                         & 0.831 & \color{red}{0.499}  &        \color{red}0.490        &  0.705         \\ 
DeepSweep       & \underline{0.939}                          &  0.907                          & \underline{0.921} & 0.744 &  \color{red}0.511  &        0.711         &       0.788 \\
 
Frequency   & 0.864                       &   0.859                            & 0.864 & 0.827 &  \color{red}0.428  & \color{red}0.540        & 0.730    \\
 \hline
Ours (data-free) & 0.936                       &   0.846                      & 0.907  & \underline{0.858} & \underline{0.893} & \underline{0.767} &  \underline{0.868}

\\
Ours (data-limited)                &  0.936                     &   0.851      & 0.907  & \textbf{0.888}   &  \textbf{0.904} &           \textbf{0.836}   &     \textbf{0.887}

\\ \bottomrule
    \end{tabular}}
    \centering
    %\vspace{-4mm}
     \label{table:result_vgg}
    \vspace{-1em}
\end{table}

\section{Additional Results under VGG Architecture}
In our main manuscript, we evaluate our method under the ResNet architecture. In this section, we conduct additional experiments under VGG-19 (BN) on Tiny ImageNet, to verify that the phenomenon of \emph{scaled prediction consistency} is valid across different model architectures.

\begin{figure}[t]
\centering
\subfigure[BadNets\label{fig:bd_vgg}]{\includegraphics[width=0.42\textwidth]{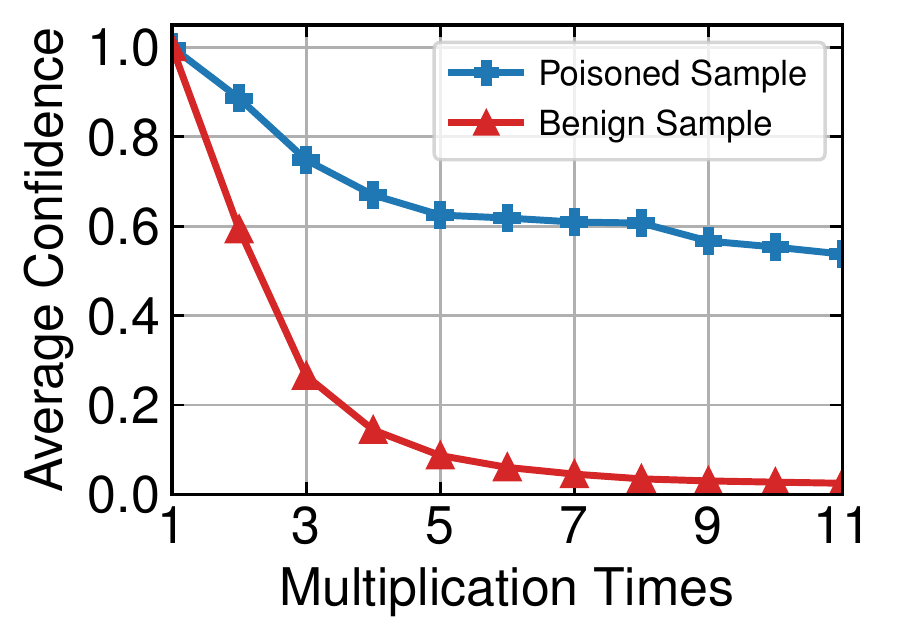}}\hspace{1.5em}
\subfigure[TUAP\label{tuap_vgg}]{\includegraphics[width=0.42\textwidth]{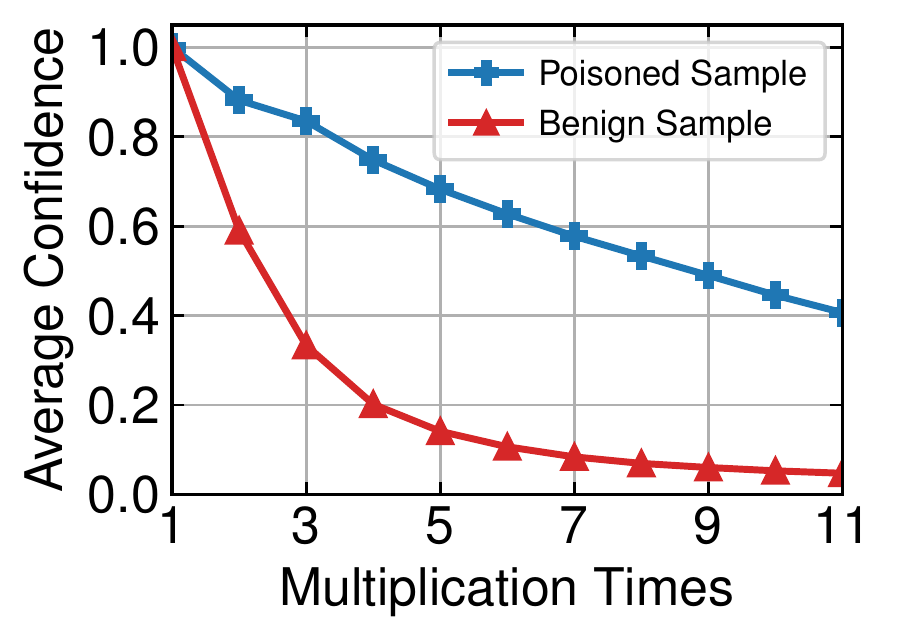}}
\subfigure[Label-Consistent\label{lc_vgg}]{\includegraphics[width=0.42\textwidth]{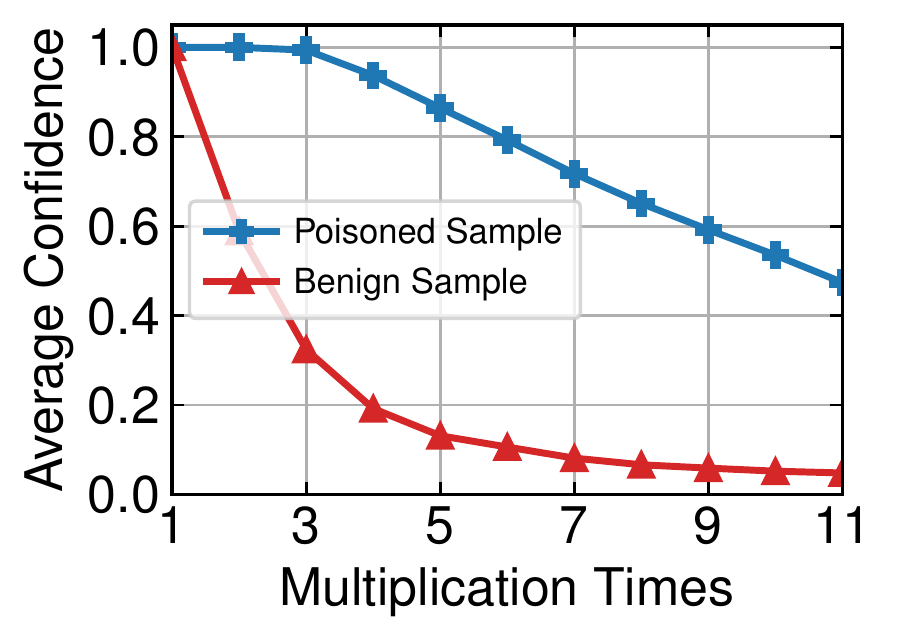}}\hspace{1.5em}
\subfigure[PhysicalBA\label{physicalBA_vgg}]{\includegraphics[width=0.42\textwidth]{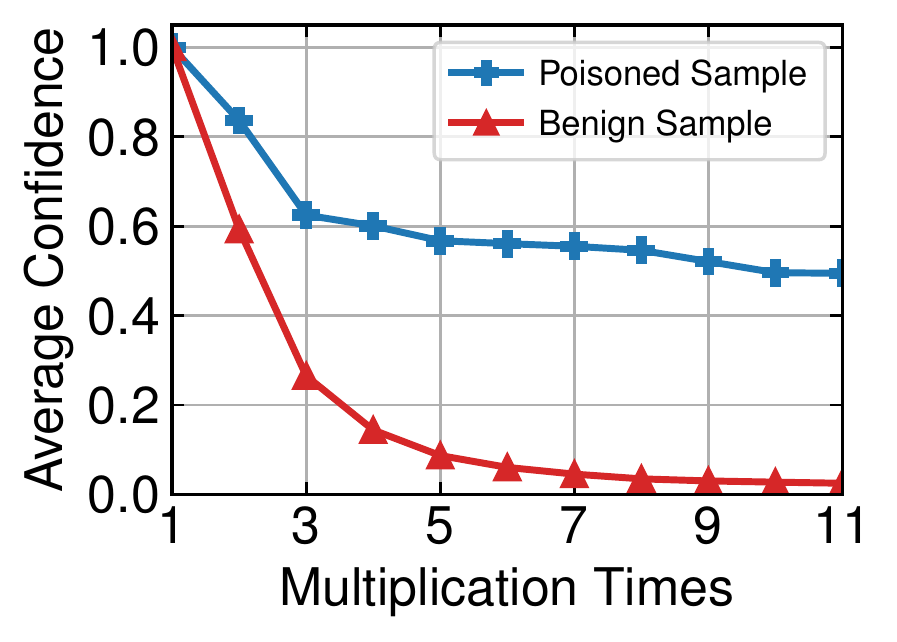}}
\subfigure[ISSBA\label{issba_vgg}]{\includegraphics[width=0.42\textwidth]{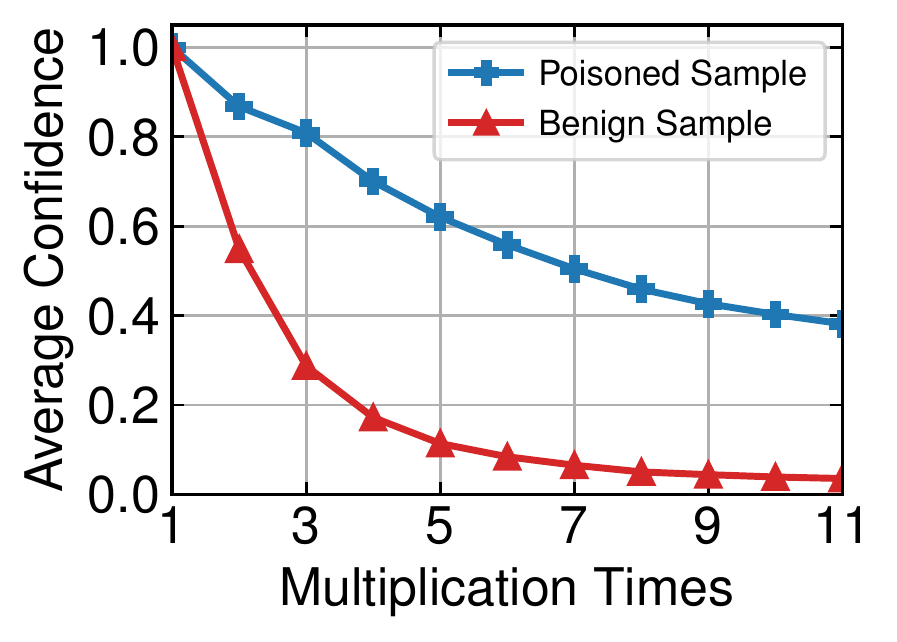}}\hspace{1.5em}
\subfigure[WaNet\label{wanet_vgg}]{\includegraphics[width=0.42\textwidth]{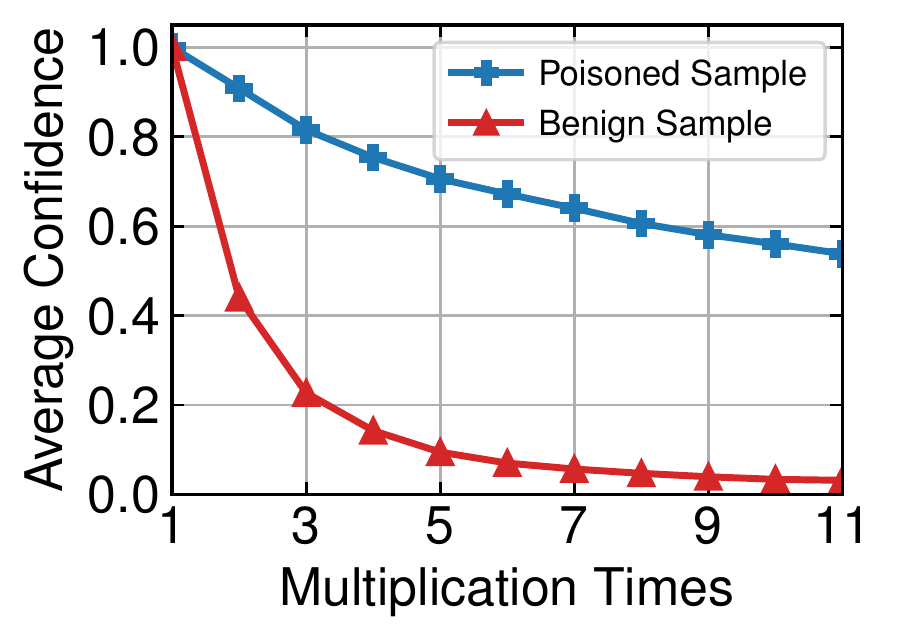}}
\vspace{-2mm}
\caption{The average confidence ($i.e.$, average probabilities on the originally predicted label) of
benign and poisoned samples $w.r.t.$ pixel-wise multiplications under benign and attacked models on the Tiny ImageNet dataset with VGG-19 (with batch normalization).}
\vspace{-3mm}
\label{fig:spc_vgg}
\end{figure}

As shown in Figure \ref{fig:spc_vgg}, the scaled prediction consistency still holds in all cases. Specifically, the average confidence of benign samples decreases significantly faster than that of poisoned ones with the increase in multiplication time. Besides, as shown in Table \ref{table:result_vgg}, our methods are still better than all baseline defenses. These results verify the effectiveness of our methods again.

%Since we have evaluated \name{} under ResNet models, we here investigate the validity of \textit{Scaled Prediction Consistency phenomenon} and the effectiveness of \name{} under the ConvNet model. We here test \name{} on VGG-19 with batch normalization implementation for Tiny-ImageNet, and the results for \textit{Scaled Prediction Consistency Phenomenon} and detection performance are shown in \cref{fig:spc_vgg} and \cref{table:result_vgg}. As seen in \cref{fig:spc_vgg} and \cref{table:result_vgg}, we find that our observation (\ie.,\textit{Scaled Prediction Consistency}) is still valid for the ConvNet architectur (\ie., VGG-Net).

\section{The ROC Curves of Defenses}
\label{ap:ROC}

\begin{figure}[!ht]
\centering
\subfigure[BadNets\label{fig:badnets_roc_cifar}]{\includegraphics[width=0.45\textwidth]{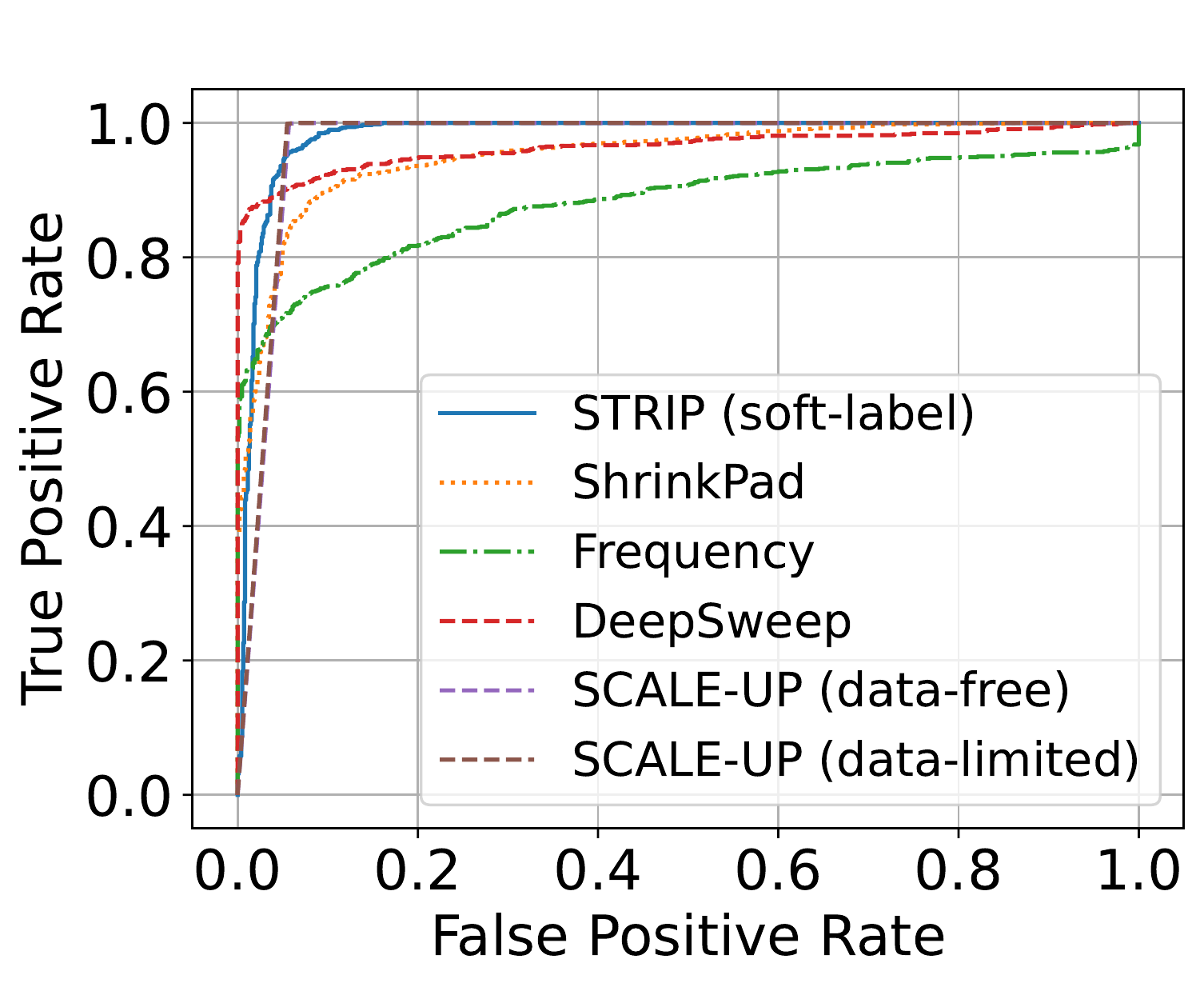}}\hspace{3em}
\subfigure[TUAP\label{tuap_roc_cifar}]{\includegraphics[width=0.45\textwidth]{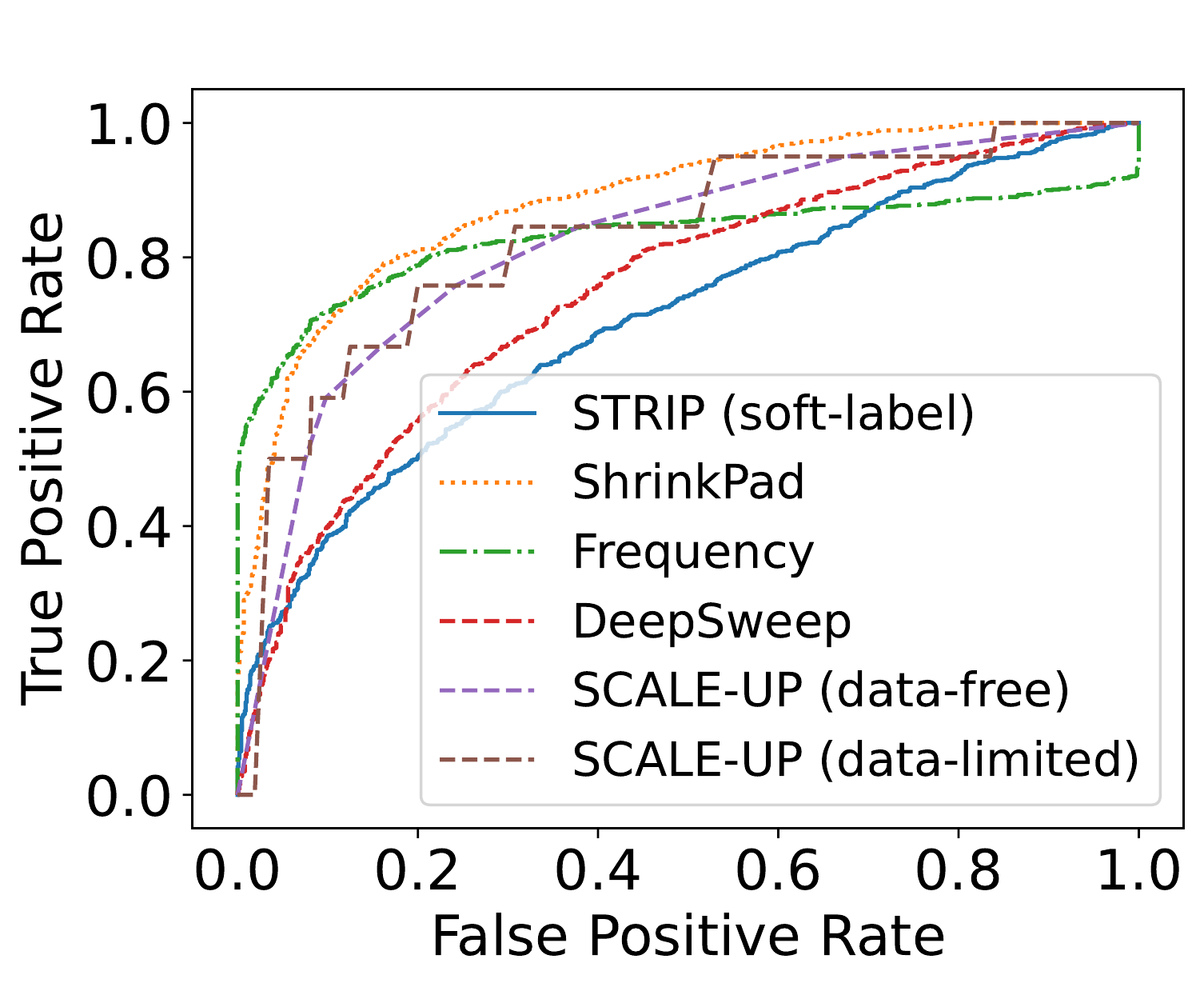}}
\subfigure[Label-Consistent\label{lc_roc_cifar}]{\includegraphics[width=0.45\textwidth]{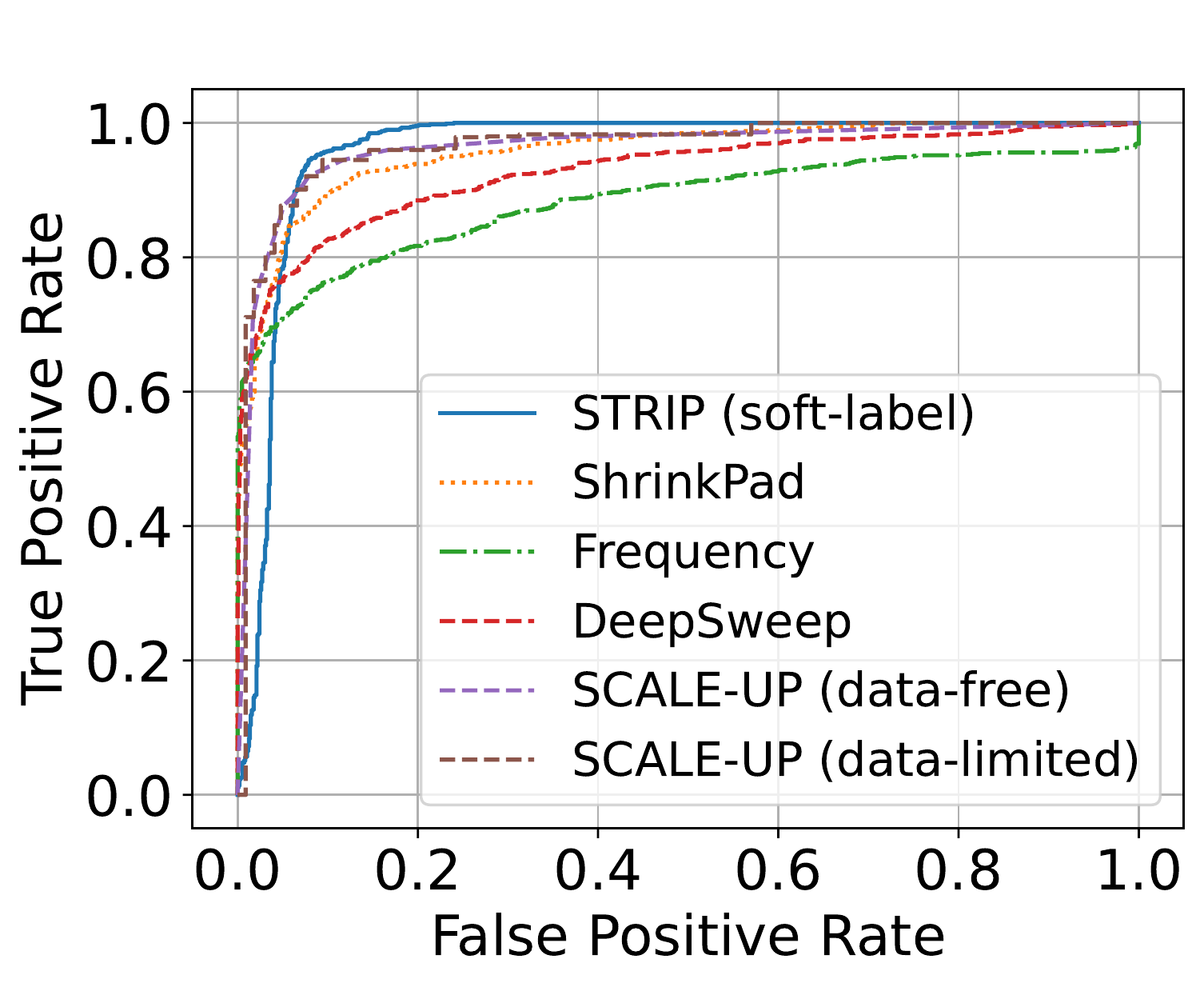}}\hspace{3em}
\subfigure[PhysicalBA\label{physicalBA_roc_phys}]{\includegraphics[width=0.45\textwidth]{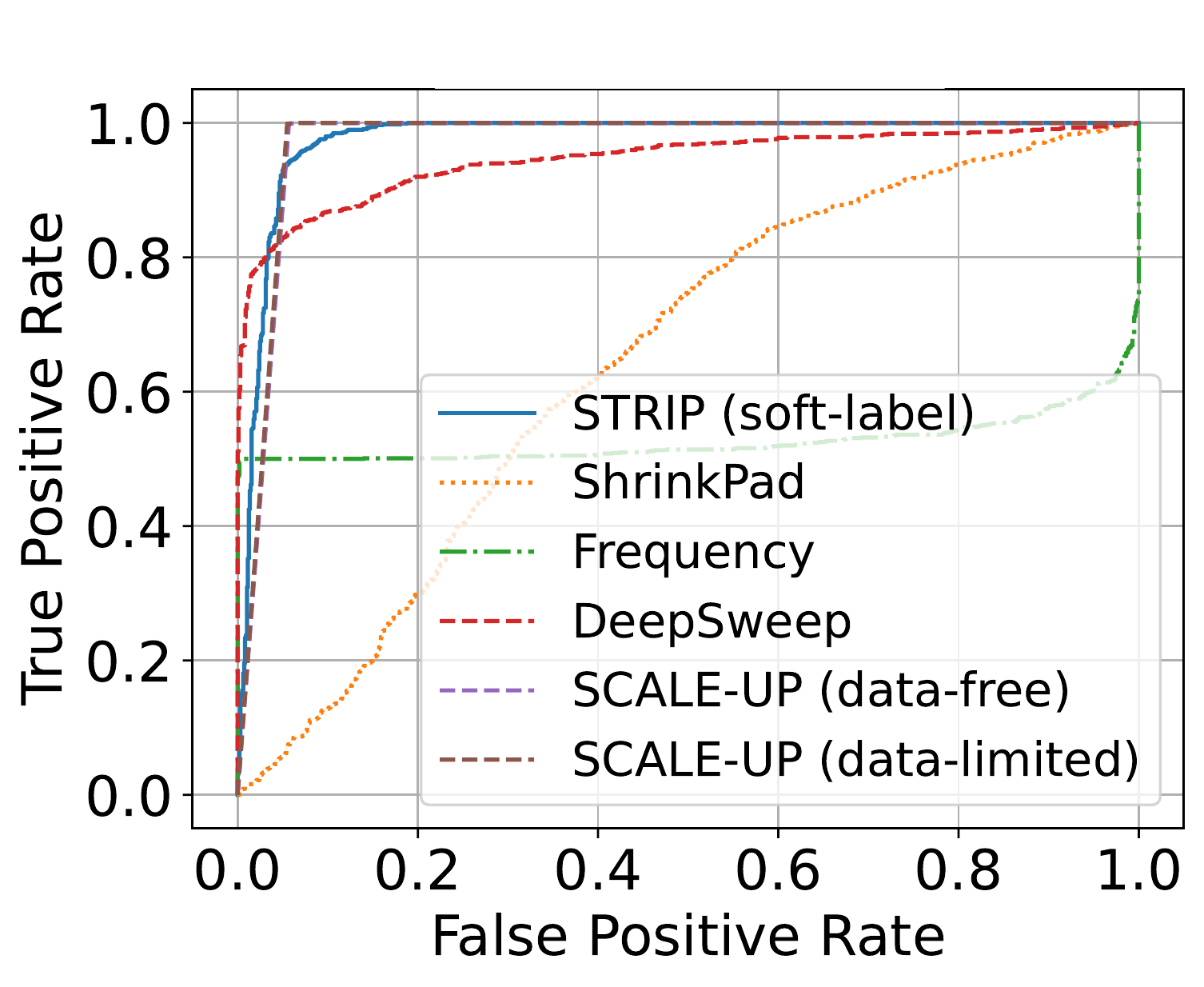}}
\subfigure[ISSBA\label{issba_roc_issba}]{\includegraphics[width=0.45\textwidth]{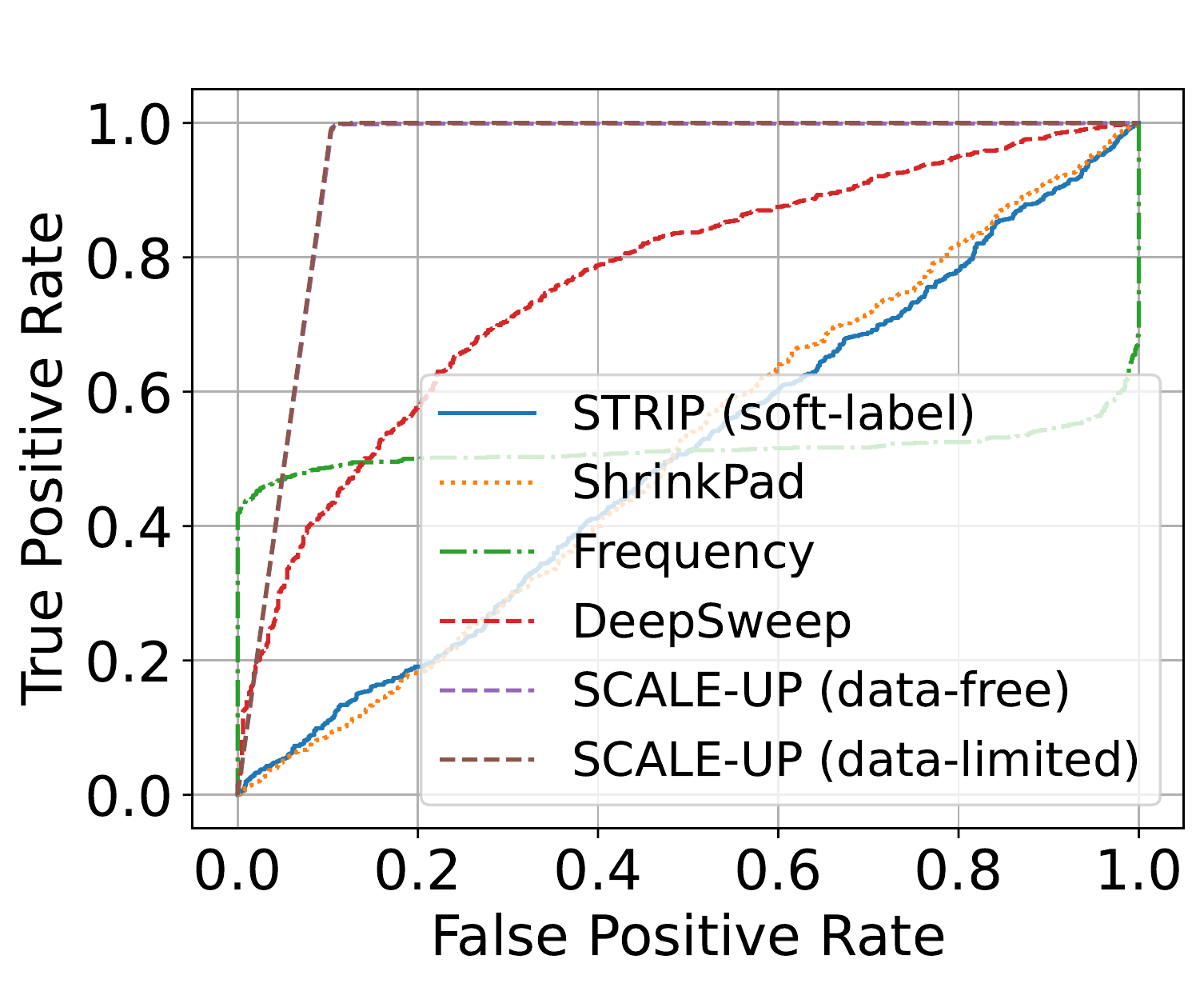}}\hspace{3em}
\subfigure[WaNet\label{wanet_roc_cifar}]{\includegraphics[width=0.45\textwidth]{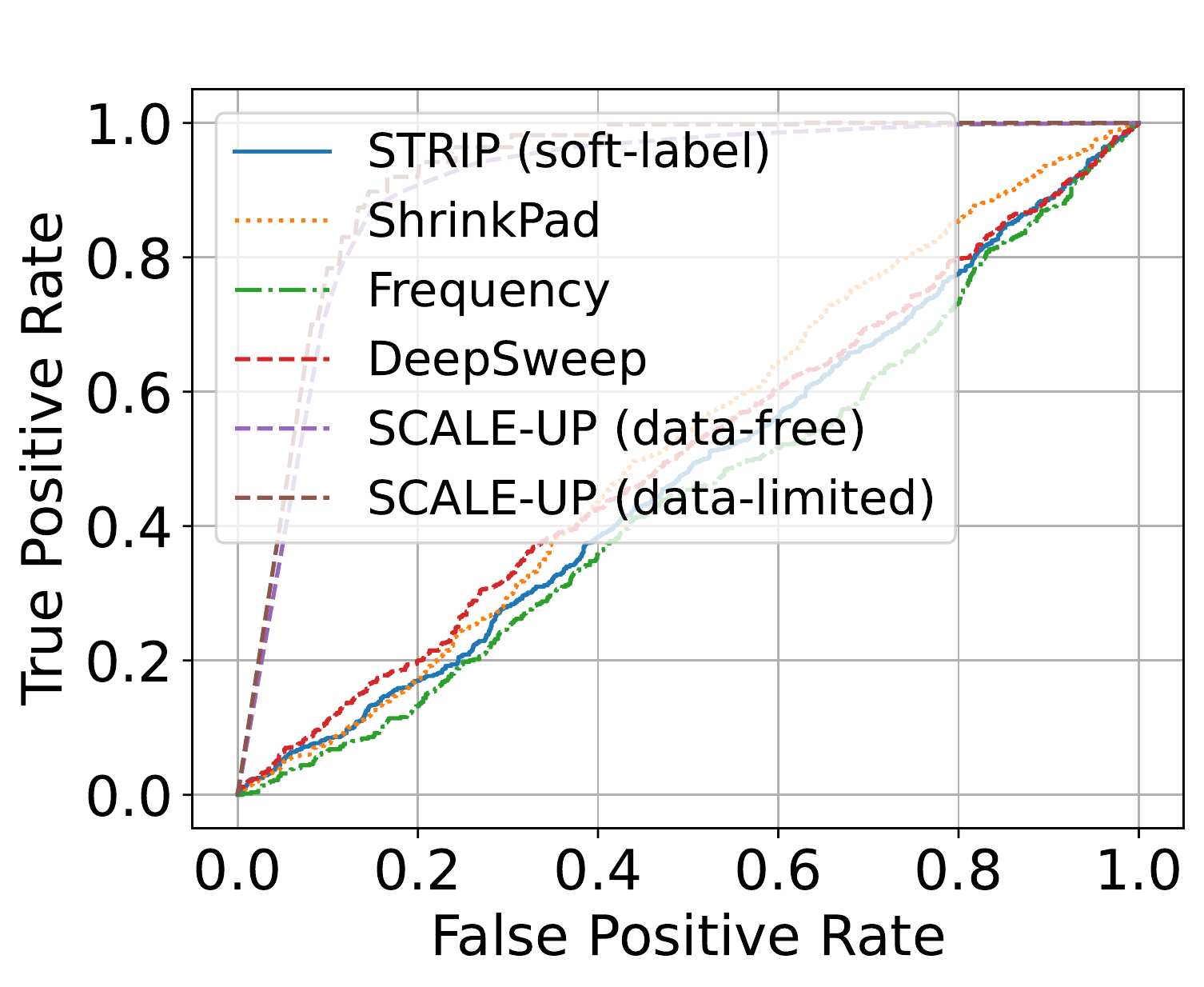}}
%\vspace{-2mm}
\caption{The ROC curves of defenses under each attack on CIFAR-10.}
\vspace{-3mm}
\label{fig:roc_curve_cifar}
\end{figure}

\begin{figure}[!ht]
\centering
\subfigure[BadNets\label{fig:badnets_roc}]{\includegraphics[width=0.45\textwidth]{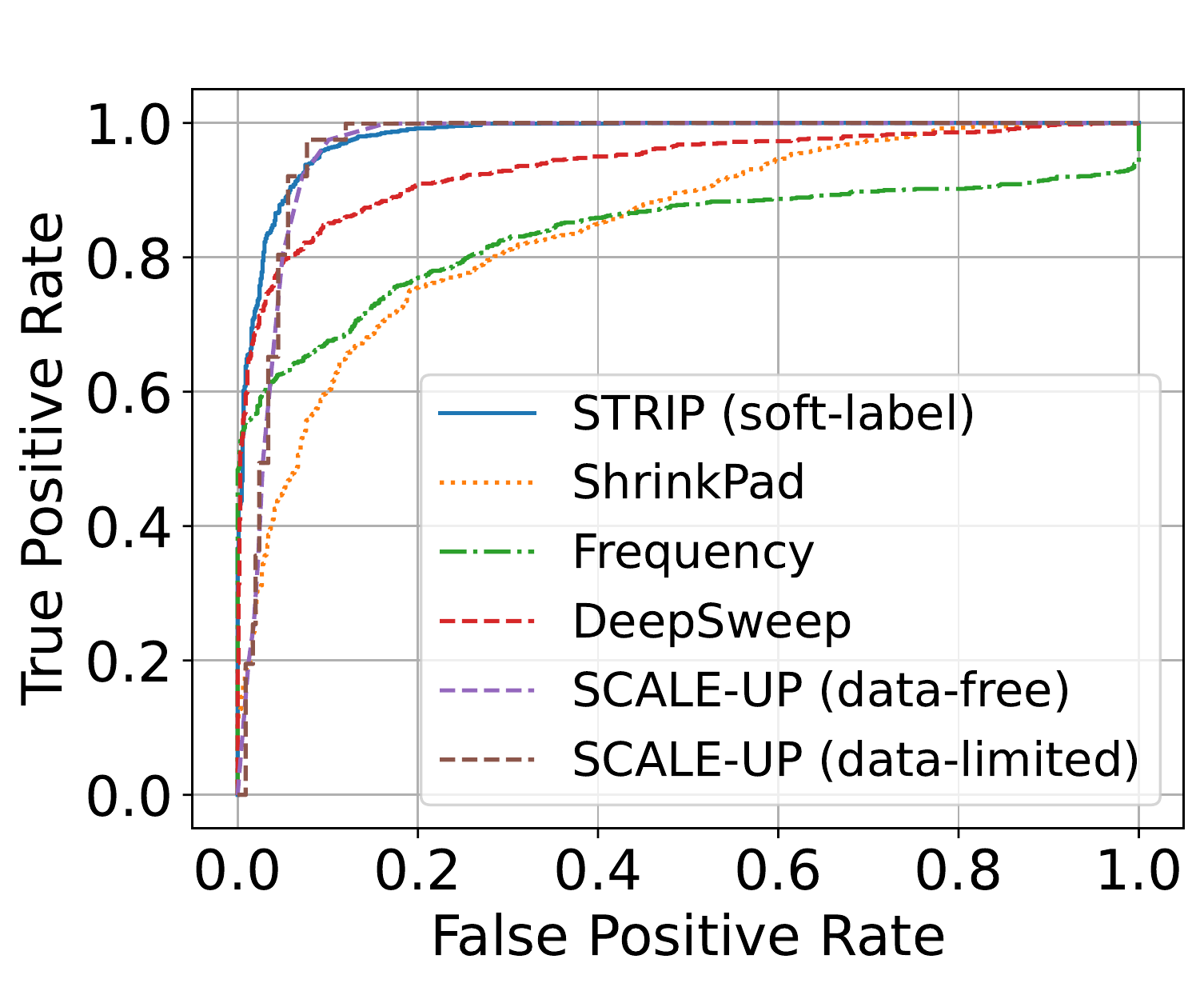}}\hspace{3em}
\subfigure[TUAP\label{tuap_roc}]{\includegraphics[width=0.45\textwidth]{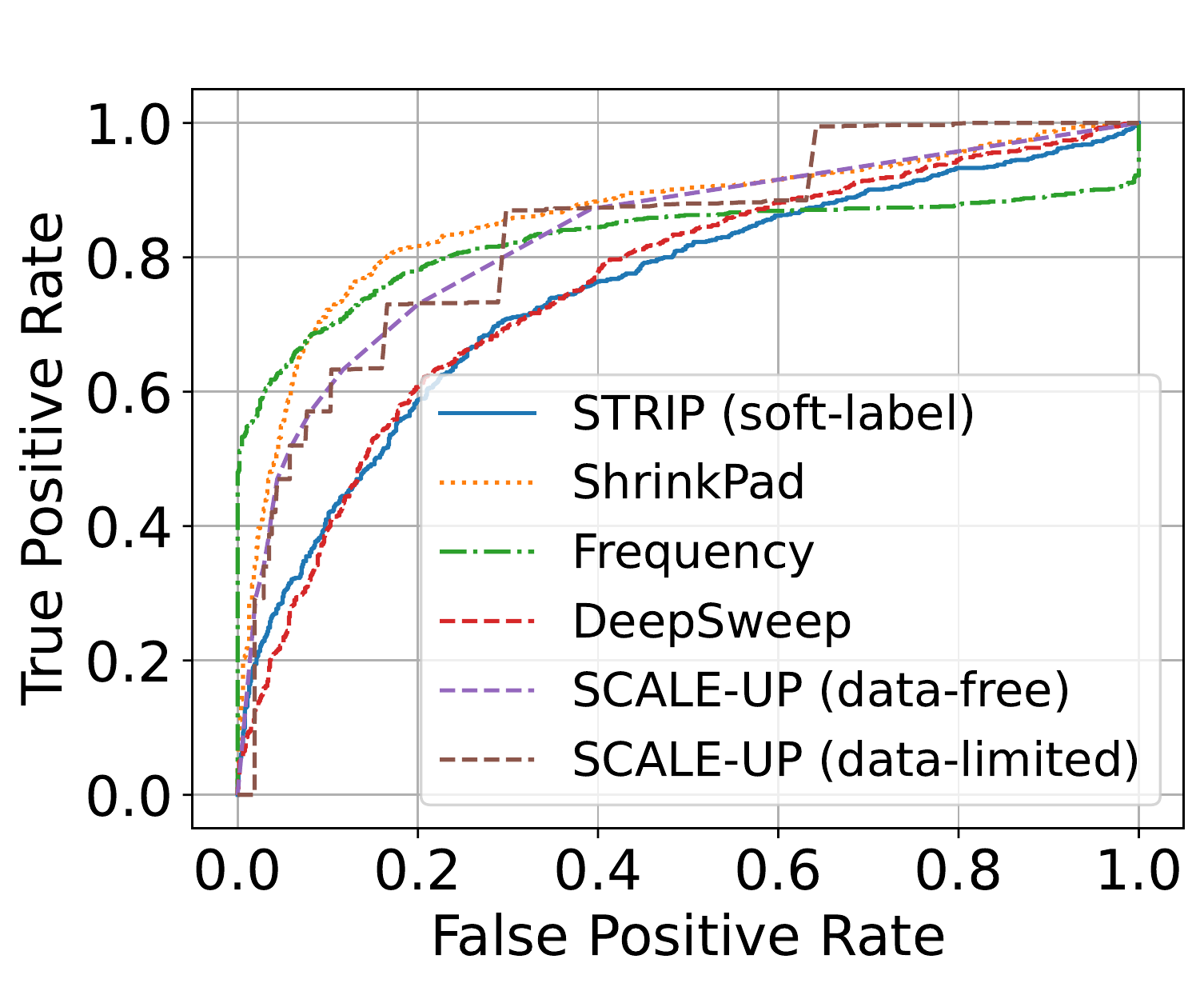}}
\subfigure[Label-Consistent\label{lc_roc}]{\includegraphics[width=0.45\textwidth]{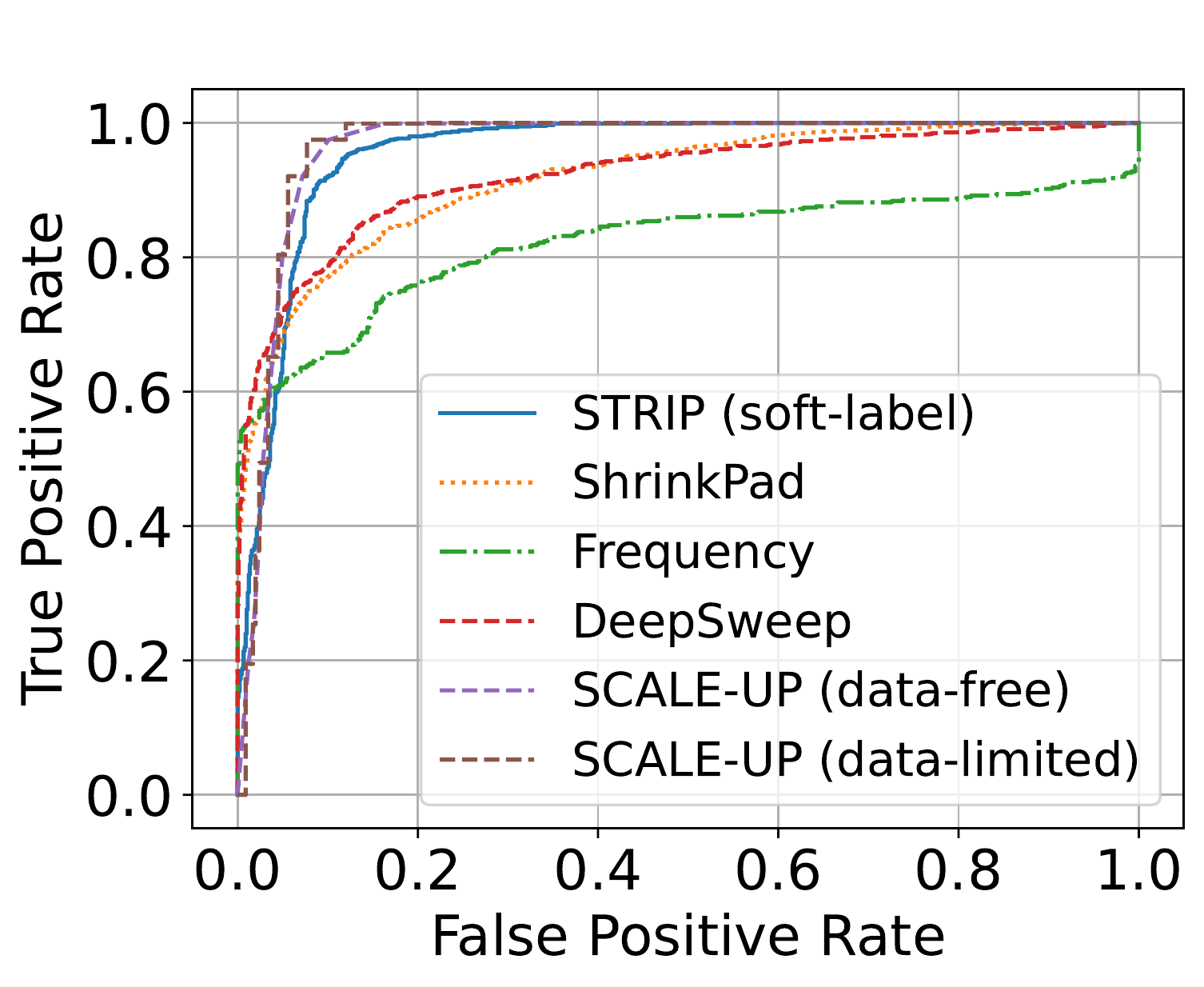}}\hspace{3em}
\subfigure[PhysicalBA\label{physicalBA_roc}]{\includegraphics[width=0.45\textwidth]{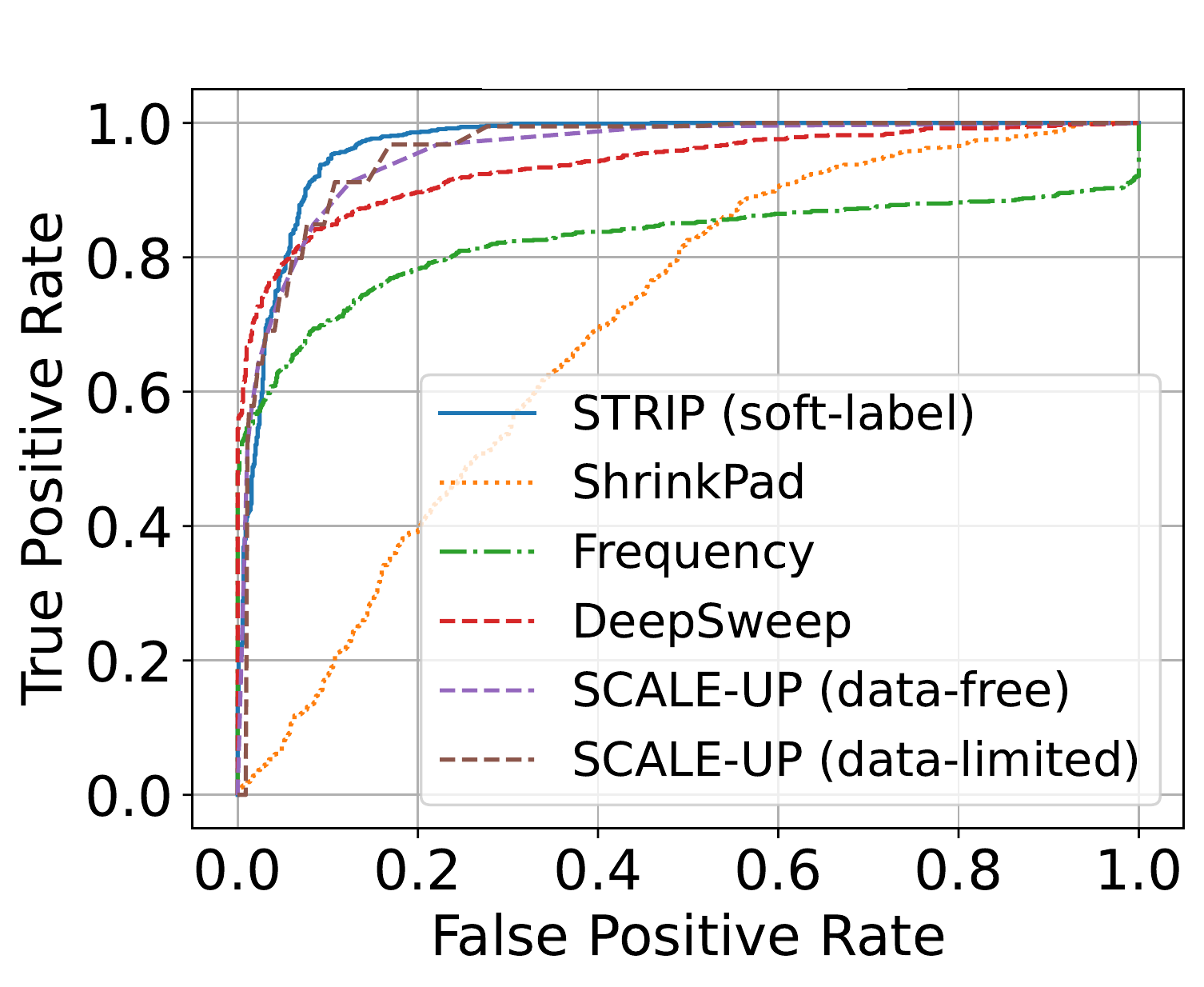}}
\subfigure[ISSBA\label{issba_roc}]{\includegraphics[width=0.45\textwidth]{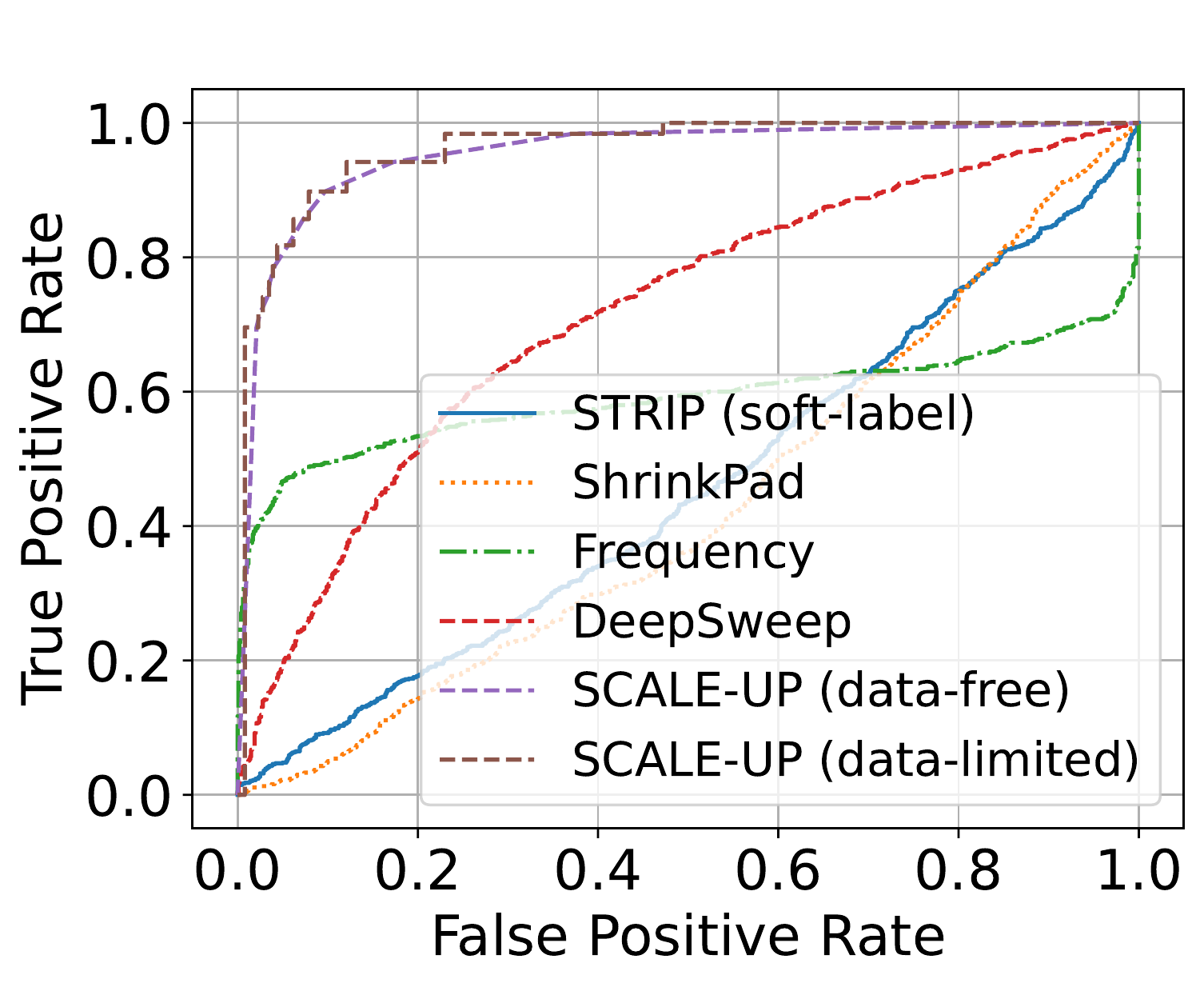}}\hspace{3em}
\subfigure[WaNet\label{wanet_roc}]{\includegraphics[width=0.45\textwidth]{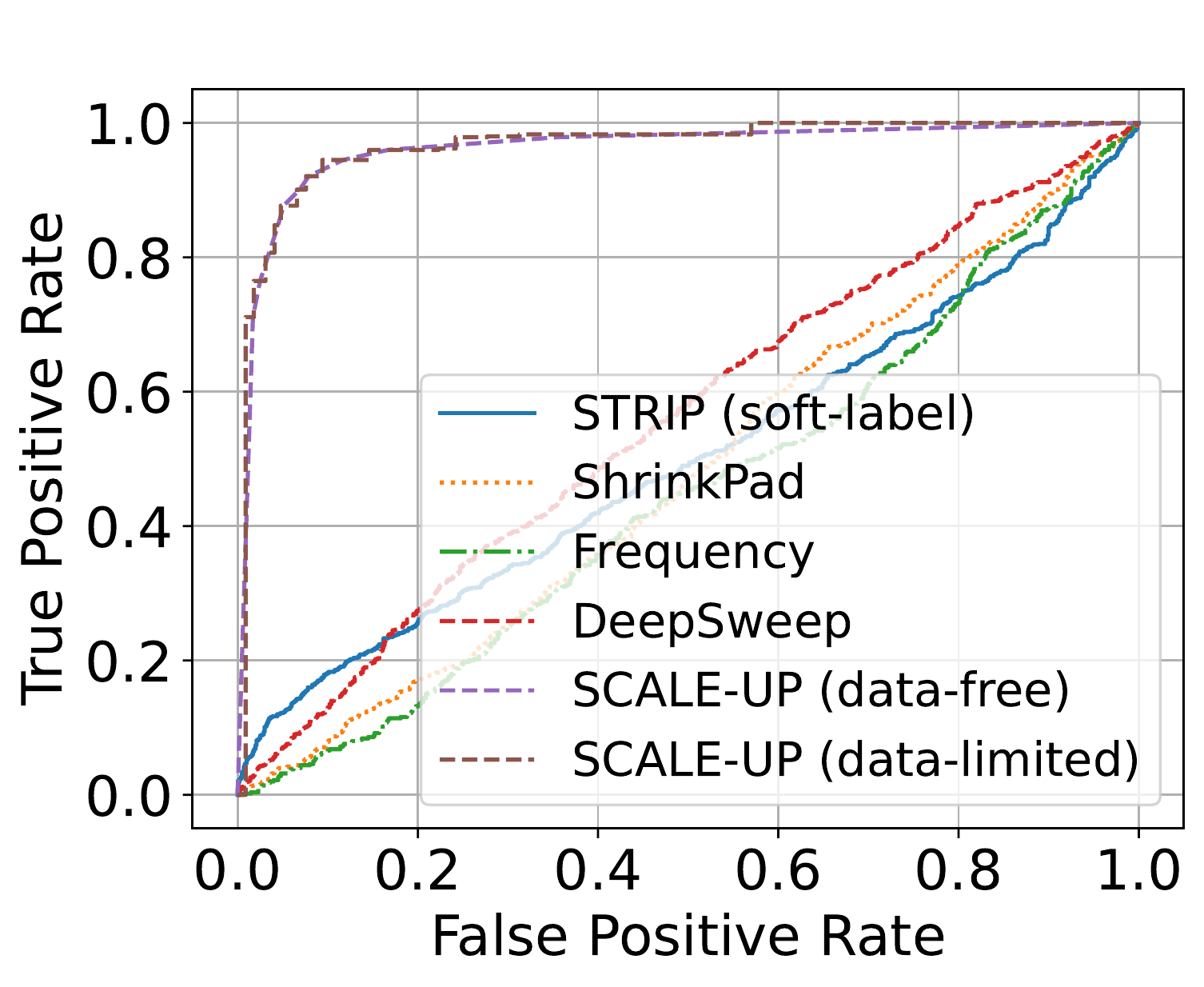}}
%\vspace{-2mm}
\caption{The ROC curves of defenses under each attack on Tiny ImageNet.}
\vspace{-3mm}
\label{fig:roc_curve_tiny}
\end{figure}

To better compare our method with baseline defenses, we also visualize the ROC curves of defenses (as shown in Figure \ref{fig:roc_curve_cifar}-\ref{fig:roc_curve_tiny}) under each attack on both CIFAR-10 and Tiny ImageNet in this section.

\begin{table}[!t]
    \centering
        \caption{The performance (AUROC) of SCALE-UP variants with random noises on CIFAR-10 and Tiny-ImageNet datasets. The failed cases ($i.e.$, AUROC $<0.55$) are marked in red.}
     \label{table:result_n}
         \scalebox{0.81}{
         \centering
        \begin{tabular}{c|c|cccccc|c}
\toprule
Dataset$\downarrow$                          & \tabincell{c}{Attack$\rightarrow$\\ Setting$\downarrow$} & BadNets & Label Consistent & PhysicalBA & TUAP  & WaNet & ISSBA & \textbf{Average} \\ \hline
\multirow{2}{*}{CIFAR-10}      & data-free      & 0.939   & 0.816            & 0.976      & 0.698 & \color{red}0.497 & \color{red}0.421 & 0.724   \\ \cline{2-9} 
                               & data-limited   & 0.939   & 0.873            & 0.981      & 0.706 & \color{red}0.432 & \color{red}0.444 & 0.729   \\ \hline
\multirow{2}{*}{Tiny ImageNet} & data-free      & 0.951   & 0.711            & 0.899      & 0.632 & \color{red}0.531 & \color{red}0.467 & 0.706   \\ \cline{2-9} 
                               & data-limited   & 0.951   & 0.761            & 0.899      & 0.644 & \color{red}0.534 & \color{red}0.501 & 0.706   \\ \bottomrule
\end{tabular}}
    %\vspace{-2mm}
    %\centering
    %\vspace{-2mm}
    
\end{table}
\section{Details for the Effectiveness of Scaling Process}
\label{ap:scale_eff}

Specifically, we design the SCALE-UP variant by replacing the scaling process with adding the same varied magnitudes of random noise to the given inputs. As shown in Table \ref{table:result_n}, using random noises is far less effective compared to the standard SCALE-UP methods, especially in detecting advanced attacks ($i.e.$, WaNet and ISSBA). We speculate that it is mostly because they adopted invisible full-image size trigger patterns and therefore the trigger-related features are less robust. Although we currently fail to provide theoretical analysis for the aforementioned phenomena, at least they verify the effectiveness of our scaling process. We will further discuss it in our future work.

\section{Potential Limitations and Future Work}
Our work is the first black-box label-only input-level backdoor detection and early-stage defenses under the black-box setting. Accordingly, we have to admit that our work still has some limitations.

Firstly, our defense requires that the attacked DNNs overfit their poisoned samples. This assumption or potential limitation is also revealed by our theoretical analysis in Section \ref{sec:motication}. In other words, if the attack success rate of a malicious model is relatively low, the detection performance of our SCALE-UP defense may degrade sharply. Secondly, we found that our SCALE-UP detection may fail when defending against attacks in some cases of simple tasks ($e.g.$, MNIST and GTSRB). We speculate that it is mostly because attacked DNNs also overfit to benign samples due to the lack of diversity and simplicity of the dataset, making them indistinguishable from some poisoned samples when analyzing the scaled prediction consistency. We will further explore the latent mechanisms of these limitations and alleviate them in our future work.

Besides, regarding another future direction of our methods, we intend to generalize and adopt them to more settings and applications, such as continual learning \citep{wzy}, non-transferable learning~\citep{wangnontransferable}, federate learning~\citep{fdl}, audio signal processing~\citep{zhaibackdoor,audio1,audio2}, and visual object tracking~\citep{lifew}. We will also evaluate our methods under other DNN structures (\eg, ViT \citep{maxvit} and GCN \citep{zxj}).     
%Since DNNs are widely deployed in a set of safety-critical applications(\eg, face recognition, autonomous driving, etc) under the Machine Learning as a Service(MLaaS) setting, where the deployed DNN is a fully black-box and the user/defender can only access the final prediction for each input. However, the threat of backdoor attacks to DNNs raises concern about third-party DNNs-driven applications. In this paper, we proposed a unified input-level black-box backdoor defense approach that can prevent suspicious backdoored samples during the inference phase. Our work is proposed to alleviate the security issues brought by backdoor attacks and we don't find any ethical issues in general. Notably, our approach can only provide robustness to DNNs in the post-training phase, while backdoor attacks can be implemented in other practical scenarios(\eg, training-phase, etc). Therefore, people should not be too optimistic about our approach.

\end{document}